\documentclass[prd,reprint,superscriptaddress,altaffillsymbol,amsmath,nofootinbib,longbibliography]{revtex4-1}

\usepackage{ifthen}
\usepackage{xcolor}
\usepackage{graphicx}
\usepackage{epsfig}
\usepackage{booktabs} 
\usepackage{natbib}
\usepackage{wasysym}
\usepackage{slashed} 
\usepackage{bbm}
\usepackage{mathrsfs}
\usepackage{amssymb}
\usepackage{dsfont}
\usepackage[export]{adjustbox}

\usepackage{color}

\definecolor{OurBlue}{rgb}{0.384,0.616,0.784}
\definecolor{OurRed}{rgb}{0.878,0.388,0.212}

\usepackage[
colorlinks=true, 
linkcolor=OurBlue,
citecolor=OurRed,
urlcolor=OurBlue,
]{hyperref}

\usepackage[compat=1.1.0]{tikz-feynman} 

\allowdisplaybreaks

\newcommand{\beqra}{\begin{eqnarray}}
\newcommand{\eeqra}{\end{eqnarray}}
\newcommand{\beq}{\begin{equation}}
\newcommand{\eeq}{\end{equation}}

\newcommand{\tr}{\mathrm{Tr}}

\renewcommand{\epsilon}{\varepsilon}

\renewcommand{\vec}[1]{\mathbf{#1}}
\renewcommand{\bar}{\overline}

\begin{document}

\title{\boldmath Linear response theory for light dark matter-electron scattering in materials}

\author{Riccardo Catena}
\email{catena@chalmers.se}
\affiliation{Chalmers University of Technology, Department of Physics, SE-412 96 G\"oteborg, Sweden}
\author{Nicola A. Spaldin}
\email{nicola.spaldin@mat.ethz.ch}
\affiliation{Department of Materials, ETH Z\"urich, CH-8093 Z\"urich, Switzerland}
\email{nicola.spaldin@mat.ethz.ch}

\begin{abstract}
We combine the non-relativistic effective theory of dark matter (DM) - electron interactions with linear response theory to obtain a formalism that fully accounts for screening and collective excitations in DM-induced electronic transition rate calculations for general DM-electron interactions.~In the same way that the response of a dielectric material to an external electric field in electrodynamics is described by the dielectric function, so in our formalism the response of a detector material to a DM perturbation is described by a set of {\it generalised susceptibilities} which can be directly related to densities and currents arising from the non-relativistic expansion of the Dirac Hamiltonian.~We apply our formalism to assess the sensitivity of non-spin-polarised detectors, and find that in-medium effects significantly affect the experimental sensitivity if DM couples to the detector's electron density, while being decoupled from other densities and currents.~Our formalism can be straightforwardly extended to the case of spin-polarised materials.
\end{abstract}

\maketitle

\section{Introduction}
The particles forming our Milky Way dark matter (DM) halo have so far stubbornly escaped detection.~A simple hypothesis that could explain this lack of detection is that the DM particle is lighter than the nucleons bound to atomic nuclei, and therefore too light to be directly detected with conventional methods based on the observation of rare nuclear recoils~\cite{Essig:2011nj}.~Indeed, an observable elastic nuclear recoil would require the incoming DM particle to carry a kinetic energy of a few keV or so, and thus to have a mass that lies above the 1 GeV threshold~\cite{Schumann:2019eaa}.~This hypothesis motivates the search for DM in electronic transitions induced by the scattering of Milky Way DM particles in detector materials, as these can be triggered by smaller energy depositions than nuclear recoils \cite{Mitridate:2022tnv}.  

Recent experimental proposals for the detection of DM particles with mass in the MeV to GeV range include the search for atomic ionisations in noble gas xenon and argon detectors~\cite{Essig:2011nj,Essig:2017kqs,DarkSide:2018ppu,Catena:2019gfa,Aprile:2019xxb,XENON:2020rca} and for electronic transitions in semiconductor crystals~\cite{Graham:2012su,Essig:2015cda,Derenzo:2016fse,Agnese:2018col,Kurinsky:2019pgb,DAMIC:2019dcn,EDELWEISS:2020fxc,SENSEI:2020dpa,Griffin:2020lgd,Catena:2021qsr,Griffin:2021znd,Knapen:2021run,Hochberg:2021pkt,Lasenby:2021wsc,Chen:2022pyd} as well as in superconductors~\cite{Hochberg:2015pha,Hochberg:2021yud} and 3D Dirac materials~\cite{Hochberg:2017wce,Geilhufe:2019ndy,Coskuner2021Jan}.~They also include the search for electron ejections from graphene layers~\cite{Hochberg:2016ntt,Catena:2023qkj} and carbon nanotubes~\cite{Cavoto:2019flp,Catena:2023awl}, as well as for excitations of collective phenomena such as phonons~\cite{Knapen:2017ekk,Trickle:2019nya} and magnons~\cite{Trickle:2019ovy}.~Further examples can be found in e.g. \cite{Kahn:2021ttr,Mitridate:2022tnv}.

The standard theoretical framework for assessing the potential of these proposals is the dark photon model, where the DM candidate, typically a spin 0 or 1/2 particle, couples to the electrically charged fermions of the Standard Model through the exchange of a ``heavy'' or ``light'' spin-1 mediator particle (i.e.~the dark photon)~\cite{Fayet:1980ad,Holdom:1985ag,Boehm:2003hm,Fayet:2004bw,Battaglieri:2017aum}.~In this context, a mediator is heavy (light) if the typical momentum transfer in a non-relativistic DM-electron scattering event is much smaller (larger) than its mass.~Within this framework, a critical theoretical input to the predicted rate of electronic transitions induced by the scattering of DM particles by the electrons bound to a given material is the overlap integral between the initial and final electron wave functions.~In the standard treatment of DM-induced atomic ionisations, the modulus squared of this integral is called the {\it atomic or ionisation form factor}, and has been computed using non-relativistic single-particle atomic wave functions~\cite{Kopp:2009et}, as well as accounting for many-body~\cite{Pandey:2018esq} and relativistic corrections~\cite{Roberts:2016xfw}.~In the case of DM-induced electronic transitions from the valence to the conduction band in crystals, the modulus squared of this overlap integral is called the {\it crystal form factor}, and has been computed in density functional theory (DFT) by expanding the Bloch states describing electrons in a crystal lattice in plane waves~\cite{Essig:2015cda}, in an atom-centered gaussian basis~\cite{Dreyer:2023ovn}, or by combining plane waves with atomic orbitals to capture higher momentum contributions \cite{Griffin:2021znd}.

An important observation that has been made recently is that, within the dark photon model, the rate of DM-induced electronic transitions in dielectric materials can be expressed in terms of the underlying dielectric function~\cite{Hochberg:2021pkt,Knapen:2021run}, i.e.~the linear response of a dielectric to an external electric field.~While the atomic/crystal form factor and dielectric function formalisms are in principle equivalent, the latter allows one to directly account for screening and collective excitation effects which would otherwise be missed by the former when electrons are described using a basis of single-particle states, and in-medium electron-electron interactions are neglected.~Notice that screening occurs when the electron density in the target material rearranges itself to partially cancel out the DM-electron interaction.~On the other hand, collective excitations occur when the momentum transferred from the DM particle to the medium is smaller than the inverse spacing between separate nuclei, or separate electrons, and the DM particle interacts with multiple particles in the target.

Going beyond the dark photon model, a variety of products of electron wave function overlap integrals can in principle contribute to the rate of DM-induced electronic transitions in materials.~We have proven this statement in a recent series of works~\cite{Catena:2023qkj,Catena:2023awl,Catena:2022fnk,Catena:2021qsr,Catena:2019gfa}, where we have used effective field theory (EFT) methods to describe the interaction between DM and electrons in materials.~EFT is a powerful method to address multi-scale physics problems involving a finite set of relevant degrees of freedom and known symmetries.~In the case of DM-electron scattering, there is a first separation of scales between the small momentum transfer in the scattering and the electron mass, and a second one between the non-relativistic DM speed in the Milky Way and the speed of light.~The relevant degrees of freedom are the DM particle and the electron, while their interactions are constrained by Galilean invariance, and momentum and energy conservation.~Combining these building blocks, EFT methods allowed us to write the amplitude for DM-electron scattering as a power series in the small momentum transfer to electron mass ratio, and DM speed to speed of light ratio.~This amplitude can describe virtually any model for sub-GeV DM in terms of a finite set of $S-$matrix elements.

Exploiting our EFT approach to DM-electron interactions, we have found that up to seven products of overlap integrals can appear in electron transition rate calculations.~These reduce to five in the case of crystals and within a simplified treatment of the local DM velocity distribution~\cite{Catena:2021qsr}.~They further reduce to four in the case of isolated atoms~\cite{Catena:2019gfa}, and to one when the final state electron is described by a plane wave~\cite{Catena:2023qkj,Catena:2023awl}.~While the framework we have developed in~\cite{Catena:2023qkj,Catena:2023awl,Catena:2022fnk,Catena:2021qsr,Catena:2019gfa} allows a rigorous description of previously intractable DM models, such as models where DM has an anapole or a magnetic dipole moment, it does not account for the aforementioned screening and collective excitation effects, as it does not include the many-body response of the remaining electrons to an electronic transition between two bound states.
~This makes it impossible to assess whether ``in-medium effects'' are important in the case of general DM-electron interactions.~Furthermore, it prevents us from properly modelling them in cases where they are actually significant.

The main purpose of this work is to extend the dielectric function formalism to the case of general DM-electron interactions in materials.~This will enable us to account for in-medium effects in theories beyond the dark photon model.~We achieve this goal through the following steps:
\begin{enumerate}
\item We start by identifying the electron densities and currents that a spin-1/2 DM particle can couple to in a material.~In the dark photon model, DM couples to the electron number density only.~In the case of general DM-electron interactions, we find that DM can couple to the electron number density, the paramagnetic current, the spin current, the scalar product of spin and paramagnetic current, and the Rashba spin-orbit current.~We then write down the time-dependent potential, $V_{\rm eff}^{ss'}(t)$ in Eq.~(\ref{eq:Vt}), which describes the scattering of DM particles by the bound electrons in any solid-state system in terms of these five densities and currents.
\item We apply linear response theory to calculate the response of a given material to the external, time dependent DM perturbation described by the potential $V_{\rm eff}^{ss'}(t)$.~As in electrodynamics the response of a dielectric material to an external electric field is described by the dielectric function, so in our formalism the response of a detector material to a DM perturbation is described by a set of {\it generalised susceptibilities}.~These susceptibilities are associated with the above densities and currents.
\item Using Fermi's golden rule, we  express the rate of DM-induced electronic transitions in detector materials in terms of our set of generalised susceptibilities. 
\item We  derive and solve a time-evolution equation for the generalised susceptibilities describing the response of a generic solid-state system to an external DM perturbation.~Focusing on non-spin-polarised and nearly isotropic materials, we evaluate the solution to this equation, and interpret it diagrammatically.
\item Combing the results from point 3) and point 4) above, we apply our formalism to reassess the sensitivity of hypothetical silicon and germanium detectors.
\end{enumerate}

The linear response theory for light DM direct detection we develop in this work enables us to study the impact of in-medium effects on electronic transition rate calculations in the presence of general DM-electron interactions.~Furthermore, it provides us with a framework where we can disentangle in a neat manner the solid state physics contribution, encoded in a set of generalised susceptibilities, from the astro- and particle physics inputs to the rate of DM-induced electronic excitations in materials.~While in this work we focus on materials used in existing detectors, our framework can straightforwardly be extended to the case of anisotropic materials, as well as to the case of spin-polarised detectors.

This work is organised as follows.~In Sec.~\ref{sec:theory}, we identify the densities and currents that DM can couple to in a material.~In Sec.~\ref{sec:formalism} we apply linear response theory to obtain the set of generalised susceptibilities describing the response of a solid-state system to a DM perturbation that couples to the aforementioned densities and currents.~We also provide an explicit expression for the rate of DM-induced electronic transitions in materials as a function of our generalised susceptibilities.~In Sec.~\ref{sec:evaluation} we derive and solve a time-evolution equation for the generalised susceptibilities identified in Sec.~\ref{sec:formalism}.~This equation enables us to perform explicit electronic transition rate calculations in the presence of general DM-electron interactions.~We apply our formalism to a sample of DM direct detection experiments in Sec.~\ref{sec:application} and conclude in Sec.~\ref{sec:conclusion}.

\section{Dark matter-electron scattering in materials}
\label{sec:theory}

\subsection{Free scattering amplitude in effective theories}
\label{sec:M}

In the non-relativistic effective theory of DM-electron interactions~\cite{Catena:2019gfa}, the amplitude for DM scattering by a free electron, $\mathcal{M}$, can be expressed in terms of the DM particle and electron spin operators, $\mathbf{S}_\chi$ and $\mathbf{S}_e$, respectively, the momentum transfer $\mathbf{q}$ and the transverse relative velocity $\mathbf{v}_{\rm el}^\perp$, i.e.~the component of the relative DM-electron velocity that is perpendicular to $\mathbf{q}$ when the scattering is elastic
\begin{align}
\mathbf{v}_{\rm el}^\perp \equiv   \mathbf{v}^{\perp}_\chi + \mathbf{v}^{\perp}_e \equiv \left( \frac{\mathbf{p} + \mathbf{p}'}{2 m_\chi} \right)  - \left( \frac{\mathbf{k} + \mathbf{k}'}{2 m_e} \right) \,.
\end{align}
Here $\mathbf{k}$ ($\mathbf{k}'$) is the initial (final) electron momentum, while $\mathbf{p}$ ($\mathbf{p}'$) is the momentum of the incoming (outgoing) DM particle.~The electron mass and DM particle mass are denoted by $m_e$ and $m_\chi$, respectively.~In the case of spin-1/2 DM, $\mathbf{S}_\chi=\boldsymbol{\sigma}_\chi/2$ and $\mathbf{S}_e=\boldsymbol{\sigma}_e/2$, where the components of the three-dimensional vectors $\boldsymbol{\sigma}_\chi$ and $\boldsymbol{\sigma}_e$ consist of the three Pauli matrices, and the indexes $\chi$ and $e$ identify the DM particle or electron spin, respectively.~For this choice of DM particle spin, 
and to first order in $\mathbf{v}_{\rm el}^\perp$, the amplitude for non-relativistic DM-electron scattering is \cite{Catena:2019gfa}
\begin{equation}
 \label{eq:Mnr0}
\mathcal{M} 
= \sum_i \left(c_i^s +c^\ell_i \frac{q_{\rm ref}^2}{|\mathbf{q}|^2} \right) \,\langle \mathcal{O}_i \rangle  \,,
\end{equation}
where the interaction operators $\mathcal{O}_i$ are defined in Tab.~\ref{tab:operators}, and $q_{\rm ref}\equiv \alpha m_e$ is a reference momentum,  
with $\alpha$ the fine structure constant.~We denote the coupling constants of the $i$-th operator in Tab.~\ref{tab:operators} by $c_i^s$ and $c_i^\ell$, where $c^s_i\neq 0$ and $c^\ell_i=0$ corresponds to the case of interactions mediated by a heavy particle, while $c^s_i=0$ and $c^\ell_i\neq0$ refers to the case of a light mediator.~Angle brackets in the amplitude $\mathcal{M}$ denote matrix elements between the two-component spinors $\xi_\chi^s$ and $\xi_\chi^{s'}$ for the DM particle, and $\xi_e^r$ and $\xi_e^{r'}$ for the electron.~For example, in the case of $\mathcal{O}_4$,
\begin{align}
\langle \mathcal{O}_4 \rangle \equiv \xi_\chi^{s'\dagger} \frac{\boldsymbol{\sigma}_\chi}{2} \xi_\chi^s \, \xi_e^{r'\dagger} \frac{\boldsymbol{\sigma}_e}{2} \xi_e^r \,.
\end{align}
By promoting the coupling constants $c^s_i$ and $c^\ell_i$ to functions of the momentum transfer, virtually any model for DM-electron interactions can be matched onto the free scattering amplitude in Eq.~(\ref{eq:Mnr0}) in the non-relativistic limit.

\begin{table}[t]
    \centering
    \begin{tabular*}{\columnwidth}{@{\extracolsep{\fill}}ll@{}}
    \toprule
      $\mathcal{O}_1 = \mathds{1}_{\chi}\mathds{1}_{e}$ & $\mathcal{O}_9 = i\mathbf{S}_\chi\cdot\left(\mathbf{S}_e\times\frac{ \mathbf{q}}{m_e}\right)$  \\
        $\mathcal{O}_3 = i\mathbf{S}_e\cdot\left(\frac{ \mathbf{q}}{m_e}\times \mathbf{v}^{\perp}_{\rm el}\right)
        \mathds{1}_{\chi}$ &   $\mathcal{O}_{10} = i\mathbf{S}_e\cdot\frac{ \mathbf{q}}{m_e}\mathds{1}_{\chi}$   \\
        $\mathcal{O}_4 = \mathbf{S}_{\chi}\cdot \mathbf{S}_e$ &   $\mathcal{O}_{11} = i\mathbf{S}_\chi\cdot\frac{ \mathbf{q}}{m_e}\mathds{1}_{e}$   \\                                                                             
        $\mathcal{O}_5 = i\mathbf{S}_\chi\cdot\left(\frac{ \mathbf{q}}{m_e}\times \mathbf{v}^{\perp}_{\rm el}\right)
        \mathds{1}_{e}$ &  $\mathcal{O}_{12} = \mathbf{S}_{\chi}\cdot \left(\mathbf{S}_e \times \mathbf{v}^{\perp}_{\rm el} \right)$ \\                                                                                                                 
        $\mathcal{O}_6 = \left(\mathbf{S}_\chi\cdot\frac{ \mathbf{q}}{m_e}\right) \left(\mathbf{S}_e\cdot\frac{\hat{{\bf{q}}}}{m_e}\right)$ &  $\mathcal{O}_{13} =i \left(\mathbf{S}_{\chi}\cdot  \mathbf{v}^{\perp}_{\rm el}\right)\left(\mathbf{S}_e\cdot \frac{ \mathbf{q}}{m_e}\right)$ \\   
        $\mathcal{O}_7 = \mathbf{S}_e\cdot  \mathbf{v}^{\perp}_{\rm el} \mathds{1}_{\chi}$ &  $\mathcal{O}_{14} = i\left(\mathbf{S}_{\chi}\cdot \frac{ \mathbf{q}}{m_e}\right)\left(\mathbf{S}_e\cdot  \mathbf{v}^{\perp}_{\rm el}\right)$  \\
        $\mathcal{O}_8 = \mathbf{S}_{\chi}\cdot  \mathbf{v}^{\perp}_{\rm el} \mathds{1}_{e}$  & $\mathcal{O}_{15} = i\mathcal{O}_{11}\left[ \left(\mathbf{S}_e\times  \mathbf{v}^{\perp}_{\rm el} \right) \cdot \frac{ \mathbf{q}}{m_e}\right] $ \\       
    \bottomrule
    \end{tabular*}
    \caption{Operators defining the non-relativistic effective theory of spin 1/2 DM-electron interactions~\cite{Catena:2019gfa} (see~\cite{Fan:2010gt,Anand:2013yka} for the case of nucleons).~$\mathbf{S}_e$ ($\mathbf{S}_\chi$) is the electron (DM) spin,  $\mathds{1}_{e}$ ($\mathds{1}_{\chi}$) the identity in the electron (DM) spin space, $\mathbf{q}$ the momentum transfer, and $\mathbf{v}_{\rm el}^\perp$ the relative DM-electron velocity component that is perpendicular to $\mathbf{q}$ when the scattering is elastic.}
\label{tab:operators}
\end{table}

Inspection of Tab.~\ref{tab:operators} shows that, after ``factorising out the electronic contribution'', Eq.~(\ref{eq:Mnr0}) can be rewritten as follows
\begin{align}
\mathcal{M}=&F_0^{ss'} \xi_e^{r'\dagger}\mathds{1}_e\xi_e^r \nonumber \\
&+ F_A^{ss'} \left(\frac{\mathbf{k} + \mathbf{k}' }{2 m_e}\right) \cdot \xi_e^{r'\dagger}\boldsymbol{\sigma}_e \xi_e^r \nonumber \\
&+ \mathbf{F}_5^{ss'} \cdot \xi_e^{r'\dagger}\boldsymbol{\sigma}_e \xi_e^r \nonumber \\
&+ \mathbf{F}_M^{ss'} \cdot \left(\frac{\mathbf{k} + \mathbf{k}' }{2 m_e}\right) \xi_e^{r'\dagger}\mathds{1}_e \xi_e^r \nonumber\\
&+ \mathbf{F}_E^{ss'} \cdot \left(-i\,\frac{\mathbf{k} + \mathbf{k}' }{2 m_e} \times \xi_e^{r'\dagger}\boldsymbol{\sigma}_e \xi_e^r \right) 
\label{eq:Mnr} \,,
\end{align}
where $\mathds{1}_e$ is the $2\times2$ identity matrix in the electron spin space.~We provide explicit expressions for the ``partial amplitudes'' $F_0^{ss'}$, $F_A^{ss'}$, $\mathbf{F}_5^{ss'}$, $\mathbf{F}_M^{ss'}$, $\mathbf{F}_E^{ss'}$ in App.~\ref{sec:BB}.~They depend on $c_i^s$ and $c_i^\ell$, the momentum transfer, the initial DM velocity, and the initial (final) DM spin configuration $s$ ($s'$).~Notice that the operator $\mathcal{O}_{13}$ in Tab.~\ref{tab:operators} constitutes an exception to the factorisation in Eq.~(\ref{eq:Mnr}).~This follows from 
\begin{align}
\mathcal{O}_{13} =&~i \left(\mathbf{S}_{\chi}\cdot  \mathbf{v}^{\perp}_{\chi}\right)\left(\mathbf{S}_e\cdot \frac{ \mathbf{q}}{m_e}\right)
+ i \left( \mathbf{v}^{\perp}_{e} \times  \mathbf{S}_e \right) \left( \mathbf{S}_\chi \times  \frac{\vec q}{m_e}\right) \nonumber\\
&+i \left( \mathbf{S}_e \cdot \mathbf{S}_\chi \right) \left(  \mathbf{v}^{\perp}_{e} \cdot  \frac{\vec q}{m_e} \right) \,.
\label{eq:O13}
\end{align}
While the first (second) term in Eq.~(\ref{eq:O13}) would contribute to the third (last) line in Eq.~(\ref{eq:Mnr}), the term in the last line of Eq.~(\ref{eq:O13}) would generate a new tensor in Eq.~(\ref{eq:Mnr}), namely $(\mathbf{S}_{el})_l (\mathbf{v}^{\perp}_{\rm el})_m$, $l,m=1,2,3$ because $\mathbf{v}^{\perp}_{e} \cdot \vec q$  is in general not zero in the inelastic DM-electron scattering.~However, since the operator $\mathcal{O}_{13}$ only arises at next-to-leading order in the non-relativistic reduction of simplified models~\cite{Catena:2022fnk}, we prefer not to introduce an additional tensor specific to the operator $\mathcal{O}_{13}$ and set simply $c_{13}^s=c_{13}^\ell=0$ in Eq.~(\ref{eq:Mnr0}).

\subsection{Effective potential: free electrons}
\label{sec:Veff}
In this section, we derive an explicit relation between the non-relativistic scattering amplitude in Eq.~(\ref{eq:Mnr}), $\mathcal{M}$, and the associated potential, $\widehat{V}$.~To this end, we start by noticing that the matrix element of $\widehat{V}$ between two DM-electron states $|\phi\rangle$ and $|\psi\rangle$,  $\langle \phi | \widehat{V} | \psi \rangle$, can be written as 
\begin{align}
\langle \phi | \widehat{V} | \psi \rangle &= \int {\rm d} \mathbf{r}_e\int {\rm d} \mathbf{r}'_e \int {\rm d} 
\mathbf{r}_\chi\int {\rm d} \mathbf{r}'_\chi \, \langle \phi | \mathbf{r}'_e, \mathbf{r}'_\chi \rangle \,\widehat{V}_{\mathcal{X}} \nonumber\\ 
&\times 
\langle  \mathbf{r}_e, \mathbf{r}_\chi | \psi \rangle \,,
\label{eq:mel}
\end{align}
where 
\begin{align}
\widehat{V}_{\mathcal{X}} \equiv \langle  \mathbf{r}'_e, \mathbf{r}'_\chi | \widehat{V} | \mathbf{r}_e, \mathbf{r}_\chi \rangle 
\label{eq:Vchi}
\end{align} 
while the one-particle states $|\mathbf{r}_e\rangle$  and $|\mathbf{r}'_e\rangle$, and $|\mathbf{r}_\chi\rangle$ and $|\mathbf{r}'_\chi\rangle$, are eigenstates of the electron and DM particle position operators, respectively.~Furthermore, we notice that 
\begin{align}
\widehat{V}_{\mathcal{X}} = \sum_{\mathbf{k},\mathbf{k}'}\sum_{\mathbf{p},\mathbf{p}'}\sum_{ss'}\sum_{rr'}
& \langle \mathbf{r}'_e, \mathbf{r}'_\chi | \mathbf{k}',r';\mathbf{p}',s'  \rangle \nonumber\\
&\times \langle \mathbf{k}',r';\mathbf{p}',s' | \widehat{V} 
| \mathbf{k},r;\mathbf{p},s \rangle \nonumber \\[3pt]
&\times \langle \mathbf{k},r;\mathbf{p},s | \mathbf{r}_e, \mathbf{r}_\chi \rangle \,,
\label{eq:Vdis}
\end{align}
where we introduced a complete set of one-particle states labelled by the initial (final) electron and DM particle momenta, $\mathbf{k}$ ($\mathbf{k}'$) and $\mathbf{p}$ ($\mathbf{p}'$), as well as by the initial and final electron and DM particle spins, $r$ ($r'$) and $s$ ($s'$), respectively.~Taking the continuum limit in Eq.~(\ref{eq:Vdis}), that is
\begin{align}
\frac{1}{V}\sum_{\mathbf{k}} \longrightarrow \frac{1}{(2\pi)^3}\int{\rm d} \mathbf{k}\,,
\end{align}
and evaluating $\langle \mathbf{k}',r';\mathbf{p}',s' | \widehat{V} 
| \mathbf{k},r;\mathbf{p},s \rangle$ in the Born approximation,
\begin{align}
\langle \mathbf{k}',r';\mathbf{p}',s' | \widehat{V} 
| \mathbf{k},r;\mathbf{p},s \rangle =& -\frac{\mathcal{M}}{4 m_e m_\chi V^2} \nonumber\\
&\times(2\pi)^3 \delta^{(3)}(\mathbf{k}'+\mathbf{p}'-\mathbf{k}-\mathbf{p}) \,,
\end{align}
we finally obtain,
\begin{align}
\widehat{V}_{\mathcal{X}} =-& \int \frac{{\rm d} \mathbf{k}}{(2\pi)^3}\int \frac{{\rm d} \mathbf{k}'}{(2\pi)^3}
\int \frac{{\rm d} \mathbf{p}}{(2\pi)^3}\int \frac{{\rm d} \mathbf{p}'}{(2\pi)^3} \,e^{i\mathbf{k}'\cdot \mathbf{r}'_e}e^{i\mathbf{p}'\cdot \mathbf{r}'_\chi}\nonumber\\[5pt]
& \times \widetilde{\mathcal{M}} \,e^{-i \mathbf{k}\cdot \mathbf{r}_e}e^{-i \mathbf{p}\cdot \mathbf{r}_\chi} \nonumber\\
&\times (2\pi)^3 \delta^{(3)}(\mathbf{k}'+\mathbf{p}'-\mathbf{k}-\mathbf{p}) \,.
\label{eq:Vcon}
\end{align}
where 
\begin{align}
\widetilde{\mathcal{M}} \equiv \sum_{ss'}\sum_{rr'}  \xi_e^{r'} \xi_\chi^{s'}\frac{\mathcal{M}}{4 m_e m_\chi} \xi_\chi^{s\dagger}\xi_e^{r\dagger}
\end{align}
and we made use of the position representation wave functions
\begin{align}
\langle \mathbf{r}'_e | \mathbf{k}',r'\rangle &= \frac{1}{\sqrt{V}} \,e^{i \mathbf{k}'\cdot \mathbf{r}'_e} \xi_e^{r'}\nonumber\\
\langle \mathbf{r}'_\chi | \mathbf{p}',s'\rangle &= \frac{1}{\sqrt{V}} \,e^{i \mathbf{p}'\cdot \mathbf{r}'_\chi} \xi_\chi^{s'}\nonumber\\
\langle \mathbf{k} , r | \mathbf{r}_e \rangle &= \frac{1}{\sqrt{V}} \,e^{- i \mathbf{k}\cdot \mathbf{r}_e} \xi_e^{r\dagger}\nonumber\\
\langle \mathbf{p} , s | \mathbf{r}_\chi \rangle &= \frac{1}{\sqrt{V}} \,e^{- i \mathbf{p}\cdot \mathbf{r}_\chi}\xi_\chi^{s\dagger} \,,
\label{eq:wavefunctions}
\end{align}
where $V=(2\pi)^2\delta^{(3)}(0)$ is the spatial volume and $\xi_\chi^s$, $\xi_\chi^{s' \dagger}$ and $\xi_e^r$, $\xi_e^{r' \dagger}$ are two-component spinors for the DM particle and electron, respectively.~For local interactions, one has
\begin{align}
\widehat{V}_{\mathcal{X}} = \widehat{V}_{\mathcal{X}}(\mathbf{r}_e,\mathbf{r}_\chi) \delta^{(3)}(\mathbf{r}_e-\mathbf{r}'_e)
 \delta^{(3)}(\mathbf{r}_\chi-\mathbf{r}'_\chi)
\label{eq:loc}
\end{align}
where
\begin{align}
\widehat{V}_{\mathcal{X}}(\mathbf{r}_e,\mathbf{r}_\chi) \equiv \frac{1}{\mathcal{N}^2} \langle  \mathbf{r}_e, \mathbf{r}_\chi | \widehat{V} | \mathbf{r}_e, \mathbf{r}_\chi \rangle 
\end{align}
and $\mathcal{N}=\delta^{3}(0)$.~In this particular case, Eq.~(\ref{eq:Vcon}) reduces to
\begin{align}
\widehat{V}_{\mathcal{X}}(\mathbf{r}_e,\mathbf{r}_\chi) =& -\frac{1}{\mathcal{N}^2} 
\int \frac{{\rm d} \mathbf{k}}{(2\pi)^3}\int \frac{{\rm d} \mathbf{k}'}{(2\pi)^3}
\int \frac{{\rm d} \mathbf{p}}{(2\pi)^3}\int \frac{{\rm d} \mathbf{p}'}{(2\pi)^3} \nonumber\\[5pt]
& \times e^{i\mathbf{k}'\cdot \mathbf{r}_e}e^{i\mathbf{p}'\cdot \mathbf{r}_\chi}\, \widetilde{\mathcal{M}} \,e^{-i \mathbf{k}\cdot \mathbf{r}_e}e^{-i \mathbf{p}\cdot \mathbf{r}_\chi} \nonumber\\
&\times (2\pi)^3 \delta^{(3)}(\mathbf{k}'+\mathbf{p}'-\mathbf{k}-\mathbf{p}) \,.
\label{eq:Vcon2}
\end{align}  
As one can see from Eq.~(\ref{eq:loc}), the only local interactions in Tab.~\ref{tab:operators} are $\mathcal{O}_1$ and $\mathcal{O}_4$, as all other interaction operators involve combinations of particle velocities.~The potential associated with a non-local interaction is in general a function of $\mathbf{r}_e$, $\mathbf{r}'_e$, $\mathbf{r}_\chi$ and $\mathbf{r}'_\chi$, as shown in Eq.~(\ref{eq:Vchi}).~As a first application of Eqs.~(\ref{eq:Vcon}) and (\ref{eq:Vcon2}), we now focus on the case of the local interaction $\mathcal{O}_1$.~In this example, the amplitude for DM-electron scattering can be written as follows
\begin{align}
\mathcal{M}=c_1\,\delta^{ss'}\delta^{rr'} \,,
\end{align}
and Eq.~(\ref{eq:Vcon2}) yields
\begin{align}
\widehat{V}_{\mathcal{X}}(\mathbf{r}_e,\mathbf{r}_\chi) = - \frac{c_1}{4m_e m_\chi}\, \delta^{(3)}(\mathbf{r}_e-\mathbf{r}_\chi)\, \mathds{1}_\chi \mathds{1}_e \,.
\label{eq:Vc1}
\end{align}
Let us next turn our attention to the non-local interaction operator $\mathcal{O}_7$.~The amplitude for DM-electron scattering is now
\begin{align}
\mathcal{M}=c_{7}\,\left[\frac{(\mathbf{p}+\mathbf{p}')}{2 m_\chi}-\frac{(\mathbf{k}+\mathbf{k}')}{2 m_e} \right]\cdot  \xi_e^{r'\dagger} \boldsymbol{\sigma}_e \xi_e^r \,\delta^{ss'} \,,
\end{align}
By applying Eq.~(\ref{eq:Vcon}) to this amplitude, for the operator $\widehat{V}_{\mathcal{X}}$ we find 
\begin{align}
\widehat{V}_{\mathcal{X}} = \frac{c_7}{4m_e m_\chi}\,\Bigg\{\frac{1}{2m_\chi}\Bigg[ &i\nabla_{\mathbf{r}'_\chi}\delta^{(3)}(\mathbf{r}'_\chi-\mathbf{r}'_e) \delta^{(3)}(\mathbf{r}'_e-\mathbf{r}_\chi) \nonumber\\ 
- &i\nabla_{\mathbf{r}_\chi} \delta^{(3)}(\mathbf{r}'_e-\mathbf{r}_\chi)\delta^{(3)}(\mathbf{r}'_\chi-\mathbf{r}'_e)\Bigg] \nonumber\\ 
\times&\delta^{(3)}(\mathbf{r}'_e-\mathbf{r}_e) \nonumber\\
-\frac{1}{2m_e}\Bigg[ &i\nabla_{\mathbf{r}'_e}\delta^{(3)}(\mathbf{r}'_e-\mathbf{r}'_\chi) \delta^{(3)}(\mathbf{r}'_\chi-\mathbf{r}_e) \nonumber\\ 
- &i\nabla_{\mathbf{r}_e} \delta^{(3)}(\mathbf{r}'_\chi-\mathbf{r}_e)\delta^{(3)}(\mathbf{r}'_e-\mathbf{r}'_\chi)\Bigg] \nonumber\\ 
\times&\delta^{(3)}(\mathbf{r}'_\chi-\mathbf{r}_\chi) \Bigg\} \cdot  \boldsymbol{\sigma}_e\mathds{1}_\chi
\label{eq:Vc7}
\end{align}
While the interaction in Eq.~(\ref{eq:Vc7}) is formally non-local, in the evaluation of matrix elements it is equivalent to a potential of the type $\widehat{V}_{\mathcal{X}} = \widehat{V}_{\mathcal{X}}(\mathbf{r}_e,\mathbf{r}_\chi) \delta^{(3)}(\mathbf{r}_e-\mathbf{r}'_e)
 \delta^{(3)}(\mathbf{r}_\chi-\mathbf{r}'_\chi)$ with
\begin{align}
\widehat{V}_{\mathcal{X}}(\mathbf{r}_e,\mathbf{r}_\chi) = \frac{c_7 \mathds{1}_\chi}{4 m_e m_\chi} \Bigg\{\frac{-i}{2m_\chi}&\Bigg[
\overleftarrow{\nabla}_{\mathbf{r}_\chi}\cdot \boldsymbol{\sigma}_e\, \delta^{(3)}(\mathbf{r}_e-\mathbf{r}_\chi) \nonumber\\
& -  \delta^{(3)}(\mathbf{r}_e-\mathbf{r}_\chi)\,\boldsymbol{\sigma}_e \cdot \overrightarrow{\nabla}_{\mathbf{r}_\chi}
\Bigg]    \nonumber\\
+ \frac{i}{2m_e} &\Bigg[
\overleftarrow{\nabla}_{\mathbf{r}_e}\cdot \boldsymbol{\sigma}_e\, \delta^{(3)}(\mathbf{r}_e-\mathbf{r}_\chi) \nonumber\\
& -  \delta^{(3)}(\mathbf{r}_e-\mathbf{r}_\chi)\,\boldsymbol{\sigma}_e \cdot \overrightarrow{\nabla}_{\mathbf{r}_e}
\Bigg] \Bigg\}
\label{eq:Vc72}
\end{align}
if we impose that $\overrightarrow{\nabla}_{\mathbf{r}_e}$ ($\overleftarrow{\nabla}_{\mathbf{r}_e}$) only acts on the initial (final) electron wave function and $\overrightarrow{\nabla}_{\mathbf{r}_\chi}$ ($\overleftarrow{\nabla}_{\mathbf{r}_\chi}$) on the initial (final) DM particle wave function.~In order to show the equivalence of the two expressions for potential, Eq.~(\ref{eq:Vc7}) and Eq.~(\ref{eq:Vc72}), we set $|\psi\rangle= | \mathbf{k},r;\mathbf{p},s \rangle $ and $|\phi\rangle=| \mathbf{k}',r';\mathbf{p}',s' \rangle $ in Eq.~(\ref{eq:mel}) and then calculate the matrix element $\langle \phi | \widehat{V} | \psi \rangle$ in two ways.~In the first one, we assume that $\widehat{V}_{\mathcal{X}}$ is given by Eq.~(\ref{eq:Vc7}).~In the second one, we take $\widehat{V}_{\mathcal{X}}$ from Eq.~(\ref{eq:loc}) and set $\widehat{V}_{\mathcal{X}}(\mathbf{r}_e, \mathbf{r}_\chi)$ as in Eq.~(\ref{eq:Vc72}).~We find that the two calculations lead to the same matrix element.~Because of this equivalence, we use Eq.~(\ref{eq:loc}) with $\widehat{V}_{\mathcal{X}}(\mathbf{r}_e, \mathbf{r}_\chi)$ given by Eq.~(\ref{eq:Vc72}) as the interaction potential associated with the $\mathcal{O}_7$ operator. The advantage of this approach is that it allows us to treat the local and non-local interactions underlying Eq.~(\ref{eq:Mnr}) in the same manner.

\subsection{Effective potential: bound electrons}
\label{sec:Veff2}
The potential $\widehat{V}$ associated with the amplitude $\mathcal{M}$ describes the non-relativistic interaction between free electrons and DM particles.~We can now evaluate matrix elements of $\widehat{V}$  between states involving a {\it bound electron} and a free DM particle to identify the {\it effective potential} that directly enters  the calculation of scattering cross sections and transition rates, where the initial (and final) electron is bound to the detector material.~In particular, we are interested in evaluating matrix elements of $\widehat{V}$ of the type below,
\begin{align}
\langle f;\mathbf{p}',s'  | \widehat{V} | \mathbf{p},s; i \rangle =& \frac{1}{V} \int{\rm d}\mathbf{r}_e   \int{\rm d}\mathbf{r}_\chi
\,\psi_f^{*}(\mathbf{r}_e) e^{- i \mathbf{p}'\cdot \mathbf{r}_\chi} 
\nonumber \\
&\times \xi_\chi^{s'\dagger} \widehat{V}_{\mathcal{X}}(\mathbf{r}_e,\mathbf{r}_\chi)  \xi_\chi^s\nonumber\\
&\times e^{i \mathbf{p}\cdot \mathbf{r}_\chi}    \psi_i(\mathbf{r}_e)\,,
\label{eq:Vmatrix}
\end{align}
where $|i\rangle$ ($|\mathbf{p},s\rangle$) is the initial electron (DM particle) state and $|f\rangle$ ($|\mathbf{p}',s'\rangle$) the  final electron (DM particle) state.~In Eq.~(\ref{eq:Vmatrix}), the overall $1/V$ factor arises from the initial and final DM particle wave functions, defined here as in Eq.~(\ref{eq:wavefunctions}).~Notice that $V=(2\pi)^3\delta^{(3)}(0)\neq \mathcal{N}$ is the spatial volume, while $\mathcal{N}=\delta^{(3)}(0)$ is the momentum space volume, i.e.~it has dimension [momentum]$^3$.~Here, we also introduce the initial and final state electron wave functions
\begin{align}
\psi_i(\mathbf{r}_e)&=\langle \mathbf{r}_e | i \rangle \nonumber\\
\psi_f(\mathbf{r}_e)&=\langle \mathbf{r}_e | f \rangle \,,
\label{eq:wavef}
\end{align}
respectively.~We can now perform the integral over the DM particle position in Eq.~(\ref{eq:Vmatrix}) by noticing that $\widehat{V}_{\mathcal{X}}(\mathbf{r}_e,\mathbf{r}_\chi)$ depends on the coordinates as follows (see Sec. \ref{sec:Veff}),
\begin{align}
\widehat{V}_{\mathcal{X}}(\mathbf{r}_e,\mathbf{r}_\chi) = \widehat{V}_{\mathcal{X}}(\mathbf{r}_e-\mathbf{r}_\chi,\overrightarrow{\nabla}_{\mathbf{r}_e},\overleftarrow{\nabla}_{\mathbf{r}_e},\overrightarrow{\nabla}_{\mathbf{r}_\chi},\overleftarrow{\nabla}_{\mathbf{r}_\chi}) \,.
\label{eq:Vr1r2}
\end{align}
This expression for $\widehat{V}_{\mathcal{X}}(\mathbf{r}_e,\mathbf{r}_\chi)$ allows us to rewrite Eq.~(\ref{eq:Vmatrix}) as 
\begin{align}
\langle f;\mathbf{p}',s'  | \widehat{V} | \mathbf{p},s; i \rangle &=  \frac{1}{V} \int{\rm d}\mathbf{r}_e \,\psi_f^{*}(\mathbf{r}_e) e^{i \mathbf{q}\cdot \mathbf{r}_e}\nonumber\\
&\times  \xi_\chi^{s'\dagger}\widetilde{V}_{\mathcal{X}}(\mathbf{q},\overrightarrow{\nabla}_{\mathbf{r}_e},\overleftarrow{\nabla}_{\mathbf{r}_e},i\mathbf{p},-i\mathbf{p}') \xi_\chi^s \nonumber\\
& \times \psi_i(\mathbf{r}_e)
\label{eq:Vmatrix2}
\end{align}
where $\vec q = \vec p - \vec p'$. In Eq.~(\ref{eq:Vmatrix2}), we changed integration variables from $(\mathbf{r}_e, \mathbf{r}_\chi)$ to $(\mathbf{r}_e, \mathbf{r}_e-\mathbf{r}_\chi)$, acted with $\overrightarrow{\nabla}_{\mathbf{r}_\chi}$ ($\overleftarrow{\nabla}_{\mathbf{r}_\chi}$) on the initial (final) DM matter plane wave, and performed the Fourier transform, 
\begin{align}
\widetilde{V}_{\mathcal{X}}(\mathbf{q},\dots) = \int {\rm d}(\mathbf{r}_e-\mathbf{r}_\chi)\, e^{- i\mathbf{q}\cdot (\mathbf{r}_e-\mathbf{r}_\chi)} 
\, \widehat{V}_{\mathcal{X}}(\mathbf{r}_e-\mathbf{r}_\chi,\dots) \,,
\label{eq:FT}
\end{align}
where the dots stand for the four nabla operators in Eq.~(\ref{eq:Vr1r2}).~Eq.~(\ref{eq:Vmatrix}) naturally leads us to define the effective potential 
\begin{align}
V^{ss'}_{\rm eff}(\overrightarrow{\nabla}_{\mathbf{r}_e},\overleftarrow{\nabla}_{\mathbf{r}_e};\mathbf{q},\mathbf{v}) &\equiv \frac{1}{V}  
e^{i \mathbf{q}\cdot \mathbf{r}_e} \nonumber\\
& \times
 \xi_\chi^{s'\dagger} \widetilde{V}_{\mathcal{X}}(\mathbf{q},\overrightarrow{\nabla}_{\mathbf{r}_e},\overleftarrow{\nabla}_{\mathbf{r}_e},i\mathbf{p},-i\mathbf{p}') \xi_\chi^s\,,
\label{eq:Veff}
\end{align}
where $\mathbf{v}=\mathbf{p}/m_\chi$.~Within this notation, we can rewrite Eq.~(\ref{eq:Vmatrix}) as 
\begin{align}
\langle f;\mathbf{p}',s'  | \widehat{V} | \mathbf{p},s; i \rangle &= \int{\rm d}\mathbf{r}_e \,\psi_f^{*}(\mathbf{r}_e) \nonumber\\
&\times V^{ss'}_{\rm eff}(\overrightarrow{\nabla}_{\mathbf{r}_e},\overleftarrow{\nabla}_{\mathbf{r}_e};\mathbf{q},\mathbf{v}) \, \psi_i(\mathbf{r}_e)\,,
\end{align}
and, therefore\footnote{Here and below, $V^{ss'}_{\rm eff}\equiv V^{ss'}_{\rm eff}(\overrightarrow{\nabla}_{\mathbf{r}_e},\overleftarrow{\nabla}_{\mathbf{r}_e};\mathbf{q},\mathbf{v}) $ to simplify the notation.},
\begin{align}
\langle f;\mathbf{p}',s'  | \widehat{V} | \mathbf{p},s; i \rangle = \langle f  | V^{ss'}_{\rm eff} |  i \rangle \,.
\end{align}
We are now ready to calculate the effective potential $V^{ss'}_{\rm eff}$ associated with the amplitude $\mathcal{M}$ in the general case, where all coupling constants are different from zero in Eq.~(\ref{eq:Mnr}).~From our analysis of the $\mathcal{O}_1$ and $\mathcal{O}_7$ operators, we find
\begin{align}
V^{ss'}_{\rm eff} = -&\frac{1}{4m_e m_\chi V} \Big\{F_0^{ss'}e^{i\mathbf{q}\cdot \mathbf{r}_e} \mathds{1}_{e}\nonumber\\
+&F_A^{ss'} \frac{i}{2m_e}\left[ \overleftarrow{\nabla}_{\mathbf{r}_e}\cdot \boldsymbol{\sigma}_e\, e^{i\mathbf{q}\cdot{\mathbf{r}_e}} -  e^{i\mathbf{q}\cdot{\mathbf{r}_e}} \,\boldsymbol{\sigma}_e\cdot\overrightarrow{\nabla}_{\mathbf{r}_e} \right] \nonumber\\
+&\mathbf{F}_5^{ss'}\cdot \boldsymbol{\sigma}_e\,e^{i\mathbf{q}\cdot \mathbf{r}_e}  \nonumber\\
+&\mathbf{F}_M^{ss'}\cdot  \frac{i}{2m_e}\left[ \overleftarrow{\nabla}_{\mathbf{r}_e} e^{i\mathbf{q}\cdot{\mathbf{r}_e}} -  e^{i\mathbf{q}\cdot{\mathbf{r}_e}}\overrightarrow{\nabla}_{\mathbf{r}_e} \right] \mathds{1}_{e}\nonumber\\
+&\mathbf{F}_E^{ss'}\cdot  \frac{1}{2m_e}\left[ \overleftarrow{\nabla}_{\mathbf{r}_e}\times\boldsymbol{\sigma}_e\, e^{i\mathbf{q}\cdot{\mathbf{r}_e}} +  e^{i\mathbf{q}\cdot{\mathbf{r}_e}} \,\boldsymbol{\sigma}_e\times\overrightarrow{\nabla}_{\mathbf{r}_e} \right] \nonumber\\ &\Big\} \,.
\label{eq:Vefftot}
\end{align}
The functions $F_0^{ss'}$, $F_A^{ss'}$, $\mathbf{F}_5^{ss'}$, $\mathbf{F}_M^{ss'}$ and $\mathbf{F}_E^{ss'}$ depend on coupling constants, $\mathbf{q}$ and $\mathbf{v}$, and are given in Eq.~(\ref{eq:M2}).~Since $n_0(\mathbf{r})\equiv\delta^{(3)}(\mathbf{r}-\mathbf{r}_e)$ is the electron density at $\mathbf{r}$, and
\begin{align}
\widetilde{n}_0(\mathbf{q}) &= \int {\rm d^3} r\,e^{-i\mathbf{q} \cdot \mathbf{r}} \, n_0(\mathbf{r}) = e^{-i\mathbf{q}\cdot \mathbf{r}_e}\,
\label{eq:n0}
\end{align}
is its Fourier transform at $\mathbf{q}$, we can rewrite the exponential factor in the first line of Eq.~(\ref{eq:Vefftot}) as $e^{i\mathbf{q}\cdot\mathbf{r}_e}=\widetilde{n}_0(-\mathbf{q})$.~Consequently, when the underlying DM-electron interaction contributes to the ``strength function'' $F^{ss'}_0$, then the DM couples to the electron density $\widetilde{n}_0(\mathbf{q})$ in the target material.~Similarly, when the DM particle contributes to the strength functions $F_A^{ss'}$, $\mathbf{F}_5^{ss'}$, $\mathbf{F}_M^{ss'}$ and $\mathbf{F}_E^{ss'}$, it couples, respectively, to the additional electron densities and currents 
\begin{align}
\widetilde{n}_A(\mathbf{q}) &=  \frac{i}{2m_e}\left[ \overleftarrow{\nabla}_{\mathbf{r}_e}\cdot \boldsymbol{\sigma}_e\, e^{-i\mathbf{q}\cdot{\mathbf{r}_e}} -  e^{-i\mathbf{q}\cdot{\mathbf{r}_e}} \,\boldsymbol{\sigma}_e\cdot\overrightarrow{\nabla}_{\mathbf{r}_e} \right]  \nonumber \\
\widetilde{\mathbf{j}}_5(\mathbf{q}) &= \boldsymbol{\sigma}_e\,e^{-i\mathbf{q}\cdot \mathbf{r}_e} \nonumber \\
\widetilde{\mathbf{j}}_M(\mathbf{q}) &= \frac{i}{2m_e}\left[ \overleftarrow{\nabla}_{\mathbf{r}_e} e^{-i\mathbf{q}\cdot{\mathbf{r}_e}} -  e^{-i\mathbf{q}\cdot{\mathbf{r}_e}}\overrightarrow{\nabla}_{\mathbf{r}_e} \right] \nonumber \\
\widetilde{\mathbf{j}}_E(\mathbf{q}) &= \frac{1}{2m_e}\left[ \overleftarrow{\nabla}_{\mathbf{r}_e}\times\boldsymbol{\sigma}_e\, e^{-i\mathbf{q}\cdot{\mathbf{r}_e}} + e^{-i\mathbf{q}\cdot{\mathbf{r}_e}} \,\boldsymbol{\sigma}_e\times\overrightarrow{\nabla}_{\mathbf{r}_e} \right] \,.
\label{eq:nj}
\end{align}
The electron densities and currents introduced in Eq.(\ref{eq:nj}) have an electromagnetic analogue.~For example, $\widetilde{\mathbf{j}}_M$ and $\widetilde{\mathbf{j}}_5$ can be identified with, respectively, the paramagnetic current and electron spin current.~In this analogy, $\mathbf{F}_M^{ss'}$ plays the role of an electromagnetic vector potential while $\mathbf{F}_5^{ss'}$ is a magnetic field.~Notice that  paramagnetic and spin current can be derived by expanding the Dirac Hamiltonian at zeroth order in $1/c$, where $c$ is the speed of light~\cite{Solyom2007}.~Within the same analogy, the density $\widetilde{n}_A$ and the current $\widetilde{\mathbf{j}}_E$ can be identified with the spin-paramagnetic current coupling and the Rashba term arising at second order in the $1/c$ expansion of the Dirac Hamiltonian~\cite{Solyom2007}.~There is also a close analogy between the densities and currents identified here and those found in the context of the effective theory for DM-nucleon interactions of~\cite{Anand:2013yka}, from which we adapted our notation.~Notice that while $\widetilde{n}_0$, $\widetilde{n}_A$, $\widetilde{\mathbf{j}}_5$ and $\widetilde{\mathbf{j}}_M$ are hermitian, $\widetilde{\mathbf{j}}_E$ is antihermitian because of the $-i$ factor in the last line of Eq.~(\ref{eq:Mnr}).

\subsection{Rate of dark matter-induced electronic transitions}
\label{sec:gamma}
Given the effective potential $V^{ss'}_{\rm eff}$, we can now calculate the total rate of DM-induced electronic transitions in a detector material, $\mathscr{R}$, by applying Fermi's golden rule.~We first rewrite the effective potential in a compact form,
\begin{align}
V^{ss'}_{\rm eff} = -\frac{1}{4 m_e m_\chi V} \sum_{\alpha} F^{ss'}_\alpha j_\alpha (-\mathbf{q}) \,,
\label{eq:Veff}
\end{align}
where the index $\alpha$ labels the components of the arrays collecting the strength functions, electron densities and currents,
\begin{align}
&(j_1,\dots,j_{11}) = (\widetilde{n}_0, \widetilde{n}_A, \widetilde{\boldsymbol{j}}_5, \widetilde{\boldsymbol{j}}_M, \widetilde{\boldsymbol{j}}_E)
\nonumber\\
&(F^{ss'}_1,\dots,F^{ss'}_{11}) = (F^{ss'}_0,F^{ss'}_A, \mathbf{F}^{ss'}_{5},\mathbf{F}^{ss'}_{M},\mathbf{F}_E^{ss'})\,.
\label{eq:notation}
\end{align}
We now apply Fermi's golden rule and obtain the differential rate of DM-induced transition in a detector material
\begin{align}
{\rm d}\Gamma &= \frac{2\pi}{V} \sum_{\alpha\beta}  \langle F_\alpha F_\beta^* \rangle \sum_{i,f} \frac{e^{-\bar{\beta} E_i}}{Z} \langle f| j_{\alpha}(-\mathbf{q}) |i\rangle \langle i| j^\dagger_{\beta}(\mathbf{q}) |f \rangle \nonumber\\
&\times  \left( \frac{1}{16 m_e^2 m_\chi^2} \right) 
\delta(E_f-E_i+\Delta E_\chi) \frac{{\rm d}\mathbf{q}}{(2 \pi)^3}\,,
\label{eq:dG}
\end{align}
where 
\begin{align}
\Delta E_\chi = \frac{q^2}{2m_\chi} -\mathbf{q}\cdot\mathbf{v}
\end{align}
is the energy deposited by the DM particle in the scattering,  $E_i$ ($E_f$) is the initial (final) electron energy, $\bar{\beta}=1/T$ is the reciprocal of the thermodynamic temperature of the material, $Z=\sum_i \exp(-\bar{\beta} E_i)$ is the partition function, and 
\begin{align}
\langle F_\alpha F_\beta^* \rangle = \frac{1}{2}\sum_{s s'} F^{ss'}_\alpha F_\beta^{ss'*} \,.
\end{align}
Recalling that the correlation function of two density or current operators, $K_{j^\dagger_\beta j_\alpha }(\mathbf{q},\omega)$, can be written as in Eq.~(\ref{eq:speK}),
\begin{align}
K_{j^\dagger_\beta j_\alpha }(\mathbf{q},\omega) &=\frac{2\pi}{V} \sum_{i,f} \frac{e^{-\bar{\beta} E_i}}{Z} 
\langle f| j_{\alpha}(-\mathbf{q}) |i\rangle \langle i| j^\dagger_{\beta}(\mathbf{q}) |f\rangle  \nonumber\\
&\times \delta(E_f-E_i-\omega)
\label{eq:corr}
\end{align}
where 
\begin{align}
K_{j^\dagger_{\beta}j_{\alpha}}(\mathbf{q},\omega) &= \int {\rm d}(t-t') e^{i\omega(t-t')}\int {\rm d(\mathbf{r}-\mathbf{r}')} e^{-i\mathbf{q}\cdot(\mathbf{r}-\mathbf{r}')} \nonumber\\ &\times K_{j^\dagger_{\beta}j_{\alpha}}(\mathbf{r}-\mathbf{r}',t-t')
\end{align}
is the double Fourier transform of $K_{j^\dagger_{\beta}j_{\alpha}}(\mathbf{r}-\mathbf{r}',t-t')$, we find
\begin{align}
{\rm d}\Gamma &=   \left( \frac{1}{16 m_e^2 m_\chi^2} \right) \int_{-\infty}^{+\infty} {\rm d}\omega \sum_{\alpha \beta}\langle F_\alpha F_\beta^* \rangle \, K_{j^\dagger_\beta j_\alpha }(\mathbf{q},\omega) \nonumber\\
&\times \delta(\omega+\Delta E_\chi) \frac{{\rm d}\mathbf{q}}{(2 \pi)^3}\,,
\label{eq:rate}
\end{align}
Here and in the following, we implicitly assume that correlation functions depend on the difference $\mathbf{r}-\mathbf{r}'$, and not on $\mathbf{r}$ and $\mathbf{r}'$ separately.~This is true in the case of translationally invariant systems, and it applies to a good approximation to the case of crystals~\cite{Weissker2010Feb}.~We will further comment on the meaning and impact of this assumption at the end of Sec.~\ref{sec:chi_intro}.

Finally, in order to obtain the total rate of DM-induced electronic transitions in a given detector material, we integrate the differential rate in Eq.~(\ref{eq:dG}) over transfer momentum, $\vec q$, and DM particle velocities in the lab frame, $\vec v$:  
\begin{align}
\mathscr{R} = n_\chi  
V \int {\rm d}\mathbf{q} \int {\rm d} \mathbf{v} \, f(\mathbf{v}) \,\frac{{\rm d}\Gamma}{\rm d \mathbf{q}}\,,
\label{eq:ratefinal}
\end{align}
where $f(\mathbf{v})$ is the DM velocity distribution in the laboratory frame, while $n_\chi$ is the local DM number density at the detector.~For $f(\vec v)$, we assume a truncated Maxwell-Boltzmann distribution with local standard of rest speed $v_0= 238$ km~s$^{-1}$~\cite{Baxter:2021pqo}, galactic escape speed $v_\mathrm{esc}=544$ km~s$^{-1}$ \cite{Baxter:2021pqo} and Earth's speed in a reference frame where the mean DM particle velocity is zero, $v_e=250.5$ km~s$^{-1}$ \cite{Baxter:2021pqo}.~For $n_\chi = \rho_\chi / m_{\rm DM}$, we assume $\rho_\chi = 0.4$ GeV cm$^{-3}$ \cite{Catena:2009mf}.

It is important to note that the effective potential, $V_{\rm eff}^{ss'}$, in Eq.~(\ref{eq:Veff}) is evaluated at a reference time, say $t=0$. In the interaction picture, we obtain the effective potential at a generic time $t$, $V_{\rm eff}^{ss'}(t)$, by replacing $\widehat{V}_{\mathcal{X}}(\vec r_e, \vec r_\chi)$ with $\exp(i H_0 t) \widehat{V}_{\mathcal{X}}(\vec r_e, \vec r_\chi) \exp(-i H_0 t)$ in Eq.~(\ref{eq:Vmatrix}), where $H_0$ is the Hamiltonian of the DM-electron system with $V_{\rm eff}^{ss'}=0$. $V_{\rm eff}^{ss'}(t)$ can then be written as follows,
\begin{align}
V_{\rm eff}^{ss'}(t) = - \sum_{\alpha}\int {\rm d}\mathbf{r} \, B_\alpha(\mathbf{r}) \, S_\alpha^{ss'}(\mathbf{r},t) \,,
\label{eq:Vt}
\end{align}
with
\begin{align}
B_\alpha(\mathbf{r})  = \int \frac{{\rm d}\mathbf{q}'}{(2\pi)^3} \, e^{i \vec q' \cdot \vec r} \, j_{\alpha}(\vec q') \,,
\end{align}
and 
\begin{align}
S_\alpha^{ss'}(\mathbf{r},t) = \frac{1}{4 m_e m_\chi V} \,F_\beta^{ss'} \, e^{i \vec q \cdot \vec r} \,e^{i\Delta E_\chi t} \,.
\end{align}

\section{The Generalised susceptibility formalism}
\label{sec:formalism}

\subsection{Generalised susceptibilities in linear response theory}
\label{sec:chi_intro}
The effective potential $V_{\rm eff}^{ss'}(t)$ in Eq.~(\ref{eq:Vt}) can be interpreted as an external perturbation affecting the physical observables of any given detector material.~Here, the physical observables of interest are the electron densities and currents in Eqs.~(\ref{eq:n0}) and (\ref{eq:nj}), which we collectively denoted by $j_\alpha$, $\alpha=1,\dots,11$.~In linear response theory, the fluctuation, $\langle \Delta j_\alpha(\mathbf{r},t)\rangle$, induced on the generic electron density or current $j_\alpha$ by the potential $V_{\rm eff}^{ss'}(t)$  
is given by 
\begin{align}
\langle \Delta j_\alpha(\mathbf{r},t)\rangle &= \sum_{\beta}\int_{-\infty}^{t} {\rm d} t' \int {\rm d}\mathbf{r}'\, \chi_{j_{\alpha}j_{\beta}}(\mathbf{r}-\mathbf{r}',t-t') \nonumber\\ &\times S_\beta^{ss'}(\mathbf{r}',t') \,,
\label{eq:dcs}
\end{align}
where
\begin{align}
\chi_{j_{\alpha}j_{\beta}}(\mathbf{r}-\mathbf{r}',t-t') = i\theta(t-t') \Big \langle \left[ j_\alpha(\mathbf{r},t),j_\beta(\mathbf{r}',t') \right] \Big \rangle 
\label{eq:chiret}
\end{align}
is the generalised susceptibility associated with $j_\alpha$ and $j_\beta$.~Since $\chi_{j_{\alpha}j_{\beta}}(\mathbf{r}-\mathbf{r}',t-t')=0$ for $t-t'<0$, $\chi_{j_{\alpha}j_{\beta}}$ expresses the response of the electron density or current $j_\alpha$ to the perturbation $S_\beta^{ss'} j_\beta$ in terms of a retarded Green function.~Similarly, one can introduce a  generalised susceptibility associated with $j_\alpha$ and $j_\beta$ that quantifies the same response in terms of an advanced correlation function, 
\begin{align}
\chi^A_{j_{\alpha}j_{\beta}}(\mathbf{r}-\mathbf{r}',t-t') = -i\theta(t'-t) \Big \langle \left[ j_\alpha(\mathbf{r},t),j_\beta(\mathbf{r}',t') \right] \Big \rangle \,.
\label{eq:chiad}
\end{align}
From the spectral representations of $\chi_{j_{\alpha}j_{\beta}}$, Eq.~(\ref{eq:spechi}), and of the correlation function $K_{j_{\alpha}j_{\beta}}$, Eq.~(\ref{eq:speK}), we also find 
\begin{align}
\chi_{j_{\alpha}j_{\beta}}(\mathbf{q},\omega) =& -\frac{1}{2\pi}  \int_{-\infty}^{+\infty} {\rm d}\omega' \,
\frac{K_{j_{\alpha}j_{\beta}}(\mathbf{q},\omega')}{\omega - \omega'+i\delta} \nonumber\\ &\times \left( 1- e^{-\bar{\beta} \omega'} \right) \,.
\end{align}
where $\delta$ is an infinitesimal parameter larger than zero, and
\begin{align}
\chi_{j_{\alpha}j_{\beta}}(\mathbf{q},\omega) &= \int {\rm d}(t-t') e^{i\omega(t-t')}\int {\rm d(\mathbf{r}-\mathbf{r}')} e^{-i\mathbf{q}\cdot(\mathbf{r}-\mathbf{r}')} \nonumber\\ &\times \chi_{j_{\alpha}j_{\beta}}(\mathbf{r}-\mathbf{r}',t-t')
\end{align}
is the double Fourier transform of $\chi_{j_{\alpha}j_{\beta}}(\mathbf{r}-\mathbf{r}',t-t')$. Similarly, 
\begin{align}
\chi^A_{j_{\alpha}j_{\beta}}(\mathbf{q},\omega) =& -\frac{1}{2\pi} \int_{-\infty}^{+\infty} {\rm d}\omega' \,
\frac{K_{j_{\alpha}j_{\beta}}(\mathbf{q},\omega')}{\omega - \omega'-i\delta} \nonumber\\ &\times \left( 1- e^{-\bar{\beta} \omega'} \right) \,.
\end{align}
Notice that,
\begin{align}
\lim_{\delta\rightarrow 0^{+}} \chi_{j_{\alpha}j_{\beta}} (\mathbf{q},\omega) =&~\frac{i}{2}\, K_{j_{\alpha}j_{\beta}} (\mathbf{q},\omega) \left( 1- e^{-\bar{\beta} \omega} \right)\nonumber\\
& -\frac{1}{2\pi}\,P\int_{-\infty}^{+\infty} {\rm d}\omega' \,
\frac{K_{j_{\alpha}j_{\beta}}(\mathbf{q},\omega')}{\omega - \omega'} \nonumber\\ &\times \left( 1- e^{-\bar{\beta} \omega'} \right)
\label{eq:P1}
\end{align}
where $P$ denotes the principal value.~Analogously, one has
\begin{align}
\lim_{\delta\rightarrow 0^{+}} \chi^A_{j_{\alpha}j_{\beta}} (\mathbf{q},\omega) = & -\frac{i}{2}\, K_{j_{\alpha}j_{\beta}} (\mathbf{q},\omega) \left( 1- e^{-\bar{\beta} \omega} \right)\nonumber\\
& -\frac{1}{2\pi}\,P\int_{-\infty}^{+\infty} {\rm d}\omega' \,
\frac{K_{j_{\alpha}j_{\beta}}(\mathbf{q},\omega')}{\omega - \omega'} \nonumber\\ &\times \left( 1- e^{-\bar{\beta} \omega'} \right)
\label{eq:P2}
\end{align}
which implies
\begin{align}
\chi_{j_{\alpha}j_{\beta}} (\mathbf{q},\omega)-\chi^A_{j_{\alpha}j_{\beta}} (\mathbf{q},\omega) = i K_{j_{\alpha}j_{\beta}} (\mathbf{q},\omega)\left( 1- e^{-\bar{\beta} \omega} \right) \,.
\label{eq:chi1}
\end{align}
For $j_\beta=j^\dagger_\alpha$, the above equation reduces to the simple relation
\begin{align}
K_{j_{\alpha}j^{\dagger}_{\alpha}} (\mathbf{q},\omega) = 2 \left( 1- e^{-\bar{\beta} \omega} \right)^{-1} \Im (\chi_{j_{\alpha}j^\dagger_{\alpha}} (\mathbf{q},\omega)) \,,
\label{eq:chi2}
\end{align}
being
\begin{align}
\chi^*_{j^\dagger_{\beta}j^{\dagger}_{\alpha}} (\mathbf{q},\omega) = \chi^A_{j_{\alpha}j_{\beta}} (\mathbf{q},\omega) \,,
\end{align}
as we show in Appendix~\ref{sec:chiK} (see Eq.~(\ref{eq:chichiA})).~Eqs.~(\ref{eq:chi1}) and (\ref{eq:chi2}) are our starting point to relate the rate of DM-induced electronic transitions to a set of generalised susceptibilities associated with the electron densities and currents $j_{\alpha}$.

As anticipated, here we assume that correlation functions and generalised susceptibilities depend on $\vec r - \vec r'$.~When $\chi_{j_{\alpha}j_{\beta}}$ depends on $\vec r$ and $\vec r'$ separately, the above equations have to be revisited by using the Fourier transform,
\begin{align}
\chi_{j_{\alpha}j_{\beta}}(\vec q,\vec q', t-t') = \int {\rm d} \vec r  \int {\rm d} \vec r' \, &e^{-\vec q \cdot \vec r} e^{-\vec q' \cdot \vec r'} 
\nonumber\\
&\times \chi_{j_{\alpha}j_{\beta}}(\mathbf{r},\mathbf{r}',t-t') \,,
\end{align}
which depends on two conjugate momenta, $\vec q$ and $\vec q'$.~The latter are such that $\vec q+\vec q'$ is a reciprocal lattice vector.~In this  case, it is customary to restrict $\vec q$ and $\vec q'$ to the first Brillouin Zone, and express the double Fourier transform of $\chi_{j_\alpha,j_\beta}$ as a matrix in reciprocal space, namely
\begin{align}
\chi^{\vec G \vec G'}_{j_\alpha,j_\beta}(\vec q,\omega) \equiv \frac{1}{V}\chi_{j_\alpha,j_\beta}(\vec q+\vec G, - \vec q - \vec G',\omega) \,,
\label{eq:cmatrix}
\end{align}
where $\vec q$ is in the first Brillouin Zone, while $\vec G$ and $\vec G'$ are reciprocal lattice vectors.~With this notation, Eq.~(\ref{eq:dcs}) implies,
\begin{align}
\langle \Delta j_\alpha(\mathbf{q}+\vec G,\omega)\rangle = \sum_{\beta} \sum_{\vec G'} &\chi^{\vec G \vec G'}_{j_\alpha,j_\beta}(\vec q,\omega) \nonumber\\ &\times S_\beta^{ss'}(\mathbf{q} + \vec G',\omega) \,.
\label{eq:sumg}
\end{align}
Notice that when the correlation functions and generalised susceptibilities depend only on $\vec r - \vec r'$ rather than $\vec r$ and $\vec r'$ separately, Eq.~(\ref{eq:cmatrix}), reduces to 
\begin{align}
\chi^{\vec G \vec G'}_{j_\alpha,j_\beta}(\vec q,\omega) \equiv \frac{1}{V}\delta_{\vec G \vec G'}\chi_{j_\alpha,j_\beta}(\vec q+\vec G, - \vec q - \vec G,\omega) \,.
\label{eq:cmatrix2}
\end{align}
Consequently, the $\vec r - \vec r'$ assumption corresponds to neglecting the $\vec G'\neq \vec G$ terms in the sum in Eq.~(\ref{eq:sumg}).~These terms account for variations of the external DM perturbation over atomic distances, and correspond to so-called local-field corrections.~Our $\vec r - \vec r'$ assumption is supported by recent works \cite{Knapen:2021run,Hochberg:2021pkt,Knapen:2021bwg} in which local-field corrections have been studied in models where DM couples to the density $n_0$, finding that they are a sub-leading effect compared to screening.~The role of the electron density-density response function in DM-electron scattering is also discussed in~\cite{Boyd:2022tcn}, with a focus on anisotropic scattering.

\subsection{Electronic transition rate and generalised susceptibilities}
\label{sec:chi_sigma}
Without restricting $\vec q$ to the first Brillouin Zone, we can now use Eqs.~(\ref{eq:chi1}) and (\ref{eq:chi2}) to rewrite the differential rate of DM-induced electronic transitions in materials as
\begin{align}
{\rm d}\Gamma=& \left(\frac{1}{8 m_e^2 m_\chi^2}\right) \int_{-\infty}^{+\infty} {\rm d}\omega \, \frac{1}{\left( 1- e^{-\bar{\beta} \omega} \right)}\,\delta(\omega+\Delta E_\chi)
\nonumber \\ &\times \sum_\beta \sum_{\alpha\le\beta}
2^{-\delta_{\alpha\beta}} \,\Re\left[ \langle F_{\alpha} F_{\beta}^* \rangle \,i \left(  \chi^A_{j^\dagger_\beta j_\alpha} - \chi_{j^\dagger_\beta j_\alpha} \right)\right]
\nonumber \\ & \times \frac{{\rm d}\mathbf{q}}{(2\pi)^3}\,.
\label{eq:rate_gensus}
\end{align}
When DM couples to the electron density $n_0$, we can apply Eq.~(\ref{eq:chi2}) to express our rate formula, Eq.~(\ref{eq:rate_gensus}), as in~\cite{Knapen:2021run,Hochberg:2021pkt},
\begin{align}
{\rm d}\Gamma&= \left( \frac{1}{8 m_e^2 m_\chi^2} \right) \int_{-\infty}^{+\infty} {\rm d}\omega \, \frac{1}{\left( 1- e^{-\bar{\beta} \omega} \right)}   \langle F_{0} F_{0}^* \rangle \Im (\chi_{n^\dagger_0 n_0} ) 
\nonumber \\ &\times  \delta(\omega+\Delta E_\chi) \frac{{\rm d}\mathbf{q}}{(2\pi)^3}\,.
\end{align}
In this particular case, we can use the relation between susceptibility, $\chi_{n^\dagger_0 n_0}$, and dielectric function of the material, 
\begin{align}
\frac{1}{\epsilon_r(\mathbf{q},\omega)} = 1 - \frac{4\pi \alpha}{q^2} \chi_{n^\dagger_0 n_0}(\mathbf{q},\omega)
\label{eq:echi}
\end{align}
to express ${\rm d}\Gamma$ in terms of the measurable quantity $\epsilon_r(\mathbf{q},\omega)$.~The minus sign on the right-hand-side arises from our definition of generalised susceptibility in Eq.~(\ref{eq:chiret}).

Similarly, if $j_\alpha$ and $j_\beta$, are the spatial components of the same current, e.g. $j_{5 \,l}$, $l=1,2,3$, Eq.~(\ref{eq:rate_gensus}) reduces to
\begin{align}
{\rm d}\Gamma=& \left(\frac{1}{16 m_e^2 m_\chi^2}\right) \int_{-\infty}^{+\infty} {\rm d}\omega \, \frac{1}{\left( 1- e^{-\bar{\beta} \omega} \right)}\,\delta(\omega+\Delta E_\chi)
\nonumber \\ &\times \sum_{\alpha \beta} 
 \,\Re\left[ \langle F_{j_{\alpha}} F_{j_{\beta}}^* \rangle \,i \left(  \chi^A_{j^\dagger_{\beta} j_{\alpha}} - \chi_{j^\dagger_{\beta} j_{\alpha}} \right)\right] 
\frac{{\rm d}\mathbf{q}}{(2\pi)^3}\,.
\label{eq:rate_gensus_simp}
\end{align}

\section{Evaluation of the generalised susceptibilities}
\label{sec:evaluation} 
Let us now focus on the evaluation of the generalised susceptibilities associated with the electron densities and currents in Eqs.~(\ref{eq:n0}) and (\ref{eq:nj}).~We start by deriving a time evolution equation for $\chi_{j_\alpha j_\beta}(\vec q,t)$ in second quantisation.~We then find a ``mean field'' solution to this equation, for which we also provide a useful diagrammatic interpretation.~This approach enables us to account for potentially important screening and collective excitation effects, which previous descriptions of general DM-electron interactions in materials~\cite{Catena:2019gfa,Catena:2021qsr,Catena:2023qkj,Catena:2023awl} could not capture.

\subsection{Second quantisation form for $j_\alpha$}

Let us start by writing the densities and currents in Eqs.~(\ref{eq:n0}) and (\ref{eq:nj}) in second quantised notation.~For $n_0$, we find
\begin{align} 
\hat{n}_0(\vec r, t)= \frac{1}{N_{\rm cell}}\sum_{ii'} \sum_{\sigma \sigma'} \hat{\psi}^{\dagger}_{i'\sigma'}(\vec r, t) \,\hat{\psi}_{i\sigma}(\vec r, t) \,,
\label{eq:jsecond_r}
\end{align}
where
\begin{align}
\hat{\psi}_{i\sigma}(\vec r, t) \equiv \frac{1}{\sqrt{V}}  \sum_{\vec k} e^{i \vec k \cdot \vec r} \, u_{i\vec k} (\vec r) \,\eta^\sigma c_{i \vec k}^{\sigma}(t) \,,
\end{align}
and $u_{i\vec k} (\vec r)$ is a periodic function with the same periodicity as the underlying lattice, and with Fourier modes $u_{i \vec k+\vec G}$:
\begin{align}
u_{i\vec k} (\vec r) = \sum_{\vec G} e^{i \vec{G}\cdot \vec r} \, u_{i \vec k+\vec G} \,,
\end{align}
where $\vec G$ is a reciprocal lattice vector.~Here $c^\sigma_{i\vec k}$ ($c^{\sigma \dagger}_{i\vec k}$) is the annihilation (creation) operator for an electron in band $i$, with reciprocal space vector in the first Brillouin zone $\vec k$ and spin configuration labelled by $\sigma$.~Spin-up electrons correspond to $\eta^{\uparrow}=(1,0)^T$, whereas spin-down electrons correspond to $\eta^\downarrow=(0,1)^T$.~Notice also that the Fourier transform of the density operator in Eq.~(\ref{eq:jsecond_r}) can be written as
\begin{align}
\hat{n}_0(\vec q, t)=&\frac{1}{N_{\rm cell}}\sum_{ii'}\sum_{\sigma\sigma'}\sum_{\vec k \vec k' \vec G \vec G'} u^*_{i'\vec k'+\vec G'}   \, u_{i\vec k+\vec G} \, \eta^{\sigma' \dagger} \eta^\sigma \nonumber\\
&\times  \frac{(2 \pi)^3}{V} \delta^{(3)}(\vec k' +\vec G'+\vec q-\vec k -\vec G) \nonumber\\
&\times c_{i' \vec k'}^{\sigma' \dagger}(t) c_{i \vec k}^{\sigma}(t)\,.
\label{eq:nhat}
\end{align}
or in a more compact form as
\begin{align}
\hat{n}_0(\vec q, t)=&\sum_{ii'\sigma\sigma'\vec k}   
\mathscr{J}_{n_0}^{ii' \sigma\sigma'}(\vec k+ \vec q,\vec k) \,
c_{i' \vec k}^{\sigma' \dagger}(t) c_{i \vec k + \vec q}^{\sigma}(t) \,,
\end{align}
where
\begin{align}
\mathscr{J}_{n_0}^{ii' \sigma\sigma'}(\vec k+ \vec q,\vec k) \equiv \sum_{\vec G} u^*_{i'\vec k +\vec G}   \, u_{i\vec k+\vec q+\vec G} \, \delta^{\sigma'\sigma} \,.
\end{align}
Here we used the definition $V=N_{\rm cell} V_{\rm cell}$, as well as $(\sum_{\vec k} 1) = N_{\rm cell}$.~Notice that the expectation value of $\hat{n}_0(\vec q, t)$ between single-particle states with $\vec{q}=\vec p-\vec p'$ gives
\begin{align}
\langle \vec{p}'j'\rho'| \hat{n}_0(\vec q, t) | \vec{p}j\rho\rangle = \mathscr{J}_{n_0}^{jj' \rho\rho'}(\vec p,\vec p-\vec q)\,,
\label{eq:me}
\end{align}
which shows the equivalence between Eq.~(\ref{eq:nhat}) and Eq.~(\ref{eq:n0}), and explains the $1/N_{\rm cell}$ factor in Eq.~(\ref{eq:jsecond_r}).
By performing an analogous calculation for the density $n_A$, we find
\begin{align}
\hat{n}_A(\vec q, t)
&=\sum_{ii'\sigma\sigma'\vec k} 
\mathscr{J}_{n_A}^{ii' \sigma\sigma'}(\vec k+ \vec q,\vec k) \,
c_{i' \vec k}^{\sigma' \dagger}(t) c_{i \vec k + \vec q}^{\sigma}(t) \,,
\end{align}
where now
\begin{align}
\mathscr{J}_{n_A}^{ii' \sigma\sigma'}(\vec k+ \vec q,\vec k) \equiv&\sum_{\vec G} u^*_{i'\vec k +\vec G}   \, u_{i\vec k+\vec q+\vec G} 
\nonumber\\
&\times (2 m_e)^{-1} \left[2(\vec k+\vec G)+\vec q \right] \cdot \eta^{\sigma' \dagger} \boldsymbol{\sigma} \eta^\sigma.
\end{align}
Similarly, writing the current $\boldsymbol{j}_5$ in second quantised notation, we find 
\begin{align}
\hat{j}_{5\,l}(\vec q, t)
=\sum_{ii'\sigma\sigma'\vec k} 
\mathscr{J}_{5 \, l}^{ii' \sigma\sigma'}(\vec k+ \vec q,\vec k) \,
c_{i' \vec k}^{\sigma' \dagger}(t) c_{i \vec k + \vec q}^{\sigma}(t) \,,
\label{eq:jcurrent}
\end{align}
where
\begin{align}
\mathscr{J}_{5 \,l}^{ii' \sigma\sigma'}(\vec k+ \vec q,\vec k) \equiv& \sum_{\vec G} u^*_{i'\vec k +\vec G}   \, u_{i\vec k+\vec q+\vec G} \,
\eta^{\sigma' \dagger} \sigma_{l} \,\eta^\sigma \,,
\end{align}
Eq.~(\ref{eq:jcurrent}) holds true for the currents $\boldsymbol{j}_M$ and $\boldsymbol{j}_E$ if one replaces $\mathscr{J}_{5 \,l}^{ii' \sigma\sigma'}(\vec k+ \vec q,\vec k)$ with, respectively, the two vectors
\begin{align}
\mathscr{J}_{M\,l}^{ii' \sigma\sigma'}(\vec k+ \vec q,\vec k) \equiv&\sum_{\vec G} u^*_{i'\vec k +\vec G}   \, u_{i\vec k+\vec q+\vec G} 
\nonumber\\
&\times (2 m_e)^{-1}\left[2(\vec k+\vec G)+\vec q \right]^l \delta^{\sigma'\sigma} \,,
\end{align}
and
\begin{align}
\mathscr{J}_{E \,l}^{ii' \sigma\sigma'}(\vec k+ \vec q,\vec k) &\equiv \frac{-i}{2 m_e} \sum_{\vec G} u^*_{i'\vec k +\vec G}   \, u_{i\vec k+\vec q+\vec G}
\nonumber\\
&\times \sum_{m,n=1}^{3} \epsilon^{lmn} \left[2(\vec k+\vec G)+\vec q \right]^m \nonumber\\
&\times \eta^{\sigma' \dagger}\sigma^{n} \eta^\sigma \,.
\end{align}
By introducing a notation similar to the one we used in Eq.~(\ref{eq:notation}), i.e.
\begin{align}
\hat{j}_{\alpha} = \left( \hat{n}_0, \hat{n}_A, \hat{\boldsymbol{j}}_5, \hat{\boldsymbol{j}}_M, \hat{\boldsymbol{j}}_E\right) \,,
\end{align}
and
\begin{align}
\mathscr{J}_{\alpha} =&~(\mathscr{J}_{n_0}, \mathscr{J}_{n_A}, \mathscr{J}_{5 \, 1},\mathscr{J}_{5 \, 2},\mathscr{J}_{5 \, 3},
 \mathscr{J}_{M \, 1}, \dots, \nonumber\\
  &\mathscr{J}_{E \, 1},  \dots  
) \,,
\end{align}
we collectively write all density and current operators as follows,
\begin{align}
\hat{j}_\alpha(\vec q, t) &=\sum_{ii'\sigma\sigma'\vec k}  
\mathscr{J}_{\alpha}^{ii' \sigma\sigma'}(\vec k+ \vec q,\vec k) \,
c_{i' \vec k}^{\sigma' \dagger}(t) c_{i \vec k + \vec q}^{\sigma}(t) \,,
\end{align}
where now $\alpha=1,\dots,11$.

\subsection{Equation of motion for $\chi_{j_\alpha j_\beta}$}
Next, we introduce the momentum-, band- and spin-resolved susceptibility,
\begin{align}
\chi_{j_\alpha j_\beta}^{ii'\sigma\sigma'}(\vec k, \vec q,t-t') &= i\theta(t-t') \sum_{\vec k'} \sum_{j j'} \sum_{\rho \rho'} \frac{1}{V} \times \nonumber\\
&\mathscr{J}_\alpha^{ii'\sigma\sigma'}(\vec k + \vec q,\vec k) \times \nonumber\\
&\mathscr{J}_\beta^{jj'\rho\rho'}(\vec k',\vec k'+\vec q)  \times\nonumber\\
&\left \langle \left[ c_{i' \vec k}^{\sigma' \dagger}(t) c_{i \vec k+\vec q}^{\sigma}(t) , c_{j' \vec k' + \vec q}^{\rho' \dagger}(t') c_{j \vec k'}^{\rho}(t') \right] \right \rangle,
\label{eq:chires}
\end{align}
such that
\begin{align}
\chi_{j_\alpha j_\beta}(\vec q,t-t') =\sum_{\vec k} \sum_{ii'} \sum_{\sigma\sigma'}\chi_{j_\alpha j_\beta}^{ii'\sigma\sigma'}(\vec k, \vec q,t-t') \,.
\end{align}
We obtain a differential time evolution equation for the susceptibility $\chi_{j_\alpha j_\beta}^{ii'\sigma\sigma'}(\vec k, \vec q,t-t')$ by acting on Eq.~(\ref{eq:chires}) with the operator 
 $i {\rm d}/{\rm d} t$, and rewriting the right-hand side of the latter as a function of $\chi_{j_\alpha j_\beta}^{ii'\sigma\sigma'}(\vec k,\vec q,t-t')$.~In the right-hand-side of Eq.~(\ref{eq:chires}), $i {\rm d}/{\rm d} t$ acts non trivially on $\theta(t-t')$ and on the product $c_{i' \vec k}^{\sigma' \dagger}(t) c_{i \vec k+\vec q}^{\sigma}(t) $.~When $i {\rm d}/{\rm d} t$ acts on $\theta(t-t')$, it generates the Dirac delta $i \delta(t-t')$, which implies $t'=t$ in the commutator in the right-hand side of Eq.~(\ref{eq:chires}).~Evaluating this equal-time commutator, we find
\begin{align}
 \left \langle \left[ c_{i' \vec k}^{\sigma' \dagger} c_{i \vec k+\vec q}^{\sigma} , c_{j' \vec k' + \vec q}^{\rho' \dagger} c_{j \vec k'}^{\rho} \right] \right \rangle 
&= \delta_{\sigma\rho'} \delta_{ij'} \delta_{\vec k \vec k'} \langle c_{i' \vec k}^{\sigma' \dagger} c_{j \vec k'}^{\rho}  \rangle \nonumber\\
&\quad\,-  \delta_{\rho\sigma'} \delta_{ji'}\delta_{\vec k \vec k'} \langle c_{j' \vec k'+\vec q}^{\rho' \dagger} c_{i \vec k + \vec q}^{\sigma}  \rangle \nonumber\\
& = \delta_{\sigma\rho'} \delta_{\sigma' \rho}  \delta_{i j'}\delta_{i' j}\delta_{\vec k \vec k'}  \nonumber\\
&\quad\, \times \Big[ f_0(\varepsilon_{i' \vec k}^{\sigma'})  - f_0(\varepsilon_{i \vec k + \vec q}^{\sigma})  \Big] \,.
\label{eq:etcom}
\end{align}
where the equilibrium occupation numbers $f_0$, e.g.
\begin{align}
f_0(\varepsilon^\sigma_{i \vec k}) \equiv  N_{\rm cell} \frac{e^{-\bar{\beta} \varepsilon^\sigma_{i \vec k}}}{Z} \,,
\end{align}
arise from
\begin{align}
 \langle c_{i' \vec k}^{\sigma' \dagger} c_{j \vec k'}^{\rho}  \rangle &= \delta_{i'j} \delta_{\sigma' \rho} \delta_{\vec k \vec k'}\, f_0(\varepsilon_{i' \vec k}^{\sigma'}) \,, \nonumber \\
   \langle c_{j' \vec k'+\vec q}^{\rho' \dagger} c_{i \vec k + \vec q}^{\sigma} \rangle &= \delta_{ij'} \delta_{\sigma \rho'} \delta_{\vec k \vec k'} f_0(\varepsilon_{i \vec k + \vec q}^{\sigma})  \,.
\label{eq:f0}   
\end{align}
When  $i {\rm d}/{\rm d} t$ acts on $c_{i' \vec k}^{\sigma' \dagger}(t) c_{i \vec k+\vec q}^{\sigma}(t) $, it generates the commutators,
\begin{align}
i\frac{\rm d}{{\rm d} t} \left[ c_{i' \vec k}^{\sigma' \dagger}(t) c_{i \vec k+\vec q}^{\sigma}(t) \right] = &- \left[ \mathcal{H}_0+\mathcal{H}_{e-e}, c_{i' \vec k}^{\sigma' \dagger}(t)  \right] c_{i \vec k+\vec q}^{\sigma}(t) \nonumber\\
&- c_{i' \vec k}^{\sigma' \dagger}(t)  \left[ \mathcal{H}_0+\mathcal{H}_{e-e},  c_{i \vec k+\vec q}^{\sigma}(t) \right] \,,
\label{eq:ddtcom}
\end{align}
where in the right-hand side of Eq.~(\ref{eq:ddtcom}) we used the Heisenberg equations for the operators $c_{i' \vec k}^{\sigma' \dagger}(t)$ and $c_{i \vec k+\vec q}^{\sigma}(t)$.~Here, $\mathcal{H}_0$ and $\mathcal{H}_{e-e}$ are the free-electron and electron-electron interaction Hamiltonians in second quantisation, which for Bloch electrons can be written as follows
\begin{align}
\mathcal{H}_0 &= \sum_{i \vec k \sigma} \varepsilon_{i \vec k}^\sigma\, c_{i \vec k}^{\sigma \dagger}(t) c_{i \vec k}^{\sigma}(t)  \,,\\
\mathcal{H}_{e-e}&= \frac{1}{2V} \sum_{\vec p \vec p' \vec q'} \sum_{\sigma_1 \sigma_2} \, \sum_{n_1 n_2 n_3 n_4} \, \sum_{\vec G_1 \vec G_2} U(\vec q') \nonumber\\
&\quad\,\,\times u^*_{n_1 \vec p + \vec q' + \vec G_1}  u^*_{n_2 \vec p' - \vec q' + \vec G_2}  \nonumber\\
&\quad\,\, \times u_{n_3 \vec p' + \vec G_2}  u_{n_4 \vec p + \vec G_1} \nonumber\\
&\quad\,\,\times c^{\sigma_1 \dagger}_{n_1 \vec p + \vec q'}(t)  c^{\sigma_2 \dagger}_{n_2 \vec p' - \vec q'}(t)   c^{\sigma_2}_{n_3 \vec p'}(t)  c^{\sigma_1}_{n_4 \vec p}(t) 
\,.
\end{align}
where $U(\vec q')$ is the Fourier transform of the Coulomb potential for electron-electron interactions.~Contrary to the external DM perturbation $V_{\rm eff}^{ss'}$, our choice for $\mathcal{H}_{e-e}$ assumes that electron-electron interactions do not induce spin-flips.~We also assume that  
$V_{\rm eff}^{ss'}$ can be neglected in the Heisenberg equations for $c_{i' \vec k}^{\sigma' \dagger}(t)$ and $c_{i \vec k+\vec q}^{\sigma}(t)$, although it is taken into account in the time evolution equation for $\chi_{j_\alpha j_\beta}^{ii'\sigma\sigma'}(\vec k,\vec q,t-t')$ via the $\mathscr{J}_\alpha$ functions in Eq.~(\ref{eq:chires}).~With these expressions for $\mathcal{H}_0$ and $\mathcal{H}_{e-e}$, we now evaluate the commutators in the right-hand-side of Eq.~(\ref{eq:ddtcom}),
\begin{align}
\left[ \mathcal{H}_0,  c_{i' \vec k}^{\sigma' \dagger}(t)  \right] &= \varepsilon_{i'\vec k}^{\sigma'} \, c_{i' \vec k}^{\sigma' \dagger}(t) \,,\nonumber\\
\left[ \mathcal{H}_0,  c_{i \vec k + \vec q}^{\sigma}(t)  \right] &= - \varepsilon_{i\vec k + \vec q}^{\sigma} \, c_{i \vec k + \vec q}^{\sigma }(t)\,,
\label{eq:H0com}
\end{align}
as well as,
\begin{align}
\left[ \mathcal{H}_{e-e},  c_{i' \vec k}^{\sigma' \dagger}(t)  \right] &=  \frac{1}{V} \sum_{\vec p' \vec q'} \sum_{\sigma_2} \, \sum_{n_1 n_2 n_3} \, \sum_{\vec G_1 \vec G_2} U(\vec q') \nonumber\\
&\quad\,\times u^*_{n_1 \vec k + \vec q' + \vec G_1}  u^*_{n_2 \vec p' - \vec q' + \vec G_2}  \nonumber\\
&\quad\,\times u_{n_3 \vec p' + \vec G_2}  u_{i' \vec k + \vec G_1} \nonumber\\
&\quad\,\times c^{\sigma'\dagger}_{n_1 \vec k + \vec q'}(t)  c^{\sigma_2 \dagger}_{n_2 \vec p' - \vec q'}(t)   c^{\sigma_2}_{n_3 \vec p'}(t) 
\,,
\label{eq:Heecom1}
\end{align}
and 
\begin{align}
\left[ \mathcal{H}_{e-e},  c_{i \vec k + \vec q}^{\sigma }(t)  \right] &= -\frac{1}{V} \sum_{\vec p' \vec q'} \sum_{\sigma_2} \, \sum_{n_2 n_3 n_4 } \sum_{\vec G_1 \vec G_2} U(\vec q') \nonumber\\
&\quad\,\times u^*_{i \vec k + \vec q + \vec G_1}  u^*_{n_2 \vec p' - \vec q' + \vec G_2}  \nonumber\\
&\quad\,\times u_{n_3 \vec p' + \vec G_2}  u_{n_4 \vec k + \vec q  - \vec q' + \vec G_1} \nonumber\\
&\quad\,\times  c^{\sigma_2 \dagger}_{n_2 \vec p' - \vec q'}(t)   c^{\sigma_2}_{n_3 \vec p'}(t) c^\sigma_{n_4 \vec k + \vec q  - \vec q'} (t)
\,.
\label{eq:Heecom2}
 \end{align}
 Inserting the commutators in Eqs.~(\ref{eq:H0com}), (\ref{eq:Heecom1}) and (\ref{eq:Heecom2}) into the right-hand-side of Eq.~(\ref{eq:ddtcom}), one generates two products of pairs of creation and annihilation operators, which, in a ``mean-field approximation'', we decouple as follows, 
\begin{align}
c^{\sigma'\dagger}_{n_1 \vec k + \vec q'} c^{\sigma_2 \dagger}_{n_2 \vec p' - \vec q'}   c^{\sigma_2}_{n_3 \vec p'} c^\sigma_{i\vec k+\vec q} &\simeq 
\langle c^{\sigma_2 \dagger}_{n_2 \vec p' - \vec q'}   c^{\sigma_2}_{n_3 \vec p'}  \rangle \, c^{\sigma'\dagger}_{n_1 \vec k + \vec q'} c^\sigma_{i\vec k+\vec q} \nonumber\\
&- \langle c^{\sigma' \dagger}_{n_1 \vec k + \vec q'}  c^{\sigma_2}_{n_3 \vec p'} \rangle \, c^{\sigma_2}_{n_2 \vec p' - \vec q'}  c^\sigma_{i\vec k+\vec q} 
\nonumber\\
&+ \langle c^{\sigma'\dagger}_{n_1 \vec k + \vec q'} c^\sigma_{i\vec k+\vec q} \rangle \, c^{\sigma_2 \dagger}_{n_2 \vec p' - \vec q'}   c^{\sigma_2}_{n_3 \vec p'} \nonumber\\
&- \langle c^{\sigma_2 \dagger}_{n_2 \vec p' - \vec q'} c^{\sigma}_{i \vec k + \vec q}   \rangle c^{\sigma'\dagger}_{n_1 \vec k + \vec q'}   c^{\sigma_2}_{n_3 \vec p'}  
\,,
\label{eq:dec1}
\end{align}
and
\begin{align}
c^{\sigma'\dagger}_{i' \vec k} c^{\sigma_2\dagger}_{n_2 \vec p' - \vec q'} c^{\sigma_2}_{n_3 \vec p'} c^{\sigma}_{n_4 \vec k + \vec q - \vec q'} &\simeq
\nonumber\\
&\hspace{-0.3 cm}+\langle c^{\sigma_2\dagger}_{n_2 \vec p' - \vec q'} c^{\sigma_2}_{n_3 \vec p'} \rangle \, c^{\sigma'\dagger}_{i' \vec k} c^{\sigma}_{n_4 \vec k + \vec q - \vec q'} \nonumber\\
&\hspace{-0.3 cm}-\langle c^{\sigma_2\dagger}_{n_2 \vec p' - \vec q'}  c^{\sigma}_{n_4 \vec k + \vec q - \vec q'}  \rangle \, c^{\sigma'\dagger}_{i' \vec k} c^{\sigma_2}_{n_3 \vec p'} \nonumber\\
&\hspace{-0.3 cm}+\langle c^{\sigma'\dagger}_{i' \vec k} c^{\sigma}_{n_4 \vec k + \vec q - \vec q'} \rangle\, c^{\sigma_2\dagger}_{n_2 \vec p' - \vec q'} c^{\sigma_2}_{n_3 \vec p'}  \nonumber\\
&\hspace{-0.3 cm}-\langle c^{\sigma'\dagger}_{i' \vec k} c^{\sigma_2}_{n_3 \vec p'} \rangle \, c^{\sigma_2\dagger}_{n_2 \vec p' - \vec q'} c^{\sigma}_{n_4 \vec k + \vec q - \vec q'} 
\label{eq:dec2}
\end{align}
where we omit terms involving the product of two expectations values as they commute with $c_{j' \vec k' + \vec q}^{\rho' \dagger}(t') c_{j \vec k'}^{\rho}(t')$, and thus do not contribute to the equation of motion for $\chi_{j_\alpha j_\beta}^{ii'\sigma\sigma'}(\vec k, \vec q,t-t')$. Notice that the expectation values in Eqs.~(\ref{eq:dec1}) and (\ref{eq:dec2}) can be expressed in terms of equilibrium occupation numbers and Kronecker deltas, as in Eq.~(\ref{eq:f0}).

The first and second lines in Eqs.~(\ref{eq:dec1}) and (\ref{eq:dec2}) 
contribute to the time derivative of $c_{i' \vec k}^{\sigma' \dagger}c_{i \vec k + \vec q}^{\sigma} $  
by renormalising the energies $\varepsilon_{i\vec k + \vec q}^\sigma$ and $\varepsilon_{i'\vec k}^{\sigma'}$, and will therefore not be considered further.~The third and fourth lines in Eqs.~(\ref{eq:dec1}) and (\ref{eq:dec2}) contribute to the time derivative in Eq.~(\ref{eq:ddtcom}) 
as follows 
\begin{widetext}
\begin{align}
 i \frac{\rm d}{{\rm d} t} \left( c_{i' \vec k}^{\sigma' \dagger}c_{i \vec k + \vec q}^{\sigma} \right) &= 
 \left( \varepsilon_{i \vec k + \vec q}^{\,\sigma} - \varepsilon_{i' \vec k}^{\,\sigma'} \right) c_{i' \vec k}^{\sigma' \dagger}c_{i \vec k + \vec q}^{\sigma} \nonumber\\
 &+\frac{\Big[  f_0( \varepsilon^{\, \sigma'}_{i' \vec k}) - f_0( \varepsilon^{\, \sigma}_{i \vec k + \vec q})  \Big]}{V}  \sum_{\vec p'}  \sum_{n_2 n_3} \sum_{\sigma_2 \sigma_3} \Bigg\{ 
 U (\vec q)  \mathscr{J}_{n_0}^{i'i\sigma' \sigma}(\vec k, \vec k + \vec q) \mathscr{J}_{n_0}^{n_3 n_2 \sigma_3 \sigma_2}(\vec p', \vec p' - \vec q)  \nonumber\\
&- U(\vec p' - \vec k - \vec q) \,\delta_{\sigma' \sigma_2} \delta_{\sigma \sigma_3}  \sum_{\vec G_1 \vec G_2} u^*_{n_2 \, \vec p' - \vec q + \vec G_1}\, u^*_{i\, \vec k + \vec q + \vec G_2} \, u_{n_3\, \vec p' + \vec G_2} \, u_{i'\,\vec k+ \vec G_1} 
\Bigg\}  \, c_{n_2 \vec p'-\vec q}^{\sigma_2 \dagger} c_{n_3 \vec p'}^{\sigma_3} 
\,,
 \label{eq:ddtcom2}
\end{align}
\end{widetext}
where the first term arises from the commutators in Eq.~(\ref{eq:H0com}), while the second (third) term originates from the third (fourth) line in Eqs.~(\ref{eq:dec1}) and (\ref{eq:dec2}).~Within the Hubbard approximation introduced in~\cite{Hubbard1958Jan}, we simplify the third line in Eq.~(\ref{eq:ddtcom2}) by neglecting the terms with $\vec G_1\neq0$ and $\vec G_2\neq0$ (i.e.~corresponding to Umklapp processes), and noticing that the largest contribution to the sum over $\vec p'$ arises from momenta with $|\vec p' - \vec k| \simeq \vec k_F$, where $\vec k_F$ is the material's Fermi momentum.~We account for this latter point by replacing $U(\vec p' - \vec k - \vec q)$ with $4\pi \alpha/(q^2 + k_F^2)$ in the above expression.~Introducing then the following function, which is called the {\it local-field factor},
\begin{align}
G(\vec q) \equiv \frac{1}{2}\frac{q^2}{q^2 + k_F^2}\,,
\end{align}
we can finally combine Eqs.~(\ref{eq:ddtcom2}) and~(\ref{eq:etcom}) with the definition in Eq.~(\ref{eq:chires}) to write down the following equation of motion
\begin{widetext}
\begin{align}
i \frac{\rm d}{{\rm d} t} \chi^{ii'\sigma\sigma'}_{j_\alpha j_\beta}(\vec k, \vec q,t-t') &= ( \varepsilon_{i \vec k + \vec q}^{\sigma} - \varepsilon_{i' \vec k}^{\sigma'} ) \chi^{ii'\sigma\sigma'}_{j_\alpha j_\beta}(\vec k, \vec q,t-t') \nonumber\\ 
&+ \frac{ f_0(\varepsilon^\sigma_{i \vec k + \vec q}) -  f_0(\varepsilon^{\sigma'}_{i' \vec k}) }{V} \Bigg\{\delta(t-t') 
\mathscr{J}_\alpha^{ii'\sigma\sigma'}(\vec k + \vec q,\vec k) 
\mathscr{J}_\beta^{i'i\sigma' \sigma}(\vec k,\vec k+\vec q) \nonumber\\
 &- U(\vec q) \left[ 1- G(\vec q) \right] \mathscr{J}_{\alpha}^{ii'\sigma\sigma'}(\vec k + \vec q,\vec k) \mathscr{J}_{n_0}^{i'i\sigma'\sigma}(\vec k, \vec k + \vec q)  \,
 \chi_{n_0 j_\beta}(\vec q, t-t') \Bigg\} \,. 
  \label{eq:tev}
\end{align}
\end{widetext}
where in the term proportional to the $G(\vec q)$ function, we used 
\begin{align}
\chi_{n_0 j_\beta}^{ii'\sigma\sigma'}(\vec k,\vec q,t-t') = \delta_{\sigma \sigma'} \frac{1}{2} \sum_{\rho \rho'}\chi_{n_0 j_\beta}^{ii'\rho\rho'}(\vec k,\vec q,t-t') \,,
\end{align}
and only accounted for the spin-diagonal contribution proportional to $\delta_{\sigma \sigma'}$.

\subsection{Solution in frequency space}
\label{sec:sol}
By rewriting $\chi_{j_\alpha j_\beta}(\vec k, \vec q,t-t')$ in terms of its Fourier transform, $\chi_{j_\alpha j_\beta}(\vec k, \vec q,\omega)$, Eq.~(\ref{eq:tev}) becomes an algebraic equation, which can be solved exactly after summing left- and right-hand sides over reciprocal space vectors $\vec k$, spin indices $\sigma$ and $\sigma'$ as well as band indices $i$ and $i'$.~Introducing,
\begin{align}
\Sigma_{j_\alpha j_\beta}(\vec q,\omega) = \frac{1}{V} &\sum_{\vec k} \sum_{ii'} \sum_{\sigma \sigma'}
\frac{f_0(\varepsilon^\sigma_{i \vec k + \vec q}) -  f_0(\varepsilon^{\sigma'}_{i' \vec k})}
{\omega- \varepsilon_{i \vec k + \vec q}^{\sigma} + \varepsilon_{i' \vec k}^{\sigma'} + i \delta} \nonumber\\
&\times\mathscr{J}_\alpha^{ii'\sigma\sigma'}(\vec k + \vec q,\vec k) \mathscr{J}_\beta^{i'i\sigma' \sigma}(\vec k,\vec k+\vec q),
\label{eq:sigmaab}
\end{align}
we find,
\begin{align}
\chi_{j_\alpha j_\beta} (\vec q , \omega) &=\Sigma_{j_\alpha j_\beta} (\vec q , \omega) \nonumber\\
&- \Sigma_{j_\alpha n_0}  (\vec q , \omega) U(\vec q) \left[ 1 - G(\vec q)\right] \chi_{n_0 j_\beta}  (\vec q , \omega) \,,
  \label{eq:tevsol}
\end{align}
Before solving Eq.~(\ref{eq:tevsol}) to obtain an explicit expression for $\chi_{j_\alpha j_\beta}  (\vec q , \omega)$, let us notice that for $j_\alpha=n_0$, Eq.~(\ref{eq:tevsol}) implies 
\begin{align}
\chi_{n_0 j_\beta}(\vec q, \omega) = \frac{\Sigma_{n_0 j_\beta} (\vec q, \omega) }{1 + U(\vec q) \left[ 1 - G(\vec q) \right] \Sigma_{n_0 n_0} (\vec q, \omega)}\,,
\label{eq:n0b}
\end{align}
which for $j_\beta=n_0$ gives the density-density response function
\begin{align}
\chi_{n_0 n_0}(\vec q, \omega) &= \frac{\Sigma_{n_0 n_0} (\vec q, \omega) }{1 + U(\vec q) \left[ 1 - G(\vec q) \right] \Sigma_{n_0 n_0} (\vec q, \omega)}
\nonumber\\
&= \frac{\Sigma_{n_0 n_0} (\vec q, \omega) }{\epsilon_r (\vec q, \omega) }\,,
\label{eq:n0n0}
\end{align} 
where in the second line we identified the dielectric function with\footnote{Strictly speaking, Eq.~(\ref{eq:eRPA}) gives the dielectric function only for $G=0$. For $G\neq 0$, the dielectric function is given by the right-hand side of Eq.~(\ref{eq:eRPA}) divided by $1-U G \Sigma_{n_0 n_0}$ \cite{Solyom2010}. This slight abuse of notation helps us keeping the relation between $\chi_{n_0 n_0}$ and $\Sigma_{n_0 n_0}$ simple, and does not affect any of the numerical results, which are based on a direct calculation of $\Sigma_{n_0 n_0}$ and on Eq.~(\ref{eq:final}).}
\begin{align}
\epsilon_r (\vec q, \omega) = 1 + U(\vec q) \left[ 1 - G(\vec q) \right] \Sigma_{n_0 n_0} (\vec q, \omega) \,.
\label{eq:eRPA}
\end{align}
Notice that the plus sign in front of $U(\vec q)$ arises from our definition of generalised susceptibility in Eq.~(\ref{eq:chiret}).~Inserting now Eq.~(\ref{eq:n0b}) into Eq.~(\ref{eq:tevsol}), we obtain our final expression for the generalised susceptibility $\chi_{j_\alpha j_\beta}  (\vec q , \omega)$, namely,
\begin{widetext}
\begin{align}
\chi_{j_\alpha j_\beta}(\vec q, \omega) &= 
 \Sigma_{j_\alpha j_\beta} (\vec q , \omega) - \frac{\Sigma_{j_\alpha n_0} (\vec q , \omega) U(\vec q) \left[ 1 - G(\vec q) \right]  \Sigma_{n_0 j_\beta} (\vec q, \omega)}{1 + U(\vec q) \left[ 1 - G(\vec q) \right] \Sigma_{n_0 n_0} (\vec q, \omega)} 
\,,
\label{eq:final}
\end{align}
\end{widetext}
which is one of the main results of our work.~Let us interpret this result by first focusing on the case $G=0$, where the generalised susceptibility $\chi_{j_\alpha j_\beta}(\vec q, \omega)$ can be written as
\begin{align}
\chi_{j_\alpha j_\beta}(\vec q, \omega) &= 
 \Sigma_{j_\alpha j_\beta} (\vec q , \omega) - \frac{\Sigma_{j_\alpha n_0} (\vec q , \omega) U(\vec q) \Sigma_{n_0 j_\beta} (\vec q, \omega)}{1 + U(\vec q)  \Sigma_{n_0 n_0} (\vec q, \omega)} \,.
 \label{eq:G0}
\end{align}
For $G=0$, Eq.~(\ref{eq:eRPA}) gives the dielectric function in the random phase approximation (RPA), Eq.~(\ref{eq:n0n0}) reproduces the RPA result for the density-density response function, while Eq.~(\ref{eq:G0}) with $j_\alpha=j_{M,\alpha}$ and $j_\beta=j_{M,\beta}$, $\alpha,\beta=1,2,3$ (i.e.~the spatial components of the paramagnetic current, see Eq.~(\ref{eq:nj})), gives the known RPA result for the current-current response function in electrodynamics.~We thus conclude that,  for $G=0$, our formalism based on linear response theory, the equation of motion method, and the mean field approximation in Eqs.~(\ref{eq:dec1}) and (\ref{eq:dec2}) provides us with generalised susceptibilities in the RPA limit.

Our RPA results capture potentially important effects related with screening and collective excitations in detector materials.~This is simple to illustrate by focusing on the generalised susceptibilities $\chi_{n_0 j_\beta}(\vec q, \omega)$, with $G=0$.~After rationalising the denominator in Eq.~(\ref{eq:n0b}), we find
\begin{align}
\chi_{n_0 j_\beta} = \frac{\Sigma_{n_0 j_\beta} \left(1 + U(\vec q) \Sigma_{n_0 n_0}  \right)^* }{\left[1 + U(\vec q) \Re \Sigma_{n_0 n_0} \right]^2
+\left[U(\vec q) \Im \Sigma_{n_0 n_0} \right]^2}\,,
\label{eq:inmed}
\end{align}
where we omitted the dependence on momentum and energy of $\Sigma_{n_0 j_\beta}$ and of the real and imaginary parts of $\Sigma_{n_0 n_0}$ to simplify the notation.~As one can see from Eq.~(\ref{eq:inmed}), for frequencies $\omega$ and momenta $\vec q$ such that $U(\vec q) \Re \Sigma_{n_0 n_0} \simeq -1$ and $U(\vec q) \Im \Sigma_{n_0 n_0} \ll 1$, the susceptibility $\chi_{n_0 j_\beta}$ is enhanced by collective excitations.~For $|U(\vec q) \Sigma_{n_0 n_0}| \gg 1$, it is suppressed by screening effects.~We will refer to these phenomena as ``in-medium'' effects.

Going beyond the RPA approximation, let us now focus on the case $G\neq0$.~In order to understand the implications of $G\neq0$, let us use Eq.~(\ref{eq:induced}) to introduce the density,
\begin{align}
n_{\rm ind}(\mathbf{r},t) &= \sum_\alpha \int_{-\infty}^{t} {\rm d} t' \int {\rm d}\mathbf{r}'\, \chi_{n_0j_\alpha}(\mathbf{r}-\mathbf{r}',t-t') \nonumber \\
&\times S_\alpha^{ss'}(\mathbf{r}',t')\,,
\label{eq:inducedn0}
\end{align}
where $n_{\rm ind}\equiv\langle \Delta n_0\rangle$ is the change in electron density in the given material induced by the external 
DM perturbation of strength $S_\alpha^{ss'}$.~In analogy with the electrostatic case~\cite{Solyom2010Dec}, we can now introduce a fictitious ``electron density'', $n_{\rm ext}(\mathbf{r},t)$, which represents the source of the external  DM perturbation, and which is defined as follows, 
\begin{align}
n_{\rm ext}(\vec q, \omega) \equiv \sum_{\beta} \Sigma_{n_0\beta} (\vec q,\omega) S_\beta^{ss'}(\vec q, \omega)
\frac{1}{U(\vec q) \Sigma_{n_0 n_0}(\vec q,\omega)}\,.
\label{eq:next}
\end{align}
Using Eq.~(\ref{eq:next}), we find that Eq.~(\ref{eq:tevsol}) implies the following relation between the induced and external electron densities,
\begin{align}
n_{\rm ind}(\vec q,\omega) &= \Big[ n_{\rm ext}(\vec q,\omega) \nonumber\\
&-(1-G(\vec q)) n_{\rm ind}(\vec q,\omega) \Big] U(\vec q) \Sigma_{n_0 n_0}(\vec q, \omega)\,.
\label{eq:indvsext}
\end{align}
Recalling now that the density-density response function is defined as the ratio of the electron density induced by the external perturbation, $n_{\rm ind}$, and the total electron density in the material, $n_{\rm eff}$~\cite{Solyom2010Dec}, we can rewrite Eq.~(\ref{eq:indvsext}) as 
\begin{align}
n_{\rm ind}(\vec q,\omega) = U(\vec q) \Sigma_{n_0 n_0}(\vec q, \omega) n_{\rm eff}(\vec q, \omega)\,,
\label{eq:neff}
\end{align}
where
\begin{align}
n_{\rm eff}(\vec q, \omega) \equiv n_{\rm ext}(\vec q,\omega) -(1-G(\vec q)) n_{\rm ind}(\vec q,\omega) \,.
\end{align}
We conclude that, for $G\neq0$, the number of electrons actually contributing to the screening of $n_{\rm ext}$ in Eq.~(\ref{eq:neff}), is reduced by a factor of $1-G(\vec q)$, e.g.~1/2 in the large $|\vec q|$ limit.
This reduction can be understood by realising that for small distances (i.e.~large $|\vec q|$) the spin-resolved electron density-density correlation function\footnote{Notice that the normalised, spin-resolved electron density-density correlation function gives the probability of finding an electron with identical or opposite spin around an electron of a given spin.}  drops to zero for electron pairs of the same spin because of the Pauli exclusion principle, as can been shown analytically within the Hartree-Fock approximation~\cite{Solyom2010}.~Consequently, in the large $|\vec q|$ limit only half of the electrons can contribute to the screening of the external electron density given in Eq.~(\ref{eq:next}).

By neglecting both Hubbard and RPA corrections, we find that the generalised susceptibility $\chi_{j_\alpha j_\beta}$ further simplifies to
\begin{align}
\chi_{j_\alpha j_\beta}(\vec q, \omega) &= 
 \Sigma_{j_\alpha j_\beta} (\vec q , \omega) \,.
 \label{eq:noin}
\end{align}
This equation neglects in-medium effects and reproduces our previous results obtained by using single-particle atomic wave functions~\cite{Catena:2019gfa} and Bloch states expanded in a plane wave basis~\cite{Catena:2021qsr}, as we will see in Secs.~\ref{sec:num} and \ref{sec:comp}.

Notice that a change in the underlying electron wave functions would primarily affect the $\mathscr{J}_{\alpha}^{ii'\sigma\sigma'}$ coefficients in the second quantisation form of the electron densities and currents. Since we have expressed the solution to the equation of motion for the relevant generalised susceptibilities in terms the $\mathscr{J}_{\alpha}^{ii'\sigma\sigma'}$ coefficients, the results presented in our manuscript are fairly material-independent, as long as we restrict ourselves to non spin-polarised materials.

\subsection{Diagrammatic interpretation}
\label{sec:diag}
The solution in Eq.~(\ref{eq:final}) admits an insightful diagrammatic representation that is valid for $|U(\vec q)(1-G(\vec q)) \Sigma_{n_0 n_0}(\vec q,\omega)| < 1$.~To illustrate this point, we first rewrite the susceptibility $\chi_{j_\alpha j_\beta}(\vec q,\omega)$ as a geometric series, 
\begin{align}
\chi_{j_\alpha j_\beta}(\vec q,\omega) &= \Sigma_{j_\alpha j_\beta}(\vec q,\omega) \nonumber\\ 
&+ \Sigma_{j_\alpha n_0} (\vec q , \omega) U(\vec q) \left[ G(\vec q) -1 \right]  \Sigma_{n_0 j_\beta} (\vec q, \omega) \nonumber\\
&\times \sum_{\ell=0}^{\infty} \left[ U(\vec q) (G(\vec q)-1) \Sigma_{n_0 n_0}(\vec q,\omega) \right]^\ell \,.
\label{eq:geom}
\end{align}
Recalling then that the susceptibility $\chi_{j_\alpha j_\beta}$ is by definition a retarded Green's function, and that it thus describes the propagation of an electron-hole pair in a medium, we can represent the first term in Eq.~(\ref{eq:geom}) as follows,

\begin{figure}[h!]
    \centering
    \includegraphics[width=0.49\textwidth]{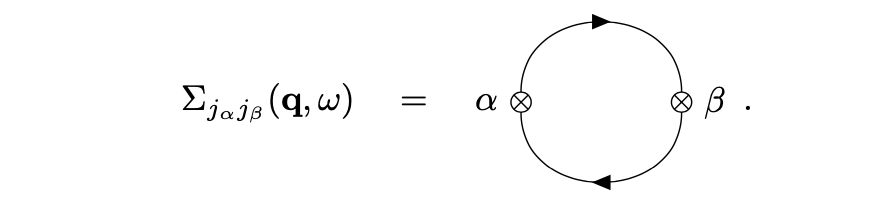}
\end{figure}    
\noindent This irreducible diagram describes the creation of an electron-hole pair in an interaction associated with the density or current $j_\alpha$ followed by its annihilation induced by $j_\beta$.\\

The geometric series in the second and third line of Eq.~(\ref{eq:geom}) describes in-medium effects that are not captured by $\Sigma_{j_\alpha j_\beta}$.~The term with $\ell=1$ can be represented by 

\begin{figure}[h!]
    \centering
    \includegraphics[width=0.49\textwidth]{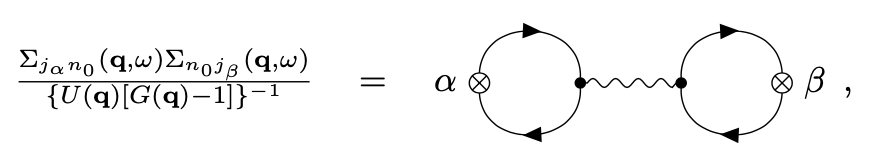}
\end{figure}  
\noindent where the Coulomb repulsion and exchange factor $U(\vec q)[G(\vec q)-1]$ has been represented diagrammatically by a wiggled line.~Here and in what follows, we denote the vertices associated with the density $n_0$ by a black dot.

By including the remaining terms with $\ell>1$, we finally obtain the desired diagrammatic representation for $\chi_{j_\alpha j_\beta}$, namely

\begin{figure}[h!]
    \centering
    \includegraphics[width=0.49\textwidth]{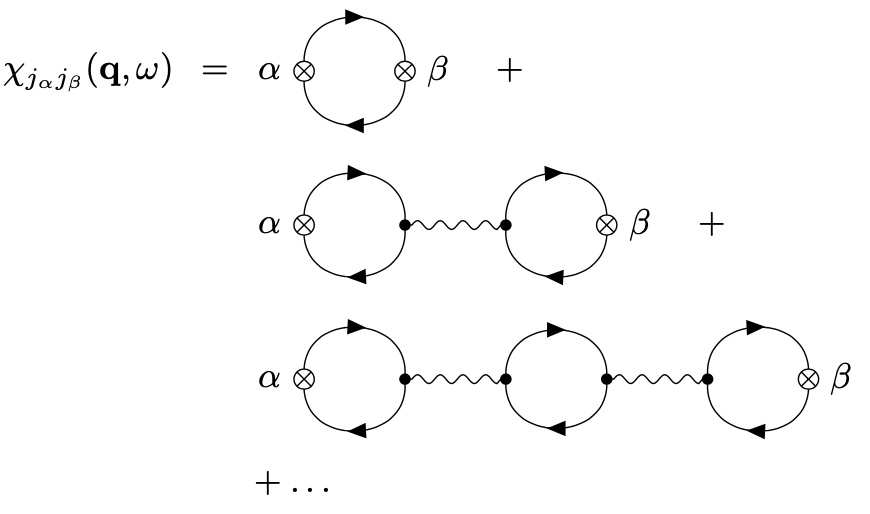}
\end{figure}
Our diagrammatic representation for $\chi_{j_\alpha j_\beta}$ clearly illustrates that the interaction between a propagating electron-hole pair and the surrounding medium is governed by the Coulomb repulsion and exchange factor $U(\vec q)[G(\vec q)-1]$, as well as by the density-density response function $\Sigma_{n_0 n_0}$, but it does not depend on the details of the underlying DM interaction, which are encoded in $j_\alpha$ and $j_\beta$. 

\section{Screened vs unscreened susceptibilities}
In this section, we focus on the numerical implementation of Eq.~(\ref{eq:final}).~In particular, we are interested in the relative size of screened and unscreened contributions to $\chi_{j_\alpha j_\beta}$.

\subsection{Unscreened susceptibilities: $\Sigma_{j_\alpha j_\beta}$}
\label{sec:num}
Let us start our study by showing that the first term in Eq.~(\ref{eq:final}) 
can be related to the ``response functions'' we computed in 
~\cite{Catena:2021qsr} by using single-particle 
Bloch states expanded in a plane wave basis.~To this end, let us introduce the scalar and vector electron wave function overlap integrals,
\begin{align}
f_{i \rightarrow f} (\mathbf{q}) &=  \int {\rm d} \mathbf{r} \,\psi^*_{f}(\mathbf{r}) e^{i \mathbf{q}\cdot \mathbf{r}}\psi_{i}(\mathbf{r})\,, \nonumber\\
\mathbf{f}_{i\rightarrow f} (\mathbf{q})& =  -\frac{i}{m_e}\int {\rm d} \mathbf{r} \,\psi^*_{f}(\mathbf{r}) e^{i \mathbf{q}\cdot \mathbf{r}}\nabla_{\mathbf{r}}\psi_{i}(\mathbf{r}) \,, 
\label{eq:f}
\end{align}
where 
\begin{align}
\psi_{i}(\mathbf{r}) &= \frac{1}{\sqrt{V}}  \sum_{\vec G} e^{i (\vec k +\vec G) \cdot \vec r} \, u_{i\,\vec k+\vec G} \eta^\sigma   \,, \nonumber \\
\psi_{f}(\mathbf{r}) &= \frac{1}{\sqrt{V}}  \sum_{\vec G'} e^{i (\vec k' +\vec G') \cdot \vec r} \, u_{i'\,\vec k'+\vec G'}  \eta^{\sigma'} \,.
\label{eq:bloch0}
\end{align}
Notice the minus sign in the equation for $\mathbf{f}_{i\rightarrow f} (\mathbf{q})$: this was missing in~\cite{Catena:2019gfa,Catena:2021qsr,Catena:2022fnk}, where the response function $W_2$ (defined below Eq.~(\ref{eq:ffsig3})) has the wrong sign. Furthermore, let us introduce the following compact notation,
\begin{align}
|f|^2 &= \frac{1}{2}\sum_{i,f} \frac{e^{-\bar{\beta} E_i}}{Z}f_{i\rightarrow f} f^*_{i\rightarrow f} \,(2\pi)\delta(E_f-E_i-\omega)\nonumber\\
f_{l} f_{m}^*&= \frac{1}{2} \sum_{i,f} \frac{e^{-\bar{\beta} E_i}}{Z}\mathbf{f}_{i\rightarrow f}\cdot \mathbf{e}_{(l)}\,\, \mathbf{f}^*_{i\rightarrow f}\cdot \mathbf{e}_{(m)} \nonumber\\ &\quad\times (2\pi)\delta(E_f-E_i-\omega)\nonumber\\
f f^*_l &= \frac{1}{2}\sum_{i,f} \frac{e^{-\bar{\beta} E_i}}{Z}f_{i\rightarrow f} \mathbf{f}^*_{i\rightarrow f}\cdot \mathbf{e}_{(l)} \,(2\pi)\delta(E_f-E_i-\omega) \,,
\end{align}
where $\mathbf{e}_l$ and $\mathbf{e}_m$ are unit vectors in the $l$-th and $m$-th direction of a cartesian coordinate system, while the sums read as
\begin{align}
\sum_{i,f}  &= \sum_{ii'\vec k \vec k' \sigma \sigma'} \,.
\end{align}
Notice that, e.g.
\begin{align}
f_{i\rightarrow f} f^*_{i\rightarrow f} &= \left|  \langle f| e^{i \vec q \cdot \vec r}| i \rangle \right|^2_{\vec k' - \vec k - \vec q + \Delta \vec G=0} \,.
\end{align}
Here, $\Delta \vec G$ is the unique reciprocal lattice vector such that, for a given $\vec q$, $\vec k - \vec k'$ is in the first Brillouin Zone. 
With this notation, we take the $\delta\rightarrow 0^+$ limit in Eq.~(\ref{eq:sigmaab}) and find,
\begin{align}
\Im (\Sigma_{n_0^\dagger n_0} ) &= \Omega |f|^2
\nonumber\\
\Im (\Sigma_{n_A^\dagger n_A} ) &=\Omega\left[ 
\frac{q^2}{4m_e^2} |f^2| + \mathbf{f} \cdot \mathbf{f}^* +\frac{q_i}{m_e} \Re(ff^*_i) \right]  \,,
\label{eq:ffsig1}
\end{align}
where
\begin{align}
\Omega =   \frac{1}{V} \left( 1-e^{-\bar{\beta} \omega} \right) \,.
\end{align}
By using the notation,
\begin{align}
\Delta \Sigma_{j_\alpha j_\beta} \equiv \left(\Sigma_{j_\alpha j_\beta} - \Sigma^A_{j_\alpha j_\beta}\right)|_{\delta\rightarrow 0^+} 
\,,
\end{align}
and combining Eq.~(\ref{eq:sigmaab}) with the spectral representation for the anticipated susceptibilities, Eq.~(\ref{eq:spechiA}), we also find
\begin{align}
\Im \Delta \Sigma_{j_{5l}^\dagger j_{5m}}  &= 2\Omega |f|^2 \,\delta_{lm}
\nonumber\\
\Im  \Delta \Sigma_{j_{Ml}^\dagger j_{Mm}} &=  2\Omega\bigg[ \Im (i f_l^* f_m) 
+ \Im \left( \frac{i q_m}{2m_e} ff^*_l + \frac{i q_l}{2m_e} f^*f_m \right) \nonumber\\
&+ \frac{q_lq_m}{4m_e^2} |f^2| \bigg]  
\nonumber\\
\Re  \Delta \Sigma_{j_{Ml}^\dagger j_{Mm}} &=  2\Omega \,\Re \bigg[i f^*_l f_m  + \left( \frac{i q_m}{2m_e} ff^*_l + \frac{i q_l}{2m_e} f^*f_m \right) \bigg] \nonumber\\
\Im  \Delta \Sigma_{j_{El}^\dagger j_{Em}} &=  2\Omega \left( \delta_{l m} \delta_{ss'} -  \delta_{ls'} \delta_{sm}   \right) \bigg[ \frac{q_sq_{s'}}{4m_e^2} |f^2| 
\nonumber\\
 &+ \Im (i f_s^*f_{s'} ) + \Im \left( \frac{i q_{s'}}{2m_e} ff^*_{s} + \frac{i q_{s}}{2m_e} f^*f_{s'} \right) \bigg] \,.
\label{eq:ffsig2}
\end{align}
Finally, for the ``off-diagonal'' susceptibilities that contribute to the rate of DM-induced electronic transitions, we find
\begin{align}
\Im \Delta \Sigma_{j^\dagger_{Ml} n_0} &= \Omega \left[ \frac{q_l}{m_e} |f^2| +2 \Re(ff^*_l) \right] \nonumber\\
\Re \Delta\Sigma_{j^\dagger_{Ml} n_0} &= -2\Omega \Im (ff^*_l) 
\nonumber\\
\Im \Delta \Sigma_{n_A^\dagger j_{5l}} &= \Im \Delta \Sigma_{j^\dagger_{Ml} n_0} \nonumber\\
\Re \Delta \Sigma_{n_A^\dagger j_{5l}}  &= \Re \Delta \Sigma_{j^\dagger_{Ml} n_0} \nonumber\\
\Im \Delta \Sigma_{j^\dagger_{5l} j_{Em}} &= -2\Omega \epsilon_{ilm} \Im(ff^*_{i}) \nonumber\\
\Re \Delta \Sigma_{j^\dagger_{5l} j_{Em}} &= \Omega \bigg[
\frac{q_i}{m_e} \epsilon_{ilm}|f|^2 + 2\epsilon_{ilm}\Re(ff^*_i) \bigg]  \nonumber\\ 
\Im \Delta \Sigma_{j^\dagger_{El} n_A} &= -2\Omega \bigg[ i
\left( \mathbf{f} \times \mathbf{f}^* \right)_l + \epsilon_{lmn}\frac{q_n}{m_e}  \Im(f f^*_m) \bigg] \,.
\label{eq:ffsig3}
\end{align}
All other susceptibilities vanish.~Eqs.~(\ref{eq:ffsig1}), (\ref{eq:ffsig2}) and (\ref{eq:ffsig3}) allow us to derive explicit relations between the trace, longitudinal and transverse parts of our generalised susceptibilities and the crystal response functions of~\cite{Catena:2021qsr}, here denoted by $W_i\equiv W_i(\vec q, \omega)$, with $i=1,\dots,5$.~Specifically,
\begin{align}
\Im (\Sigma_{n_0^\dagger n_0} ) &= \frac{\pi^2 \widetilde{\Omega}}{ \omega }  \, W_{1} \nonumber\\
\Im (\Sigma_{n_A^\dagger n_A} ) &= \frac{\pi^2 \widetilde{\Omega}}{ \omega }  \left[ 
\frac{q^2}{4m_e^2} W_{1} + W_{3} +\Re(W_{2}) \right] \,,
\label{eq:wsig1}
\end{align}
where $\widetilde{\Omega} =N_{\rm cell} \Omega$.~Furthermore,
\begin{align}
\frac{q_l q_m}{m_e^2} \Im \Delta \Sigma_{j_{Ml}^\dagger j_{Mm}}  
&=  \frac{2\pi^2 \widetilde{\Omega}}{ \omega } \bigg[ \frac{q^4}{4m_e^4} W_{1}
+ W_{4} \nonumber \\ &+ \frac{q^2}{m^2_e}\Re(W_{2}) \bigg] \nonumber\\
\delta_{lm} \Im \Delta \Sigma_{j_{Ml}^\dagger j_{Mm}} &= \frac{2\pi^2 \widetilde{\Omega}}{ \omega} \bigg[ \frac{q^2}{4m_e^2} W_{1}
+ W_{3} + \Re(W_{2}) \bigg] \nonumber\\
\epsilon_{lmi} \frac{q_i}{m_e}\Re \Delta \Sigma_{j_{Ml}^\dagger j_{Mm}}&= -
\frac{2\pi^2 \widetilde{\Omega}}{ \omega } \,W_{5} \nonumber\\
\frac{q_l q_m}{m_e^2} \Im \Delta \Sigma_{j_{El}^\dagger j_{Em}} &= \frac{2\pi^2 \widetilde{\Omega}}{ \omega } \bigg[  \frac{q^2}{m_e^2} W_{3}  - W_{4} \bigg] \nonumber\\
\delta_{lm} \Im \Delta \Sigma_{j_{El}^\dagger j_{Em}} &= \frac{2\pi^2\widetilde{\Omega}}{ \omega} \bigg[ \frac{q^2}{2m_e^2} W_{1} 
+ 2 W_{3} + \Re(W_{2}) \bigg] 
\label{eq:wsig2}
\end{align}
and finally,
\begin{align}
\frac{q_l}{m_e} \Im \Delta \Sigma_{j_{Ml}^\dagger n_0} &=  \frac{\pi^2 \widetilde{\Omega}}{ \omega} \bigg[ \frac{q^2}{m_e^2}  W_{1} 
+2 \Re(W_{2}) \bigg] \nonumber\\
\frac{q_l}{m_e} \Re \Delta \Sigma_{j_{Ml}^\dagger n_0}  &= 
-  \frac{2 \pi^2 \widetilde{\Omega}}{ \omega} \,\Im(W_{2}) \nonumber\\
\frac{q_j}{m_e} \epsilon_{jlm}\Im \Delta \Sigma_{j_{5l}^\dagger j_{Em}} &= -\frac{4 \pi^2 \widetilde{\Omega}}{ \omega }  \Im(W_{2})\nonumber\\
\frac{q_j}{m_e} \epsilon_{jlm}\Re \Delta \Sigma_{j_{5l}^\dagger j_{Em}}  &= \frac{\pi^2 \widetilde{\Omega}}{ \omega }  \bigg[ \frac{2 q^2}{m_e^2}W_{1}
+ 4 \Re(W_{2})\bigg]  \nonumber\\
\frac{q_l}{m_e} \Im \Delta \Sigma_{j_{El}^\dagger n_A} &= -2 \frac{\pi^2 \widetilde{\Omega}}{ \omega } \, W_{5} \,.
\label{eq:wsig3}
\end{align}
In the numerical results presented in Sec.~\ref{sec:results}, we use Eqs.~(\ref{eq:wsig1}), (\ref{eq:wsig2}) and (\ref{eq:wsig3}) and the crystal response functions $W_i$, $i=1,\dots,5$, we previously computed for silicon and germanium in~\cite{Catena:2021qsr} to evaluate the first term in Eq.~(\ref{eq:final}), $\Sigma_{\alpha\beta}$.~In~\cite{Catena:2021qsr}, the numerical evaluation of the $W_i$ functions was implemented in {\sffamily QEdark-EFT}~\cite{Urdshals2021May}, an extension of the {\sffamily QEdark} code~\cite{Essig:2015cda}, which interfaces with the plane-wave self-consistent field (PWscf) DFT code {\sffamily Quantum ESPRESSO}~\cite{Giannozzi2009Sep}.~We refer to~\cite{Catena:2021qsr} for further details.

As a last point, we emphasize that starting from atomic wave functions, rather than the Bloch wave functions in Eq.~(\ref{eq:bloch0}), analogous relations could be established between the generalised susceptibilities identified in this work and the atomic response functions we introduced in~\cite{Catena:2019gfa}.

\subsection{Screened susceptibilities}
\label{sec:num2}
Let us now focus on the numerical evaluation of the in-medium corrections to the susceptibilities $\chi_{j_\alpha j_\beta}$, restricting ourselves to the case of non-spin-polarised materials.~Spin-polarised materials will be studied elsewhere in a separate work. 

In-medium corrections to the generalised susceptibilities $\chi_{j_\alpha j_\beta}$ are encoded in the second term in Eq.~(\ref{eq:final}).~The latter depends on the ``off-diagonal'' susceptibilities $\Sigma_{j_\alpha n_0}$ and $\Sigma_{n_0 j_\beta}$ which, for non-spin-polarised materials, are different from zero only when $j_\alpha$ and $j_\beta$ coincide with $\boldsymbol{j}_{M}$ or $n_0$.~In all other cases, $\Sigma_{j_\alpha n_0}$ and $\Sigma_{n_0 j_\beta}$ are proportional to the trace of a Pauli matrix, and therefore vanish.~Consequently, for non-spin-polarised materials, in-medium corrections are only relevant to the  susceptibilities $\chi_{n_0 n_0}$, $\chi_{j_{Ml} j_{Mm}}$ and $\chi_{j_{Ml} n_0}$\footnote{Recall that $n_0$, and $\mathbf{j}_M$ are hermitian operators}. As far as the density-density response function $\chi_{n_0 n_0}$ is concerned, Eq.~(\ref{eq:final}) implies,
\begin{align}
\Im \chi_{n_0 n_0} &= \frac{1}{U(1-G)}\frac{\Im \epsilon_r}{|\epsilon_r|^2} \nonumber\\
&= \frac{1}{U(1-G)} \left[ \Im \epsilon_r + \frac{1-|\epsilon_r|^2}{|\epsilon_r|^2} \Im \epsilon_r \right] \,,
\label{eq:n0n0in}
\end{align}
in agreement with previous works on the dielectric function \cite{Knapen:2021run,Hochberg:2021pkt}.~Notice, however, that here we account for the exchange correction $G$ which was neglected in previous works.~In the second line of Eq.~(\ref{eq:n0n0in}), we separated the screened contribution to $\Im \chi_{n_0 n_0}$ from the unscreened one. 

In order to simplify the evaluation of in-medium corrections to $\chi_{j_{Mm} j_{Ml}}$ and $\chi_{j_{Ml} n_0}$, we assume that there are no screening corrections to the transverse response. This approximation is exact in isotropic materials, and a good approximation in high-symmetry bulk crystals such as silicon and germanium~\cite{Catena:2021qsr}.~This allows us to write 
\begin{align}
\Sigma_{j_{Ml} n_0}(\vec q,\omega) = \Sigma_{j_{Mm} n_0}(\vec q,\omega) \hat{q}_m \hat{q}_l \,,
\label{eq:iso}
\end{align}
where $\hat{q}_l = q_l/q$ and repeated spatial indices are summed over.~Next, we use the electron number continuity equation,
\begin{align}
\omega \, \hat{n}_0(\vec q,\omega) = \vec q \cdot \hat{\boldsymbol{j}}_M(\vec q,\omega)
\end{align}
to obtain
\begin{align}
\Sigma_{j_{Ml} n_0} (\vec q,\omega) &= \frac{\omega}{q} \Sigma_{n_0 n_0} (\vec q, \omega) \,\hat{q}_l \,, \nonumber\\
\Sigma_{n_0 j_{Ml}} (\vec q,\omega) &= \frac{\omega}{q} \Sigma_{n_0 n_0} (\vec q, \omega) \,\hat{q}_l \,.
\label{eq:sigiso}
\end{align}
Finally, by using Eq.~(\ref{eq:sigiso}) for $\Sigma_{j_{Ml} n_0}$ and $\Sigma_{n_0 j_{Ml}}$, we find
\begin{align}
\chi_{j_{Ml} j_{Mm}} = \Sigma_{j_{Ml}j_{Mm}} - \frac{\omega^2}{q^2}\hat{q}_l \hat{q}_m \,U(1-G) \Sigma^2_{n_0n_0} 
\frac{\epsilon^*_r}{ |\epsilon_r|^2} \,.
\label{eq:MMin}
\end{align}
Notice that for $\Im \langle F_{j_{Ml}}^* F_{j_{Mm}}\rangle=0$
, only the imaginary part of Eq.~(\ref{eq:MMin}) contributes to the transition rate.~This applies to the case of magnetic dipole, electric dipole and anapole DM, as well as in simplified DM models with a single scalar or vector mediator. Consequently, in most of the numerical implementations we only need 
\begin{align}
\Im(\chi_{j_{Ml} j_{Mm}}) = \Im(\Sigma_{j_{Ml} j_{Mm}}) + \frac{\omega^2}{q^2}\hat{q}_l \hat{q}_m \,\frac{1-|\epsilon_r|^2}{U(1-G)} 
\frac{\Im (\epsilon_r)}{ |\epsilon_r|^2} \,,
\label{eq:MMin2}
\end{align}
where we used Eq.~(\ref{eq:eRPA}) to rewrite the density-density correlation function, $\Sigma_{n_0 n_0}$, as $(\epsilon_r-1)/[U(1-G)]$.~Interestingly, the in-medium corrections to $\Im(\chi_{j_{Ml} j_{Mm}})$ can be expressed entirely in terms of the dielectric function $\epsilon_r$.~Furthermore, these corrections are longitudinal, i.e.~proportional to $\hat{q}_l \hat{q}_m$, which is a direct consequence of Eq.~(\ref{eq:iso}).~In contrast, the unscreened susceptibility in the first term of Eq.~(\ref{eq:MMin2}) has both a longitudinal and a transverse component, as one can see by acting with $\hat{q}_m \hat{q}_l$ and $(\delta_{lm} - \hat{q}_l \hat{q}_m )$ on $\Im(\Sigma_{j_{Ml} j_{Mm}})$ using Eq.~(\ref{eq:ffsig2}).~Focusing on the longitudinal component of the unscreened susceptibility $\Im(\Sigma_{j_{Ml} j_{Mm}})$, we find
\begin{align}
\Im(\Sigma_{j_{Ml} j_{Mm}}) = \frac{\omega^2}{q^2} \, \Im \Sigma_{n_0 n_0}  \hat{q}_l \hat{q}_m\,,
\label{eq:long}
\end{align}
which cancels exactly the term proportional to $|\epsilon_r|^2$ in Eq.~(\ref{eq:MMin2}).~In order to obtain Eq.~(\ref{eq:long}), we used $\mathbf{f}_{i\rightarrow f} \cdot \vec q = f_{i\rightarrow f} \omega - f_{i\rightarrow f} q^2/(2 m_e)$, which follows from the continuity equation.

Performing an analogous calculation, for $\Delta \chi_{j_{Ml} n_0}$, namely,
\begin{align}
\Delta \chi_{j_{Ml} n_0}\equiv \left(\chi_{j_{Ml} n_0} - \chi^A_{j_{Ml} n_0}\right)\,,
\end{align}
we obtain
\begin{align}
\frac{q_l}{m_e} \Im \Delta \chi_{j_{Ml} n_0} &=  
\frac{q_l}{m_e} \Im \Delta \Sigma_{j_{Ml} n_0} + \frac{2 \omega}{m_e} \,\frac{1-|\epsilon_r|^2}{U(1-G)} 
\frac{\Im (\epsilon_r)}{ |\epsilon_r|^2} \,,
\label{eq:Mn0in}
\end{align}
where in-medium corrections are also expressed in terms of $\epsilon_r$.~When we also apply Eq.~(\ref{eq:iso}) to the first term in Eq.~(\ref{eq:Mn0in}), the latter reduces to 
\begin{align}
\frac{q_l}{m_e} \Im \Delta \chi_{j_{Ml} n_0} &=  \frac{2 \omega}{m_e} \,\frac{1}{U(1-G)} 
\frac{\Im (\epsilon_r)}{ |\epsilon_r|^2} \,.
\label{eq:Mn0in2}
\end{align}
For the numerical evaluation of the dielectric function, here we use tabulated results provided with the {\sffamily DarkELF} code~\cite{Knapen:2021bwg}, that were obtained using the time-dependent DFT capability of the {\sffamily GPAW}~\cite{Mortensen2005Jan} code. The values used here were obtained using the TB09 exchange-correlation functional \cite{PhysRevLett.102.226401}, with a scissors correction applied to match the zero-kelvin band gaps to the experimental values, and with the Ge $3d$ electrons frozen in the core.

\begin{figure*}[t]
    \centering
    \includegraphics[width=0.49\textwidth]{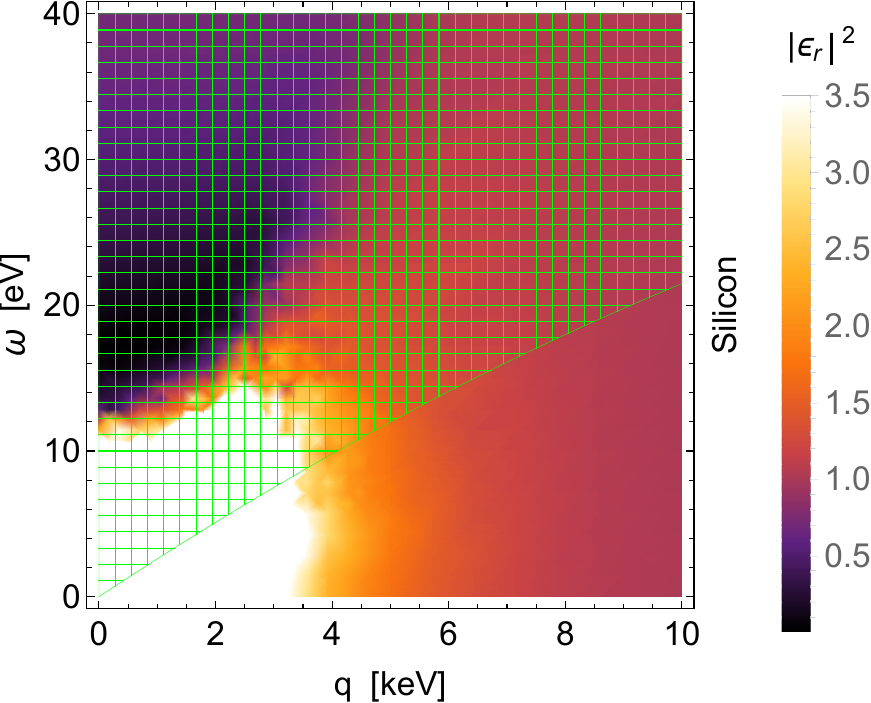}
     \includegraphics[width=0.49\textwidth]{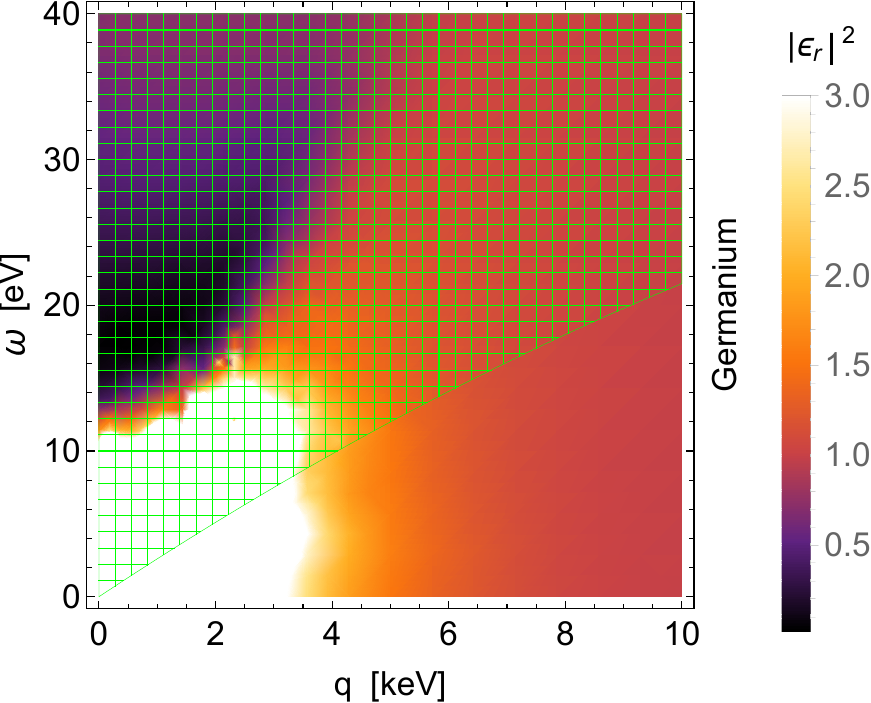}
    \caption{Modulus squared of the dielectric function, $|\epsilon_r|^2$, as a function of the momentum transfer, $|\vec q|$, and of the deposited energy $\omega$ for Si ({\bf left panel}) and Ge ({\bf right panel}).~For the generalised susceptibilities that receive in-medium corrections ($\chi_{n_0 n_0}$, $\chi_{j_{Mm} j_{Ml}}$ and $\Delta \chi_{j_{Ml} n_0}$ in the case of non-spin-polarised materials), $|\epsilon_r|^2$ determines the size and nature of such corrections.~For example, $|\epsilon_r|^2>1$ corresponds to a suppression of the material response to an external DM perturbation associated with screening, whereas $|\epsilon_r|^2<1$ implies an amplification of the material response due to collective excitations.~In both panels, we superimpose a green grid over the points that fulfil $v_{\rm min}>v_{\rm max}$, and which are thus not kinematically accessible for a DM particle mass of 10~MeV.~Collective excitations correspond to energies and momenta in the black regions, and are thus kinematically inaccessible.}
    \label{fig:Mod}
\end{figure*}

Eqs.~(\ref{eq:n0n0in}), (\ref{eq:MMin2}) and (\ref{eq:Mn0in}) allow us to compare the screened and unscreened contributions to the susceptibilities $\Im(\chi_{n_0 n_0})$, $\Im(\Delta \chi_{j_{Ml} n_0})$ and $\Im(\chi_{j_{Ml} j_{Mm}})$.~In the case of $\Im(\chi_{n_0 n_0})$, in-medium corrections are expected to be important, because $\left|1-|\epsilon_r|^2 \right|/|\epsilon_r|^2 \sim \mathcal{O}(1)$.~This is shown in Fig.~\ref{fig:Mod}, where we report 
$|\epsilon_r|^2$
 as a function of the momentum transfer, $|\vec q|$, and of the deposited energy, $\omega$, for Si (left panel) and Ge (right panel) crystals.~Here, the dielectric function is defined as in Eq.~(\ref{eq:eRPA}), and should not be confused with the direct outcome of {\sffamily GPAW}, $\epsilon_r^{\rm GPAW}$~\cite{Mortensen2005Jan}, which is Eq.~(\ref{eq:eRPA}) with $G=0$.~In Fig.~\ref{fig:Mod}, we have accounted for the $G\neq 0$ corrections to the relation between $\epsilon_r$ and the density-density response function $\chi_{n_0 n_0}$, or, equivalently, between $\epsilon_r$ and $\epsilon_r^{\rm GPAW}$.

For the same reason, namely $\left|1-|\epsilon_r|^2 \right|/|\epsilon_r|^2 \sim \mathcal{O}(1)$, in-medium corrections to $\Im(\Delta \chi_{j_{Mm} n_0})$ are also expected to be significant, as one can see explicitly from Eqs.~(\ref{eq:Mn0in}) and (\ref{eq:Mn0in2}). 

In contrast to $\Im(\chi_{n_0 n_0})$ and $\Im(\Delta \chi_{j_{Ml} n_0})$, the generalised susceptibility $\Im(\chi_{j_{Ml} j_{Mm}})$ has both longitudinal and transverse components.~In the isotropic limit,  
in-medium corrections only affect the longitudinal component of this current-current response function, leaving the transverse component unchanged, as one can see from Eq.~(\ref{eq:MMin2}).~This latter point will have an important impact on electron transition rate calculations, as we will see next.

\begin{figure*}[t]
    \centering
    \includegraphics[width=0.488\textwidth]{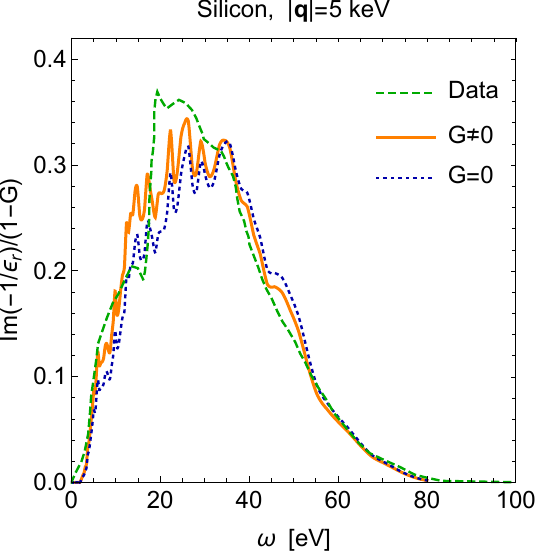}
      \includegraphics[width=0.5\textwidth]{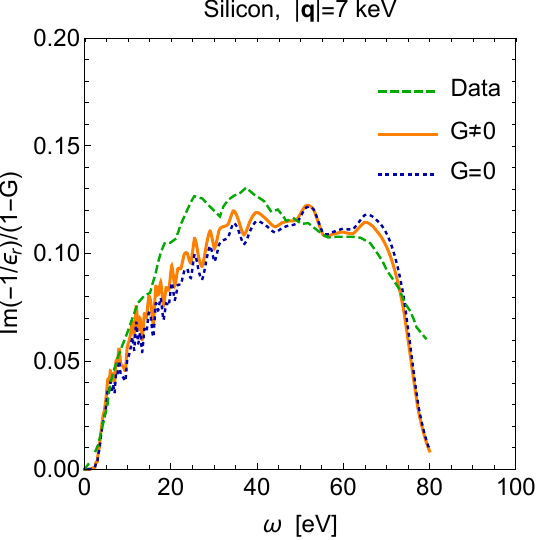}
    \caption{Silicon loss function, $\Im(-1/\epsilon_r)/(1-G)$, versus deposited energy $\omega$ for $|\vec q|=5$~keV ({\bf left panel}) and $|\vec q|=7$~keV ({\bf right panel}).~In both panels, the dashed green lines represent experimental data from~\cite{Weissker2010Feb}, while the dotted blue lines correspond to theoretical predictions based on Eq.~(\ref{eq:eRPA}) with $G=0$ and using the density-density response function $\chi_{n_0 n_0}$ computed in~\cite{Knapen:2021bwg} with the  {\sffamily GPAW} code~\cite{Mortensen2005Jan} in the RPA limit.~With these settings, the dotted blue lines account for exchange and correlation in the calculation of $\chi_{n_0 n_0}$, but not in the relation between $\chi_{n_0 n_0}$ and $\epsilon_r$.~The solid orange lines correspond to our theoretical predictions based on Eq.~(\ref{eq:eRPA}) with $G\neq0$.~They thus account for exchange and correlation both in the calculation of $\chi_{n_0 n_0}$ and in the relation between $\chi_{n_0 n_0}$ and $\epsilon_r$.~While $G\neq0$ implies a relatively small correction to $\Im(-1/\epsilon^{\rm GPAW}_r)$, it improves the agreement between theory and observations by increasing the loss function at small $\omega$, while decreasing the latter for intermediate values of $\omega$.}
    \label{fig:ELF}
\end{figure*}

\begin{figure*}[t]
    \centering
    \includegraphics[width=0.49\textwidth]{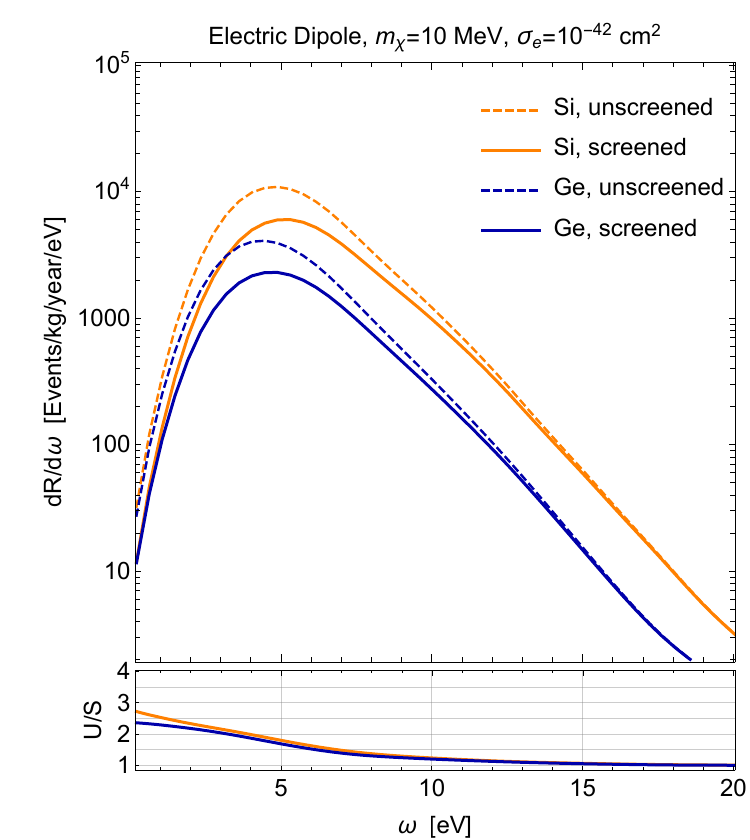}
      \includegraphics[width=0.49\textwidth]{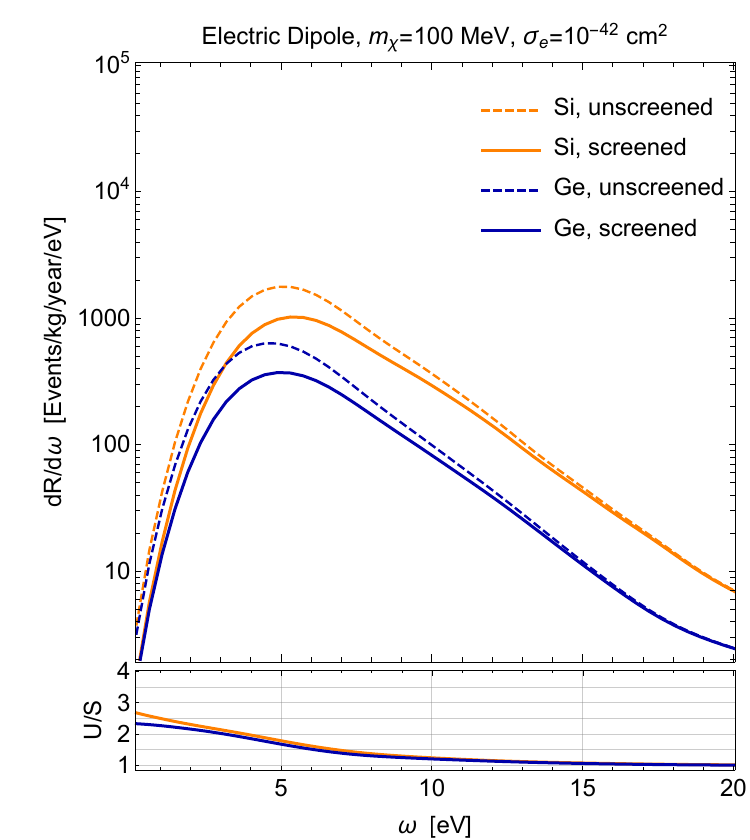}
    \caption{Differential rate of electronic transition per unit detector mass as a function of the deposited energy $\omega$ in Si (orange) or Ge (blue) crystals for a reference DM-electron scattering cross section of $10^{-42}$~cm$^2$.~We assume that the DM particle is characterised by an electric dipole and has a mass of either 10 MeV ({\bf left panels}) or 100 MeV ({\bf right panels}).~Solid lines correspond to screened interactions, i.e. $|\epsilon_r|^2\neq 1$, whereas dashed lines neglect in-medium effects, i.e.~$|\epsilon_r|^2=1$.~The bottom panels report the unscreened to screened rate ratio as function of $\omega$ for the germanium and silicon curves in the corresponding top panel.
    }
    \label{fig:EA10}
\end{figure*}

\begin{figure*}[t]
    \centering
    \includegraphics[width=0.49\textwidth]{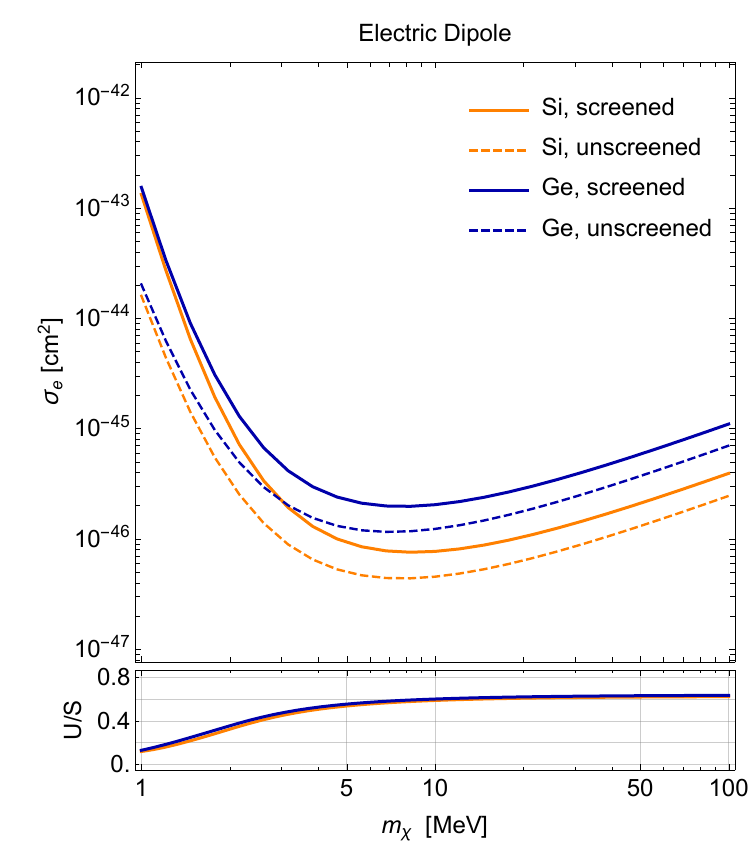}
      \includegraphics[width=0.49\textwidth]{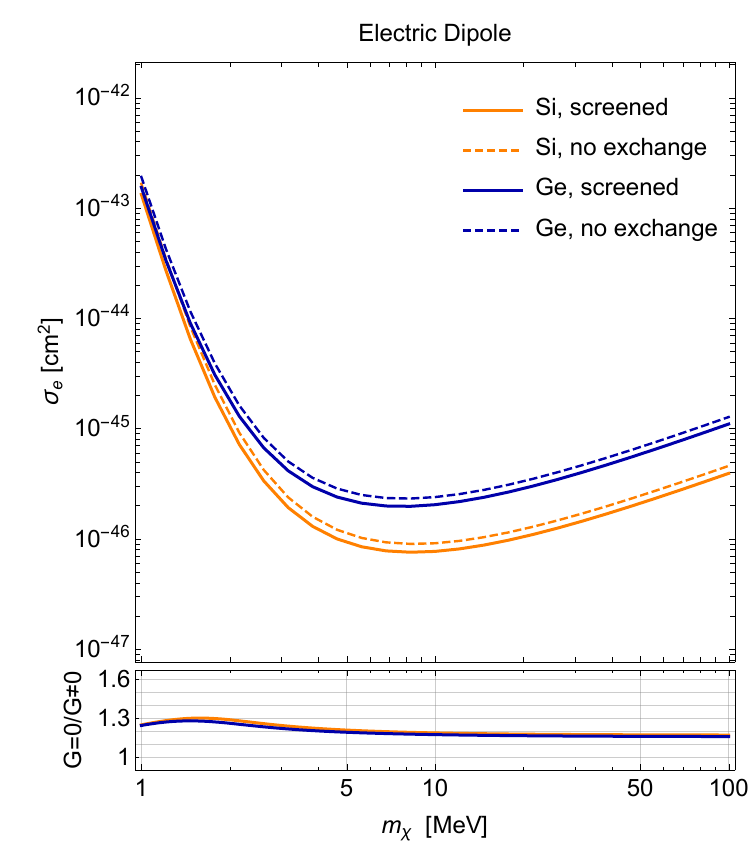}
    \caption{Projected 90\% C.L. exclusion limits on the reference cross section $\sigma_e$ as a function of the DM particle mass $m_\chi$ for electric dipole DM.~We assume a kg-year exposure in hypothetical background-free detectors made of either Si (orange lines) or Ge (blue lines) crystals.~Solid lines correspond to predictions obtained accounting for in-medium as well as exchange and correlation effects ($G\neq0$), whereas dashed lines neglect either the former ({\bf left panels}) or the latter ({\bf right panels}).~The bottom panels report the ratios between dashed and solid lines of the same colour (i.e. same target) in the corresponding top panels.}
    \label{fig:Lim}
\end{figure*}

\section{Application to dark matter direct detection}
\label{sec:application}

An important result we have derived from Eq.~(\ref{eq:final}) is that only the three generalised susceptibilities $\Im(\chi_{n_0 n_0})$, $\Im(\chi_{j_{Ml} j_{Mm}})$ and $\Im(\Delta \chi_{j_{Ml} n_0})$ receive in-medium corrections for non-spin-polarised detector materials, i.e. materials where spin-up and spin-down electrons have the same wave functions for a given band index and reciprocal space vector.~Focusing on DM models that generate these susceptibilities, we now apply the formalism developed in the previous sections to calculate the expected rates of DM-induced electronic transitions and the associated sensitivity of future experiments based on Ge and Si crystals.~We refer to App.~\ref{sec:BB} for an explicit relation between densities and currents, the associated susceptibilities, and the EFT operators in Tab.~\ref{tab:operators}.

\subsection{Electronic transition rates and exclusion limits}
\label{sec:results}
In this analysis, we focus on models where the DM particle is characterised by either an anapole or an electric dipole moment.~This allows us to place the emphasis on the impact of in-medium effects, as well as of a non-zero local field factor $G$ on the calculated electron transition rates.~In the case of DM candidates with an anapole moment, we are interested in external DM perturbations described by the potential in Eq.~(\ref{eq:Vt}) with \cite{Catena:2019gfa},
\begin{align}
c^s_8 &= 8 e m_e m_\chi \frac{g}{\Lambda^2} \nonumber\\
c^s_9 &= -8 e m_e m_\chi \frac{g}{\Lambda^2} \,,
\label{eq:canapole}
\end{align}
and all other coupling constants set to zero.~In the case of DM candidates with an electric dipole moment, we assume 
\begin{align}
c^\ell_{11} = & \frac{16 e m_\chi  m_e^2}{q_{\rm ref}^2} \frac{g}{\Lambda} \,,
\label{eq:celectric}
\end{align}
with no other coupling constants different from zero.~The dimensionless constant $g$ and the mass scale $\Lambda$ are in general different in Eqs.~(\ref{eq:canapole}) and (\ref{eq:celectric}), although here we denote them with the same symbol for simplicity.~By analogy with previous studies of anapole DM in the context of DM-nucleon scattering~\cite{DelNobile:2013cva}, we express $c^s_8$ and $c^s_9$ in terms of a reference DM-electron scattering cross section defined by
\begin{align}
\sigma_e \equiv 2 \alpha \, \frac{g^2 \mu^2}{\Lambda^4} \,.
\end{align}
Similarly, in the case of electric dipole DM, e.g.~\cite{DelNobile:2014eta}, we introduce the reference DM-electron scattering cross section,
\begin{align}
\sigma_e \equiv  4 \alpha \, \frac{g^2}{\Lambda^2} \,.
\end{align}
In terms of generalised susceptibilities, the electric dipole DM model involves the density-density response function only, $\Im(\chi_{n_0 n_0})$, whereas the anapole DM model is associated with the generalised susceptibilities $\Im(\chi_{n_0 n_0})$, $\Im(\chi_{j_{Ml} j_{Mm}})$ and $\Im(\Delta\chi_{j_{Ml} n_0})$.~As shown in Sec.~\ref{sec:num2}, for materials described by Eq.~(\ref{eq:iso}) in-medium corrections to $\Im(\chi_{n_0 n_0})$, $\Im(\chi_{j_{Ml} j_{Mm}})$ and $\Im(\Delta\chi_{j_{Ml} n_0})$ depend on the ratio $\Im (\epsilon_r)/ [|\epsilon_r|^2 (1-G)]$, known as the loss function.~We have also seen that in-medium effects vanish in the $|\epsilon_r|^2 \rightarrow 1$ limit, which motivates a study of how $|\epsilon_r|^2$ varies with $\vec q$ and $\omega$.

Fig.~\ref{fig:Mod} shows $|\epsilon_r|^2$ (not to be confused with $|\epsilon^{\rm GPAW}_r|^2$), as a function of the momentum transfer, $|\vec q|$, and of the deposited energy $\omega$ for Si (left panel) and Ge (right panel) crystals.~As anticipated in Sec.~\ref{sec:sol}, $|\epsilon_r|^2>1$ corresponds to a suppression of the generalised susceptibilities that receive in-medium corrections ($\chi_{n_0 n_0}$, $\chi_{j_{Ml} j_{Mm}}$ and $\Delta \chi_{j_{Ml} n_0}$ in the case of non-spin-polarised materials) that is due to the screening of $n_{\rm ext}$ in Eq.~(\ref{eq:neff}).~Similarly, $|\epsilon_r|^2<1$ implies an amplification of the material response due to collective excitations.~From Fig.~1, we thus expect collective excitations to be important in a region around $|\vec q|\sim$0~keV and $\omega\sim$20~eV.~This region corresponds to quasi-particle states with energies and momenta at which the real and imaginary parts of the dielectric function vanish. Since the momenta of these states are infinitesimal, they can only be excited by DM particles with De Broglie wavelengths that are much larger than the typical inter-atomic separation, which explains why they are referred to as collective excitations. Unfortunately, this region in the $(|\vec q|,\omega)$ plane is not kinematically accessible in the non-relativistic scattering of DM particles in conventional semiconductor crystals. However, in semiconductors with narrow band gaps driven by spin-orbit coupling, collective effects are expected to be much more important, as shown in Refs.~\cite{Inzani:2020szg,Chen:2022pyd}. The exploration of this class of materials would require relaxing our  assumption of spin degeneracy of bands, which we leave for future work.~To visualise this point, in both panels of Fig.~\ref{fig:Mod}, we superimpose a green grid over the points that fulfil the inequality $v_{\rm min}>v_{\rm max}$ for $m_\chi=10$~MeV, where $v_{\rm min}=\omega/q + q/(2 m_\chi)$ and $v_{\rm max}=v_e+v_\mathrm{esc}$.~The same conclusion applies to different values of the DM particle mass.

Let us now focus on the Si loss function directly.~Fig.~\ref{fig:ELF} shows $\Im (\epsilon_r)/ [|\epsilon_r|^2 (1-G)]$ as a function of the deposited energy $\omega$ for two representative values of the momentum transfer, namely $|\vec q|=5$~keV (left panel) and $|\vec q|=7$~keV (right panel).~The dashed green lines in the two panels of Fig.~\ref{fig:ELF} represent experimental data from~\cite{Weissker2010Feb}, extracted from Fig.~1 of~\cite{Knapen:2021bwg}.~In the same panels, the dotted blue lines correspond to theoretical predictions based on Eq.~(\ref{eq:eRPA}) with $G=0$ and the density-density response function $\chi_{n_0 n_0}$ computed in~\cite{Knapen:2021bwg} with the  {\sffamily GPAW} code~\cite{Mortensen2005Jan} in the RPA limit.~Consequently, the dotted blue lines in the figure account for exchange in the calculation of $\chi_{n_0 n_0}$, but not in the relation between $\chi_{n_0 n_0}$ and $\epsilon_r$.~In contrast, the solid orange lines in the two panels of Fig.~\ref{fig:ELF} correspond to our theoretical predictions based on Eq.~(\ref{eq:eRPA}) with $G\neq0$.~They thus account for exchange both in the calculation of $\chi_{n_0 n_0}$ and in the relation between $\chi_{n_0 n_0}$ and $\epsilon_r$.~While $G\neq0$ implies a relatively small correction to the Si loss function, it improves the agreement between theory and experiment by increasing the loss function at small $\omega$, while decreasing it for larger $\omega$ values. We find a qualitatively similar behaviour for the Ge loss function (not shown).

Focusing on Ge and Si crystals, let us now calculate the differential rate of DM-induced electronic transitions per unit detector mass, ${\rm d}R/{\rm d}\omega$, within our generalised susceptibility formalism.~Fig.~\ref{fig:EA10} shows the differential rate ${\rm d}R/{\rm d} \omega$ as a function of $\omega$ for the case of electric dipole DM and a reference DM-electron scattering cross section of $10^{-42}$~cm$^2$.~The left panels refer to a DM particle mass of 10~MeV, while the right panels correspond to $m_\chi=$~100 MeV.~While the top panels show the rate ${\rm d}R/{\rm d}\omega$ for a given DM-electron scattering cross section in different targets with and without in-medium effects, the bottom panels in the figure report the corresponding rate ratios to facilitate the comparison of distinct calculations.~Specifically, the solid lines in the top panels account for in-medium effects in Si (orange) and Ge (blue) crystals, while the dashed lines assume $|\epsilon_r|^2=1$.~At the same time, the bottom panels in Fig.~\ref{fig:EA10} report the unscreened to screened rate ratio as a function of $\omega$ for the crystals and mass in the corresponding top panel.~As one can see from Fig.~\ref{fig:EA10}, in-medium corrections to $\Im(\chi_{n_0 n_0})$ suppress the rate of DM-induced electronic transitions in crystals by a factor of 2 or 3 for $\omega$ below about $5$~eV, while they are negligible for $\omega$ larger than $15$~eV.~The amplitude of the in-medium corrections for dipole DM is comparable with what was found in~\cite{Knapen:2021bwg} focusing on models where DM couples to the density $n_0$ via the exchange of a heavy or light mediator, which, within our notation, would correspond to $\mathcal{M} = c^s_1 \langle \mathcal{O}_1 \rangle$ and $\mathcal{M} = c^\ell_1 (q_{\rm ref}/q)^2 \langle \mathcal{O}_1 \rangle$, respectively.

Let us now focus on the impact of in-medium effects and electron exchange on the expected sensitivity of Ge and Si detectors.~Fig.~\ref{fig:Lim} shows the expected 90\% confidence level (C.L.) exclusion limits on the reference cross section $\sigma_e$ as a function of the DM mass $m_\chi$ for electric dipole DM.~We assume a kg-year exposure in hypothetical background-free detectors made of either Si (orange lines) or Ge (blue lines) crystals.~The solid lines in the top panels correspond to predictions obtained accounting for in-medium as well as exchange effects ($G\neq0$), whereas the dashed lines neglect either the former (left panel) or the latter (right panel).~The bottom panels in Fig.~\ref{fig:Lim} report the ratios between dashed and solid lines of the same colour in the corresponding top panels.~From Fig.~\ref{fig:Lim}, we conclude that neglecting screening effects in the calculation of the expected 90\% C.L. exclusion limits for electric dipole DM leads to an order one error on $\sigma_e$, whereas neglecting the exchange factor $G$ in the relation between the dielectric function $\epsilon_r$ and $\chi_{n_0 n_0}$ induces a $10\%$ error on $\sigma_e$. 

In contrast, in-medium corrections are found to be negligible in the case of anapole DM, where the rate of electron transitions receives large contributions from the transverse components of $\Im(\chi_{j_{Ml} j_{Mm}})$, which are unscreened in nearly isotropic materials, and from $\Im \chi_{j_{5l}^\dagger j_{5m}}$, which is unscreened in non-spin-polarised detectors.~For this reason, we do not report here the corresponding differential rate per unit detector mass and expected sensitivity studies.

Since the unscreened transverse components of $\Im(\chi_{j_{Ml} j_{Mm}})$ generically tend to ``wash out'' in-medium effects in electron transition rate calculations, and the generalised susceptibility $\Im(\Delta\chi_{j_{Ml} n_0 })$ always appears together with $\Im(\chi_{j_{Ml} j_{Mm}})$, we arrive at the important conclusion that DM has to couple to the electron density $n_0$ alone for in-medium effects to be important in the DM-electron scattering in non-spin-polarised and nearly isotropic materials.

\subsection{Comparison with previous results}
\label{sec:comp}
We now compare our expression for the rate of DM-induced electronic transitions in materials, Eq.~(\ref{eq:ratefinal}), with the results found in~\cite{Catena:2021qsr}  for electronic transitions in semiconductor crystals assuming Bloch wave functions of the type
\begin{align}
\psi_{i}(\mathbf{r}_e)&= \phi_i(\vec r_1) \, \eta^\sigma \nonumber\\
\psi_{f}(\mathbf{r}_e)&= \phi_f(\vec r_1) \, \eta^{\sigma'}
\,,
\label{eq:bloch}
\end{align}
for the initial and final state electrons, respectively.~The $\eta^\sigma$ and $\eta^{\sigma'}$ spinors are defined in the text above Eq.~(\ref{eq:nhat}), whereas the $\phi_i(\vec r_1)$ and $\phi_f(\vec r_1)$ spatial wave functions are given in Eq.~(\ref{eq:bloch0}). When the electron spin wave function factorises as in Eq.~(\ref{eq:bloch}) and as assumed in~\cite{Catena:2021qsr}, the matrix elements in Eq.~(\ref{eq:corr}) can be evaluated as in the following example,
\begin{align}
|\langle f| e^{-i\mathbf{q}\cdot \mathbf{r}_e} \boldsymbol{\sigma}_{e} |i\rangle|^2 
=&
\sum_{\sigma \sigma'}\eta^{\sigma'\dagger}\boldsymbol{\sigma}_{e}\eta^\sigma\cdot\eta^{\sigma \dagger}\boldsymbol{\sigma}_{e}\eta^{\sigma'}  
 \langle f|| e^{-i\mathbf{q}\cdot \mathbf{r}_e}  ||i\rangle |^2 \nonumber\\
=&
\tr(\boldsymbol{\sigma}_{e}\cdot\boldsymbol{\sigma}_{e}) \, \langle f|| e^{-i\mathbf{q}\cdot \mathbf{r}_e}  ||i\rangle |^2
\end{align}
where  
\begin{align}
\langle f| \sigma_{e,l}e^{-i\mathbf{q}\cdot \mathbf{r}_e}  |i\rangle \equiv \eta^{\sigma'\dagger}\sigma_{e,l}\eta^{\sigma}\langle f|| e^{-i\mathbf{q}\cdot \mathbf{r}_e}  ||i\rangle \,,
\end{align}
while $\sigma_{e,l}$ is the $l$-th Pauli matrix and $\boldsymbol{\sigma}_e$ a spatial vector.~By inserting Eqs.~(\ref{eq:bloch}) and (\ref{eq:f}) and the explicit expressions for the electron densities and currents, Eq.~(\ref{eq:nj}), into our differential rate formula, Eq.~(\ref{eq:rate}), we finally obtain
\begin{align}
{\rm d}\Gamma = \frac{{\rm d}\mathbf{q}}{(2 \pi)^3} & \int{\rm d}\omega  \left( \frac{1}{8 m_e^2 m_\chi^2V}  \right)  \delta(\omega+\Delta E_\chi) \nonumber\\
&\times \Big[ A |f|^2  + B_{(lm)} \,\Im (i f_{l}^* f_{m})  + B_{[lm]} \,\Re (i f_{l}^* f_{m}) \nonumber\\
& + C_{l} \,\Re(f f^*_l) + \widetilde{C}_{l} \,\Im(f f^*_l) \Big] \,,
\label{eq:rate_expand}
\end{align}
where 
\begin{align}
A =&~\langle F^*_0 F_0 \rangle +  \frac{q^2}{4 m_e^2} \langle F^*_A F_A \rangle +
\langle \mathbf{F}^*_5  \cdot\mathbf{F}_5 \rangle  \nonumber \\
&+\frac{1}{4 m_e^2} \left[ \langle | \mathbf{q} \cdot \mathbf{F}_M |^2 \rangle +  \left( q^2\delta_{lm}  - q_lq_m \right)  
\langle F^*_{El}F_{Em} \rangle\right] \nonumber \\ 
&+ \frac{q_l}{m_e}\Re\langle F^*_{Ml} F_0 \rangle + \frac{q_l}{m_e} \Re \langle F_A^* F_{5l} \rangle \nonumber\\
&+\frac{q_i}{m_e} \epsilon_{ilm} \Im\langle F_{5l}^* F_{Em} \rangle \,,
\end{align}
while
\begin{align}
B_{(lm)} =&~\langle F_A^* F_A \rangle \delta_{lm}
+  \Re \langle F_{Ml}^* F_{Mm} \rangle \nonumber\\
&+  (\delta_{ij}\delta_{lm}-\delta_{im}\delta_{jl}) \langle F_{Ei}^* F_{Ej} \rangle \nonumber\\
B_{[lm]} =&~2\epsilon_{ilm} \langle F_{Ei}^* F_A \rangle + \Im \langle F_{Ml}^* F_{Mm} \rangle \,,
\end{align}
and
\begin{align}
C_{l} =&~\frac{q_l}{m_e} \langle F_A^* F_A \rangle +  \frac{q_m}{m_e} \Re \langle F_{M l}^* F_{Mm} \rangle \nonumber\\
&+ \frac{q_m}{m_e} (\delta_{ij}\delta_{lm}-\delta_{im}\delta_{jl})  \langle F_{Ei}^* F_{Ej} \rangle \nonumber\\
&+ 2 \Re \langle F_{Ml}^* F_0 \rangle + 2 \Re \langle F_{A}^* F_{5l} \rangle + 2 \epsilon_{lij} \Im \langle F_{5i}^* F_{Ej} \rangle \,,\nonumber\\ 
\widetilde{C}_{l} =&~-2 \Im \langle F_{Ml}^* F_0 \rangle  -2 \Im \langle F_{A}^* F_{5l} \rangle  - 2 \epsilon_{lij} \Re \langle F_{5i}^* F_{Ej} \rangle \,. 
\label{eq:dGs}
\end{align}
In all equations, a sum over repeated three-dimensional indices is understood.~Obtaining Eq.~(\ref{eq:rate_expand}), we use the identities
\begin{align}
&\tr (\sigma_{e,i}) = 0 \nonumber\\
&\tr (\sigma_{e,i} \sigma_{e,j}) = 2 \delta_{ij} \,.
\label{eq:sigma}
\end{align}
Eq.~(\ref{eq:sigma}) implies that many of the correlation functions that could in principle contribute to the differential rate ${\rm d}\Gamma$ are actually zero.~In particular, all correlation functions linear in $\boldsymbol{\sigma}_e$ vanish.~This is in general not true when spin up and spin down electrons have different wave functions, in contrast with Eq.~(\ref{eq:bloch}), or many-body wave functions are used in the evaluation of the correlation functions $K_{j^\dagger_\beta j_\alpha }$.

By inserting the explicit expressions for the quadratic ``strength functions'' given in Appendix~\ref{sec:BB} into Eq.~(\ref{eq:dGs}), we find that the total rate $\mathscr{R}$ resulting from Eqs.~(\ref{eq:ratefinal}), (\ref{eq:rate_expand}) and (\ref{eq:dGs}) coincides with that given in~\cite{Catena:2021qsr}, with the exception of the coefficient in front of the $c_{14} c_{15} W_5$ term, which we find here to be $-1/8$, but which is erroneously reported to be $-1/2$ in \cite{Catena:2021qsr}.

The formalism developed here could be extended to be applicable to phonon and magnon excitations. This extension would proceed along the lines of Trickle {\it et al.}, Ref.~\cite{Trickle:2020oki}. Specifically, one would first quantise the ion displacement field in the effective potential in our Eq.~(\ref{eq:Veff}). Then one would take the matrix elements between the vacuum and single- or multi-phonon states of the extended potential, before finally using it in Fermi’s golden rule. This procedure would establish an explicit mapping between the operators and generalised susceptibilities introduced in our work and the response functions for phonons and magnons computed in Ref.~\cite{Trickle:2020oki}. We leave this interesting calculation for future work, as it goes beyond the scope of the present study.

\section{Summary and outlook}
\label{sec:conclusion}
Within the non-relativistic effective theory of DM-electron interactions, we identified the densities and currents a spin-1/2 DM particle can couple to in a material, and found their corresponding electromagnetic analogues in a $1/c$ expansion of the Dirac Hamiltonian.~Specifically, we found that DM can in general perturb a solid state system by coupling to the electron density, the paramagnetic current, the spin current, the scalar product of spin and paramagnetic current, as well as the Rashba spin-orbit current in the material.~In the $(1/c)$ expansion of the Dirac Hamiltonian, the first, second and third couplings arise at order $(1/c)^0$, while the last two couplings originate at order $(1/c)^2$.~We then wrote down the explicit expression for the time dependent effective potential that describes the scattering of DM particles by the electrons bound to a solid-state system, $V_{\rm eff}^{ss'}(t)$ in Eq.~(\ref{eq:Vt}), in terms of the five densities and currents listed above.

We interpreted the effective potential $V_{\rm eff}^{ss'}(t)$ as an external perturbation in linear response theory, and identified the generalised susceptibilities that describe the response of a generic solid-state system to the perturbation $V_{\rm eff}^{ss'}(t)$ by extending the Kubo formula to the case of DM-electron scattering in materials.~We then combined the extended Kubo formula, Eq.~(\ref{eq:dcs}), with Fermi's golden rule to express the rate of DM-induced electronic transitions in a solid state system in terms of the generalised susceptibilities associated with the external perturbation $V_{\rm eff}^{ss'}(t)$.

This expression for the electronic transition rate allowed us to factorise in a neat manner the material physics contribution, encoded in a set of generalised susceptibilities, from the particle physics input.~Interestingly, this factorisation enables the use of existing experimental data on the generalised susceptibilities associated with $V_{\rm eff}^{ss'}(t)$ to calibrate theoretical predictions for the rate of DM-induced electronic transitions in a given detector.

In order to evaluate our transition rate formula, i.e. Eq.~(\ref{eq:rate_gensus}), we applied the equation of motion method.~This approach allowed us to express the set of generalised susceptibilities we identified in this work as the mean-field solution to a time-evolution equation, for which we also provided a useful diagrammatic interpretation.~This solution, see Eq.~(\ref{eq:final}), is one of the main results of our work.

An important conclusion we drew from Eq.~(\ref{eq:final}) is that only three generalised susceptibilities receive corrections that are associated with screening or collective excitations in the case of non-spin-polarised materials, i.e. materials where spin up and spin down electrons have the same wave functions for a given band index and reciprocal space vector.~These generalised susceptibilities are $\Im(\chi_{n_0^\dagger n_0})$, $\Im(\chi_{j^\dagger_{Ml} j_{Mm}})$ and $\Im(\Delta\chi_{j^\dagger_{Ml} n_0})$, where $n_0$ is the electron density and $ j_{Ml}$, $l=1,2,3$ is the paramagnetic current.~We also found that the in-medium corrections to $\Im(\chi_{n_0^\dagger n_0})$, $\Im(\Delta\chi_{j^\dagger_{Ml} n_0})$  and $\Im(\chi_{j^\dagger_{Ml} j_{Mm}})$  can be expressed in terms of the electron loss function in the case of isotropic materials.~Another conclusion we drew from Eq.~(\ref{eq:final}), is that it captures exchange effects that would be missed in the random phase approximation (RPA).

Finally, we applied the formalism developed in this work to calculate the expected electronic transition rate and sensitivity of hypothetical DM direct detection using Ge and Si crystals as detector materials.~This calculation was performed by the combined use of the computer programmes {\sffamily QEdark-EFT}~\cite{Urdshals2021May}, {\sffamily QEdark}~\cite{Essig:2015cda} and {\sffamily DarkELF}~\cite{Knapen:2021bwg} as explained in Sec.~\ref{sec:results}.~Emphasis was placed on quantifying the importance of in-medium corrections as well as exchange effects.~For example, we found that neglecting screening effects in the calculation of the expected 90\% C.L. exclusion limits for DM candidates with an electric dipole (modelled via  the potential $V_{\rm eff}^{ss'}(t)$) leads to an $\mathcal{O}(1)$ error in the reference DM-electron scattering cross section, $\sigma_e$, whereas neglecting electron exchange in the relation between dielectric function, $\epsilon_r$, and density-density response function, $\chi_{n_0^\dagger n_0}$, induces a $10\%$ error in $\sigma_e$. 

In contrast, in-medium corrections were found to be negligible in the case of anapole DM, because the rate of electron transitions in detector materials receives in this case large contributions from the transverse components of the current-current response functions $\Im(\chi_{j_{Ml}^\dagger j_{Mm}})$ and from $\Im \chi_{j_{5l}^\dagger j_{5m}}$.~The former is unscreened in nearly isotropic materials, while the latter is unscreened in non-spin-polarised detectors.

More generally, we arrived at the important conclusion that, if screening of the transverse responses is neglected, then in-medium effects are significant in the DM scattering from non-spin-polarized materials only when the DM couples to the electron density $n_0$ alone. ~This is due to the fact that the unscreened transverse components of $\Im(\chi_{j_{Ml}^\dagger j_{Mm}})$ generically tend to ``wash out'' in-medium effects in electron transition-rate calculations, and that $\Im(\Delta\chi_{j_{Ml}^\dagger n_0})$ always appears together with $\Im(\chi_{j_{Ml}^\dagger j_{Mm}})$.
 
Summarising, the linear response theory for light DM direct detection we developed in this work paves the way for the study of in-medium effects in the presence of general DM-electron interactions.~It provides a framework for using existing or future experimental measurements of the generalised susceptibilities $\chi_{j_\alpha j_\beta}$ to calibrate theoretical predictions of the rate of DM-induced electronic transitions in detector materials.~Finally, it can be straightforwardly extended to the case of spin-polarised detectors, as well as to highly inhomogeneous or anisotropic materials.~We leave these investigations for future work.

\acknowledgments 
It is a pleasure to thank Einar Urdshals for providing a re-binned digitised version of the crystal response functions we computed in~\cite{Catena:2021qsr}, and Micha\l{} Iglicki for a careful reading of the manuscript.~RC acknowledges support from an individual research grant from the Swedish Research Council (Dnr.~2022-04299) and from the Knut and Alice Wallenberg Foundation project ``Light Dark Matter'' (Dnr. KAW 2019.0080).~NAS was supported by the ETH Zurich, and by the European Research Council (ERC) under the European Union's Horizon 2020 research and innovation programme project HERO Grant Agreement No. 810451.

\appendix

\section{Quadratic strength functions}
\label{sec:BB}
In this appendix, we list explicit expressions for the functions $F_0^{ss'}$, $F_A^{ss'}$, $\mathbf{F}_5^{ss'}$, $\mathbf{F}_M^{ss'}$, and $\mathbf{F}_E^{ss'}$. They are given by
\begin{align}
F_0^{ss'} = &~\xi_\chi^{s'\dagger}\left[ c_1 + i \left( \frac{\mathbf{q}}{m_e}  \times \vec{v}_\chi^\perp \right) \cdot  \mathbf{S}_\chi  c_5
+ \vec{v}_\chi^\perp \cdot \mathbf{S}_\chi \right.  c_8 \nonumber\\
&+ \left. i \frac{\mathbf{q}}{m_e} \cdot \mathbf{S}_\chi c_{11} \right] \xi^{s}\nonumber \\ 
F_A^{ss'} =&~-\frac{1}{2} \xi_\chi^{s'\dagger} \left[ c_7   +i \frac{\mathbf{q}}{m_e} \cdot \mathbf{S}_\chi~ c_{14} \right] \xi_\chi^s\nonumber \\
\mathbf{F}_5^{ss'} =&~\frac{1}{2}\xi_\chi^{s'\dagger} \left[ i \frac{\mathbf{q}}{m_e} \times \vec{v}_\chi^\perp~ c_3 + \mathbf{S}_\chi c_4
+  \frac{\mathbf{q}}{m_e}~\frac{\mathbf{q}}{m_e} \cdot \mathbf{S}_\chi c_6  \right. \nonumber\\
&+ \left. \vec{v}_\chi^\perp c_7 + i \frac{\mathbf{q}}{m_e} \times \mathbf{S}_\chi c_9 + i \frac{\mathbf{q}}{m_e}c_{10}
 +  \vec{v}_\chi^\perp \times \mathbf{S}_\chi c_{12}  
 \right.  \nonumber \\&+  \left. 
i \vec{v}_\chi^\perp \frac{\mathbf{q}}{m_e} \cdot \mathbf{S}_\chi ~ c_{14}  
+\frac{\mathbf{q}}{m_e} \times \vec{v}_\chi^\perp~ \frac{\mathbf{q}}{m_e} \cdot \mathbf{S}_\chi ~ c_{15}  \right] \xi_\chi^s
\nonumber \\
\mathbf{F}_M^{ss'} =&~\xi_\chi^{s'\dagger} \left[  i \frac{\mathbf{q}}{m_e}  \times \mathbf{S}_\chi c_5 - \mathbf{S}_\chi c_8 \right] \xi_\chi^s \nonumber \\
\mathbf{F}_E^{ss'} =&~\frac{1}{2} \xi_\chi^{s'\dagger}\left[  \frac{\mathbf{q}}{m_e} ~ c_3 +i \mathbf{S}_\chi c_{12} 
-i {\vec{q} \over  m_e} \frac{\mathbf{q}}{m_e} \cdot \mathbf{S}_\chi  c_{15} \right] \xi_\chi^s
\,,
\label{eq:M2}
\end{align}
where
\begin{align}
\vec{v}_\chi^\perp = \left(\frac{\mathbf{p} + \mathbf{p}'}{2 m_\chi}\right) = \mathbf{v} - \frac{\mathbf{q}}{2 m_\chi}\,,
\end{align}
$\mathbf{v}=\mathbf{p}/m_\chi$, $\mathbf{q}=\mathbf{p}-\mathbf{p}'$ is the momentum transferred to the electron, and, finally, we shortened the notation by defining, 
\begin{align}
c_i\equiv \left(c_i^s +c^\ell_i \frac{q_{\rm ref}^2}{|\mathbf{q}|^2} \right) \,.
\end{align}
Furthermore, we list the quadratic strength functions used in Sec.~\ref{sec:formalism} to calculate the rate of DM-induced electronic transitions in materials.~They can be written as follows:
\begin{align}
\langle F_0^* F_0 \rangle &= c_1^2 +\frac{1}{4} \left| \frac{\mathbf{q}}{m_e} \times \mathbf{v}_\chi^\perp \right|^2 c_5^2  
+\frac{v_\chi^{\perp 2}}{4} c_8^2 +\frac{q^2}{4 m_e^2} c_{11}^2 \,, \\
\langle F_A^* F_A \rangle &= \frac{1}{4}\left( c_7^2 +\frac{q^2}{4 m_e^2} c_{14}^2 \right) \,, \\
\langle \mathbf{F}_5^*  \cdot \mathbf{F}_5 \rangle &=  \frac{1}{4}\bigg( \left| \frac{\mathbf{q}}{m_e} \times \mathbf{v}_\chi^\perp \right|^2 c_3^2 
+\frac{3}{4} c_4^2 + \frac{q^4}{4 m_e^4} c_{6}^2   \nonumber\\
&+ v_\chi^{\perp 2} c_7^2 +\frac{q^2}{2 m_e^2} c_{9}^2 +\frac{q^2}{m_e^2} c_{10}^2 + \frac{v_\chi^{\perp 2}}{2} c_{12}^2 
  \nonumber\\ &+   \frac{q^2}{4 m_e^2} v_\chi^{\perp 2} c_{14}^2 +  \left| \frac{\mathbf{q}}{m_e} \times \mathbf{v}_\chi^\perp \right|^2 \frac{q^2}{4m^2_e}c_{15}^2 
\nonumber\\ &+ \frac{q^2}{2m_e^2} c_4 c_6 -\frac{1}{2} \left| \frac{\mathbf{q}}{m_e} \times \mathbf{v}_\chi^\perp \right|^2 c_{12} c_{15} \bigg) \,, \\
\langle F^*_{Ml}F_{Mm} \rangle &= \frac{1}{4m_e^2} \left( q^2 \delta_{lm} - q_l q_m\right) c_5^2 + \frac{1}{4} c_8^2 \delta_{lm} 
\nonumber\\ &-\frac{i}{2} \epsilon_{lmi} \frac{q_i}{m_e} c_5 c_8\,,\\
\langle F^*_{El}F_{Em} \rangle &=  \frac{1}{4}\bigg( \frac{q_l q_m}{m_e^2} c_3^2 + \frac{1}{4} \delta_{l m}c_{12}^2 
+ \frac{q^2}{4 m_e^2} \frac{q_l q_m}{m_e^2} c_{15}^2 
\nonumber \\ &-\frac{q_lq_m}{2m_e^2} c_{12} c_{15}  \bigg) \,.
\end{align}
In addition, we made use of the following off-diagonal terms:
\begin{align}
\langle F_{El}^* F_{A} \rangle &=  \frac{1}{4} \left( -\frac{q_l}{m_e} c_3 c_7 - \frac{q_l}{4m_e} c_{12}c_{14} +\frac{q_l}{4 m_e} \frac{q^2}{m_e^2} c_{14}c_{15}\right) \\
\langle F_{Ml}^* F_{0} \rangle &=-\frac{1}{4}(v_\chi^\perp)_l c_8^2  -\frac{i}{2} \left| \frac{\mathbf{q}}{m_e} \times \mathbf{v}_\chi^\perp \right|_l c_5 c_8 
-\frac{i}{4} \frac{q_l}{m_e} c_8 c_{11}\\
\langle F_{A}^* F_{5l} \rangle &= -\frac{1}{4} (v_\chi^\perp)_l c_7^2 - \frac{q^2}{16 m_e^2} (v_\chi^\perp)_lc_{14}^2 
\nonumber\\ &-\frac{i}{4} \left( \frac{\mathbf{q}}{m_e}\times \mathbf{v}_\chi^\perp \right)_l c_3c_7
-\frac{i}{4}\frac{q_l}{m_e} c_7 c_{10} 
\nonumber\\ & +\frac{i}{16} \frac{q_l}{m_e} c_4 c_{14} +\frac{i}{16} \frac{q^2}{m_e^2} \frac{q_l}{m_e} c_6 c_{14}
\nonumber\\ &+\frac{i}{16} \frac{q^2}{m_e^2}\left( \frac{\mathbf{q}}{m_e}\times \mathbf{v}_\chi^\perp  \right)_l c_{14} c_{15}\,,
\label{eq:FAF5}
\end{align}
and, finally,
\begin{align}
\epsilon_{ilm} \langle F_{5l}^* F_{Em} \rangle &= \frac{1}{4} \bigg[  \frac{i}{m_e^2}\left(q_i q_m-q^2\delta_{mi}\right) (v_\chi^\perp)_m c_3^2
\nonumber\\ &-\frac{i}{2} (v_\chi^\perp)_i c_{12}^2
\nonumber\\ &+\frac{i}{4m_e^2}\left( \mathbf{q}\cdot\mathbf{v}_\chi^\perp q_i -q^2 (v_\chi^\perp)_i\right)\frac{q^2}{m_e^2}c_{15}^2
\nonumber\\ & +\epsilon_{ilm} (v_\chi^\perp)_l \frac{q_m}{m_e} c_3 c_7 -\frac{q_i}{2m_e} c_9 c_{12}
\nonumber\\ &-\frac{5i}{4m_e^2}\left( \mathbf{q}\cdot\mathbf{v}_\chi^\perp q_i -q^2 (v_\chi^\perp)_i\right)c_{12}c_{15}
\nonumber\\ &+\epsilon_{ilm} (v_\chi^\perp)_l \frac{q_m}{m_e} c_{12}c_{14}
\nonumber\\ &-\frac{q^2}{4m_e^2}\epsilon_{ilm} (v_\chi^\perp)_l\frac{q_m}{m_e} c_{14}c_{15} 
\bigg]\,.
\end{align}

\section{Kubo formula for dark matter-electron scattering}
\label{sec:kubo}
We are interested in DM-induced perturbations to detector materials that can be described by the effective potential
\begin{align}
V_{\rm eff}^{ss'}(t) \equiv - \int {\rm d}\mathbf{r} \, B(\mathbf{r}) S^{ss'}(\mathbf{r},t)
\label{eq:ext}
\end{align}
where $B(\mathbf{r})$ is an operator acting on the wave functions of the electrons in the material and $S^{ss'}(\mathbf{r},t)$ is the strength of the perturbation.~$S^{ss'}(\mathbf{r},t)$ depends on the initial and final DM particle spin configurations, $s$ and $s'$.~Each term in the effective potential actually used in this work, Eq.~(\ref{eq:Vefftot}), has the form assumed here in Eq.~(\ref{eq:ext}) for illustrative purposes.~Under such perturbations, the density matrix $\rho$ of the given detector material evolves according to
\begin{align}
\frac{{\rm d}\rho(t)}{{\rm d}t} = i \left[ \rho(t), H_{0} + V^{ss'}_{\rm eff}(t) \right]
\label{eq:master}
\end{align}
where $H_{0}$ is the Hamiltonian of the system in the absence of external perturbations.~By imposing $\rho(t\rightarrow -\infty) = \rho_{0}$, where $\rho_{0}$ is the density matrix when $V^{ss'}_{\rm eff}=0$, and using
\begin{align}
\frac{{\rm d}}{{\rm d} t} \Big[ e^{i H_{0} t} \rho(t) \,e^{-i H_{0} t} \Big] =
i e^{iH_0 t} \left[ \rho(t), V^{ss'}_{\rm eff}(t)\right] e^{-iH_0 t} \,,
\end{align}
Eq.~(\ref{eq:master}) can conveniently be rewritten in an integral form,
\begin{align}
\rho(t) = \rho_0 + i\int_{-\infty}^t {\rm d}t'e^{-i H_0(t-t') } \left[ \rho(t'), V^{ss'}_{\rm eff}(t')\right]e^{i H_0(t-t') }\,.
\label{eq:int}
\end{align}
At first order in $V^{ss'}_{\rm eff}$, Eq.~(\ref{eq:int}) admits the following solution
\begin{align}
\rho(t) = \rho_0 + i\int_{-\infty}^t {\rm d}t'e^{-i H_0(t-t') } \left[ \rho_0, V^{ss'}_{\rm eff}(t')\right]e^{i H_0(t-t') }\,.
\label{eq:sol}
\end{align}
We can now use Eq.~(\ref{eq:sol}) to calculate the expectation value of any observable $A(\mathbf{r},t=0)\equiv A(\mathbf{r})$.~This is given by
\begin{align}
\langle A \rangle &= \tr \left\{ A\rho \right\} \nonumber\\ &= \tr \left\{ A \rho_0 \right\} + \langle \Delta A \rangle
\end{align}
where $\langle \Delta A \rangle$ is the induced perturbation in the observable $A$, namely
\begin{align}
\langle \Delta A(\mathbf{r},t)\rangle  
=& i\tr \left\{ \int_{-\infty}^{t} {\rm d} t' \, A(\mathbf{r},t-t') \left[ \rho_0, V^{ss'}_{\rm eff}(t')\right] \right\} \nonumber\\
=& -i\int_{-\infty}^{t} {\rm d} t' \,\tr \left\{ \rho_0 \left[ A(\mathbf{r},t-t'),V^{ss'}_{\rm eff}(t') \right] \right\}\nonumber\\
=& i\int_{-\infty}^{t} {\rm d} t' \int {\rm d}\mathbf{r}'\, \Big \langle \left[ A(\mathbf{r},t-t'),B(\mathbf{r}') \right] \Big \rangle \nonumber\\ &\times S^{ss'}(\mathbf{r}',t') \nonumber\\
=& i\int_{-\infty}^{t} {\rm d} t' \int {\rm d}\mathbf{r}'\, \Big \langle \left[ A(\mathbf{r},t),B(\mathbf{r}',t') \right] \Big \rangle \nonumber\\ &\times S^{ss'}(\mathbf{r}',t') \,,
\end{align}
where $A(\mathbf{r},t)=\exp(iH_0t)A(\mathbf{r})\exp(-iH_0 t)$, and similarly for $B(\mathbf{r}',t')$.~Introducing now the generalised susceptibility 
\begin{align}
\chi_{AB}(\mathbf{r}-\mathbf{r}',t-t') = i\theta(t-t') \Big \langle \left[ A(\mathbf{r},t),B(\mathbf{r}',t') \right] \Big \rangle 
\end{align}
we obtain the Kubo formula for DM-electron scattering, namely
\begin{align}
\langle \Delta A(\mathbf{r},t)\rangle = \int_{-\infty}^{t} {\rm d} t' \int {\rm d}\mathbf{r}'\, \chi_{AB}(\mathbf{r}-\mathbf{r}',t-t') S^{ss'}(\mathbf{r}',t')
\label{eq:induced}
\end{align}
which describes the response to the DM-induced perturbation $V^{ss'}_{\rm eff}$ of a given observable $A$ in a detector material.

\section{Spectral representation of generalised susceptibilities}
\label{sec:chiK}
In this appendix, we derive the spectral representations for the correlation function $K_{j_\alpha j_\beta}$ and the generalised susceptibility $\chi_{j_\alpha j_\beta}$ that we use in Sec.~\ref{sec:formalism}.~We treat the case of translationally invariant systems, in which both $K_{j_\alpha j_\beta}$ and $\chi_{j_\alpha j_\beta}$ depend on the relative distance $(\mathbf{r}-\mathbf{r}')$ between the spatial points at which the densities or current densities $j_\alpha$ and $j_\beta$ are evaluated, and not on $\mathbf{r}$ and $\mathbf{r}'$ separately.~Consequently, the Fourier transform with respect to $(\mathbf{r}-\mathbf{r}')$ of the correlation function $K_{j_\alpha j_\beta}(\mathbf{r}-\mathbf{r}',t-t')$ can be written as follows
\begin{align}
K_{j_\alpha j_\beta}(\mathbf{q},t-t') &= \int {\rm d}(\mathbf{r}-\mathbf{r}') e^{-i\mathbf{q}\cdot (\mathbf{r}-\mathbf{r}')} \langle j_\alpha(\mathbf{r},t) 
j_\beta(\mathbf{r}',t') \rangle \nonumber\\
&=\frac{1}{V} \langle j_\alpha(\mathbf{q},t)  j_\beta(-\mathbf{q},t') \rangle\,.
\end{align}
Furthermore, the Fourier transform of $K_{j_\alpha j_\beta}(\mathbf{q},t-t')$ with respect to $t-t'$ can be expressed in terms of a complete set of energy eigenstates, denoted here by $|\psi_n\rangle$.~One finds,
\begin{align}
K_{j_\alpha j_\beta}(\mathbf{q},\omega) =& \int_{-\infty}^{+\infty} {\rm d}(t-t') e^{i\omega(t-t')} K_{j_\alpha j_\beta}(\mathbf{q},t-t') \nonumber\\
 =& \frac{1}{V}\int_{-\infty}^{+\infty} {\rm d}(t-t') e^{i\omega(t-t')} \sum_{n,m} \frac{e^{-\bar{\beta} E_n }}{Z}
\nonumber\\ & \times \langle \psi_n | j_\alpha(\mathbf{q},t) | \psi_m \rangle \langle \psi_m | j_\beta(-\mathbf{q},t') | \psi_n \rangle 
\end{align}
Translating now the operators $j_\alpha$ and $j_\beta$ to time $t=0$, we find
\begin{align}
K_{j_\alpha j_\beta}(\mathbf{q},\omega) =&  \frac{1}{V} \sum_{n,m}\int_{-\infty}^{+\infty}{\rm d}(t-t') e^{i\omega(t-t')} 
\nonumber\\ & \times \frac{e^{-\bar{\beta} E_n}}{Z} e^{i(E_n-E_m)(t-t')} 
\nonumber\\ & \times \langle \psi_n | j_\alpha(\mathbf{q}) | \psi_m \rangle \langle \psi_m | j_\beta(-\mathbf{q}) | \psi_n \rangle \,,
\end{align}
with $j_\alpha(\mathbf{q})=j_\alpha(t=0,\mathbf{q})$ and $j_\beta(\mathbf{q})=j_\beta(t'=0,\mathbf{q})$.~Performing the integral over $(t-t')$ explicitly, we finally obtain
\begin{align}
K_{j_\alpha j_\beta}(\mathbf{q},\omega) =&  \frac{2\pi}{V}  \sum_{n,m}\frac{e^{-\bar{\beta} E_n}}{Z} \delta(E_n-E_m+\omega)
\nonumber\\ &\times \langle \psi_n | j_\alpha(\mathbf{q}) | \psi_m \rangle \langle \psi_m | j_\beta(-\mathbf{q}) | \psi_n \rangle
\label{eq:speK}
\end{align}
which is the spectral representation for the correlation function $K_{j_\alpha j_\beta}$ used in Sec.~\ref{sec:formalism}.~Similarly, the double Fourier transform of the generalised susceptibility $\chi_{j_\alpha j_\beta}$, namely
\begin{align}
\chi_{j_\alpha j_\beta}(\mathbf{q},\omega) =& \int_{-\infty}^{+\infty} {\rm d}(t-t') e^{i\omega(t-t')} \chi_{j_\alpha j_\beta}(\mathbf{q},t-t') 
\end{align}
can be written as
\begin{align}
\chi_{j_\alpha j_\beta}(\mathbf{q},\omega) =&  \frac{i}{V} \int_{-\infty}^{+\infty} {\rm d}(t-t') \theta(t-t') e^{i\omega(t-t')} 
\nonumber\\ & \times \sum_{n,m}\frac{e^{-\bar{\beta} E_n}}{Z} e^{i(E_n-E_m)(t-t')} 
\nonumber\\ & \times \langle \psi_n | j_\alpha(\mathbf{q}) | \psi_m \rangle \langle \psi_m | j_\beta(-\mathbf{q}) | \psi_n \rangle 
\nonumber\\ &\times \left( 1 - e^{-\bar{\beta}(E_m-E_n)}\right) \,.
\end{align}
Using now the integral representation for the step function,
\begin{align}
\theta(t-t') = - \frac{1}{2\pi i} \int^{+\infty}_{-\infty} {\rm d} \omega \, \frac{e^{-i\omega (t-t')}}{\omega+i\delta} \,,
\end{align}
we find
\begin{align}
\chi_{j_\alpha j_\beta}(\mathbf{q},\omega) =&  -\frac{1}{V} \sum_{n,m}\frac{e^{-\bar{\beta} E_n}}{Z} 
\nonumber\\ & \times \langle \psi_n | j_\alpha(\mathbf{q}) | \psi_m \rangle \langle \psi_m | j_\beta(-\mathbf{q}) | \psi_n \rangle 
\nonumber\\ &\times \frac{\left( 1 - e^{-\bar{\beta}(E_m-E_n)}\right)}{\omega+E_n-E_m+i\delta}\,,
\label{eq:spechi}
\end{align}
which is the spectral representation for $\chi_{j_\alpha j_\beta}$ we use in Sec.~\ref{sec:formalism}.~The spectral representation for $\chi^A_{j_\alpha j_\beta}$ can be derived using
\begin{align}
\theta(t'-t) = \frac{1}{2\pi i} \int^{+\infty}_{-\infty} {\rm d} \omega \, \frac{e^{i\omega (t'-t)}}{\omega-i\delta} \,.
\end{align}
One finds
\begin{align}
\chi^A_{j_\alpha j_\beta}(\mathbf{q},\omega) =&  -\frac{1}{V} \sum_{n,m}\frac{e^{-\bar{\beta} E_n}}{Z} 
\nonumber\\ & \times \langle \psi_n | j_\alpha(\mathbf{q}) | \psi_m \rangle \langle \psi_m | j_\beta(-\mathbf{q}) | \psi_n \rangle 
\nonumber\\ &\times \frac{\left( 1 - e^{-\bar{\beta}(E_m-E_n)}\right)}{\omega+E_n-E_m-i\delta} \,.
\label{eq:spechiA}
\end{align}
Before concluding, we notice that
\begin{align}
\chi^*_{j^\dagger_\beta j^\dagger_\alpha}(\mathbf{q},\omega) =& -\frac{1}{V} \sum_{n,m}\frac{e^{-\bar{\beta} E_n}}{Z} 
\nonumber\\ & \times \langle \psi_n | j^\dagger_\beta(\mathbf{q}) | \psi_m \rangle^* \langle \psi_m | j^\dagger_\alpha(-\mathbf{q}) | \psi_n \rangle^* 
\nonumber\\ &\times \frac{\left( 1 - e^{-\bar{\beta}(E_m-E_n)}\right)}{\omega+E_n-E_m-i\delta} \,.
\end{align}
Since $\langle \psi_n | j^\dagger_\beta(\mathbf{q}) | \psi_m \rangle^* = \langle \psi_m | j_\beta(-\mathbf{q}) | \psi_n \rangle $, we finally obtain
\begin{align}
\chi^*_{j^\dagger_\beta j^\dagger_\alpha}(\mathbf{q},\omega) = \chi^{A}_{j_\alpha j_\beta}(\mathbf{q},\omega) \,.
\label{eq:chichiA}
\end{align}


\begin{thebibliography}{64}%
\makeatletter
\providecommand \@ifxundefined [1]{%
 \@ifx{#1\undefined}
}%
\providecommand \@ifnum [1]{%
 \ifnum #1\expandafter \@firstoftwo
 \else \expandafter \@secondoftwo
 \fi
}%
\providecommand \@ifx [1]{%
 \ifx #1\expandafter \@firstoftwo
 \else \expandafter \@secondoftwo
 \fi
}%
\providecommand \natexlab [1]{#1}%
\providecommand \enquote  [1]{``#1''}%
\providecommand \bibnamefont  [1]{#1}%
\providecommand \bibfnamefont [1]{#1}%
\providecommand \citenamefont [1]{#1}%
\providecommand \href@noop [0]{\@secondoftwo}%
\providecommand \href [0]{\begingroup \@sanitize@url \@href}%
\providecommand \@href[1]{\@@startlink{#1}\@@href}%
\providecommand \@@href[1]{\endgroup#1\@@endlink}%
\providecommand \@sanitize@url [0]{\catcode `\\12\catcode `\$12\catcode
  `\&12\catcode `\#12\catcode `\^12\catcode `\_12\catcode `\%12\relax}%
\providecommand \@@startlink[1]{}%
\providecommand \@@endlink[0]{}%
\providecommand \url  [0]{\begingroup\@sanitize@url \@url }%
\providecommand \@url [1]{\endgroup\@href {#1}{\urlprefix }}%
\providecommand \urlprefix  [0]{URL }%
\providecommand \Eprint [0]{\href }%
\providecommand \doibase [0]{http://dx.doi.org/}%
\providecommand \selectlanguage [0]{\@gobble}%
\providecommand \bibinfo  [0]{\@secondoftwo}%
\providecommand \bibfield  [0]{\@secondoftwo}%
\providecommand \translation [1]{[#1]}%
\providecommand \BibitemOpen [0]{}%
\providecommand \bibitemStop [0]{}%
\providecommand \bibitemNoStop [0]{.\EOS\space}%
\providecommand \EOS [0]{\spacefactor3000\relax}%
\providecommand \BibitemShut  [1]{\csname bibitem#1\endcsname}%
\let\auto@bib@innerbib\@empty
\bibitem [{\citenamefont {Essig}\ \emph {et~al.}(2012)\citenamefont {Essig},
  \citenamefont {Mardon},\ and\ \citenamefont {Volansky}}]{Essig:2011nj}%
  \BibitemOpen
  \bibfield  {author} {\bibinfo {author} {\bibfnamefont {Rouven}\ \bibnamefont
  {Essig}}, \bibinfo {author} {\bibfnamefont {Jeremy}\ \bibnamefont {Mardon}},
  \ and\ \bibinfo {author} {\bibfnamefont {Tomer}\ \bibnamefont {Volansky}},\
  }\bibfield  {title} {\enquote {\bibinfo {title} {{Direct Detection of Sub-GeV
  Dark Matter}},}\ }\href {\doibase 10.1103/PhysRevD.85.076007} {\bibfield
  {journal} {\bibinfo  {journal} {Phys. Rev.}\ }\textbf {\bibinfo {volume}
  {D85}},\ \bibinfo {pages} {076007} (\bibinfo {year} {2012})},\ \Eprint
  {http://arxiv.org/abs/1108.5383} {arXiv:1108.5383 [hep-ph]} \BibitemShut
  {NoStop}%
\bibitem [{\citenamefont {Schumann}(2019)}]{Schumann:2019eaa}%
  \BibitemOpen
  \bibfield  {author} {\bibinfo {author} {\bibfnamefont {Marc}\ \bibnamefont
  {Schumann}},\ }\bibfield  {title} {\enquote {\bibinfo {title} {{Direct
  Detection of WIMP Dark Matter: Concepts and Status}},}\ }\href {\doibase
  10.1088/1361-6471/ab2ea5} {\bibfield  {journal} {\bibinfo  {journal} {J.
  Phys. G}\ }\textbf {\bibinfo {volume} {46}},\ \bibinfo {pages} {103003}
  (\bibinfo {year} {2019})},\ \Eprint {http://arxiv.org/abs/1903.03026}
  {arXiv:1903.03026 [astro-ph.CO]} \BibitemShut {NoStop}%
\bibitem [{\citenamefont {Mitridate}\ \emph {et~al.}(2023)\citenamefont
  {Mitridate}, \citenamefont {Trickle}, \citenamefont {Zhang},\ and\
  \citenamefont {Zurek}}]{Mitridate:2022tnv}%
  \BibitemOpen
  \bibfield  {author} {\bibinfo {author} {\bibfnamefont {Andrea}\ \bibnamefont
  {Mitridate}}, \bibinfo {author} {\bibfnamefont {Tanner}\ \bibnamefont
  {Trickle}}, \bibinfo {author} {\bibfnamefont {Zhengkang}\ \bibnamefont
  {Zhang}}, \ and\ \bibinfo {author} {\bibfnamefont {Kathryn~M.}\ \bibnamefont
  {Zurek}},\ }\bibfield  {title} {\enquote {\bibinfo {title} {{Snowmass white
  paper: Light dark matter direct detection at the interface with condensed
  matter physics}},}\ }\href {\doibase 10.1016/j.dark.2023.101221} {\bibfield
  {journal} {\bibinfo  {journal} {Phys. Dark Univ.}\ }\textbf {\bibinfo
  {volume} {40}},\ \bibinfo {pages} {101221} (\bibinfo {year} {2023})},\
  \Eprint {http://arxiv.org/abs/2203.07492} {arXiv:2203.07492 [hep-ph]}
  \BibitemShut {NoStop}%
\bibitem [{\citenamefont {Essig}\ \emph {et~al.}(2017)\citenamefont {Essig},
  \citenamefont {Volansky},\ and\ \citenamefont {Yu}}]{Essig:2017kqs}%
  \BibitemOpen
  \bibfield  {author} {\bibinfo {author} {\bibfnamefont {Rouven}\ \bibnamefont
  {Essig}}, \bibinfo {author} {\bibfnamefont {Tomer}\ \bibnamefont {Volansky}},
  \ and\ \bibinfo {author} {\bibfnamefont {Tien-Tien}\ \bibnamefont {Yu}},\
  }\bibfield  {title} {\enquote {\bibinfo {title} {{New Constraints and
  Prospects for sub-GeV Dark Matter Scattering off Electrons in Xenon}},}\
  }\href {\doibase 10.1103/PhysRevD.96.043017} {\bibfield  {journal} {\bibinfo
  {journal} {Phys. Rev.}\ }\textbf {\bibinfo {volume} {D96}},\ \bibinfo {pages}
  {043017} (\bibinfo {year} {2017})},\ \Eprint
  {http://arxiv.org/abs/1703.00910} {arXiv:1703.00910 [hep-ph]} \BibitemShut
  {NoStop}%
\bibitem [{\citenamefont {Agnes}\ \emph {et~al.}(2018)\citenamefont {Agnes}
  \emph {et~al.}}]{DarkSide:2018ppu}%
  \BibitemOpen
  \bibfield  {author} {\bibinfo {author} {\bibfnamefont {P.}~\bibnamefont
  {Agnes}} \emph {et~al.} (\bibinfo {collaboration} {DarkSide}),\ }\bibfield
  {title} {\enquote {\bibinfo {title} {{Constraints on Sub-GeV
  Dark-Matter\textendash{}Electron Scattering from the DarkSide-50
  Experiment}},}\ }\href {\doibase 10.1103/PhysRevLett.121.111303} {\bibfield
  {journal} {\bibinfo  {journal} {Phys. Rev. Lett.}\ }\textbf {\bibinfo
  {volume} {121}},\ \bibinfo {pages} {111303} (\bibinfo {year} {2018})},\
  \Eprint {http://arxiv.org/abs/1802.06998} {arXiv:1802.06998 [astro-ph.CO]}
  \BibitemShut {NoStop}%
\bibitem [{\citenamefont {Catena}\ \emph {et~al.}(2020)\citenamefont {Catena},
  \citenamefont {Emken}, \citenamefont {Spaldin},\ and\ \citenamefont
  {Tarantino}}]{Catena:2019gfa}%
  \BibitemOpen
  \bibfield  {author} {\bibinfo {author} {\bibfnamefont {Riccardo}\
  \bibnamefont {Catena}}, \bibinfo {author} {\bibfnamefont {Timon}\
  \bibnamefont {Emken}}, \bibinfo {author} {\bibfnamefont {Nicola~A.}\
  \bibnamefont {Spaldin}}, \ and\ \bibinfo {author} {\bibfnamefont {Walter}\
  \bibnamefont {Tarantino}},\ }\bibfield  {title} {\enquote {\bibinfo {title}
  {{Atomic responses to general dark matter-electron interactions}},}\ }\href
  {\doibase 10.1103/PhysRevResearch.2.033195} {\bibfield  {journal} {\bibinfo
  {journal} {Phys. Rev. Res.}\ }\textbf {\bibinfo {volume} {2}},\ \bibinfo
  {pages} {033195} (\bibinfo {year} {2020})},\ \Eprint
  {http://arxiv.org/abs/1912.08204} {arXiv:1912.08204 [hep-ph]} \BibitemShut
  {NoStop}%
\bibitem [{\citenamefont {Aprile}\ \emph {et~al.}(2019)\citenamefont {Aprile}
  \emph {et~al.}}]{Aprile:2019xxb}%
  \BibitemOpen
  \bibfield  {author} {\bibinfo {author} {\bibfnamefont {E.}~\bibnamefont
  {Aprile}} \emph {et~al.} (\bibinfo {collaboration} {XENON}),\ }\bibfield
  {title} {\enquote {\bibinfo {title} {{Light Dark Matter Search with
  Ionization Signals in XENON1T}},}\ }\href {\doibase
  10.1103/PhysRevLett.123.251801} {\bibfield  {journal} {\bibinfo  {journal}
  {Phys. Rev. Lett.}\ }\textbf {\bibinfo {volume} {123}},\ \bibinfo {pages}
  {251801} (\bibinfo {year} {2019})},\ \Eprint
  {http://arxiv.org/abs/1907.11485} {arXiv:1907.11485 [hep-ex]} \BibitemShut
  {NoStop}%
\bibitem [{\citenamefont {Aprile}\ \emph {et~al.}(2020)\citenamefont {Aprile}
  \emph {et~al.}}]{XENON:2020rca}%
  \BibitemOpen
  \bibfield  {author} {\bibinfo {author} {\bibfnamefont {E.}~\bibnamefont
  {Aprile}} \emph {et~al.} (\bibinfo {collaboration} {XENON}),\ }\bibfield
  {title} {\enquote {\bibinfo {title} {{Excess electronic recoil events in
  XENON1T}},}\ }\href {\doibase 10.1103/PhysRevD.102.072004} {\bibfield
  {journal} {\bibinfo  {journal} {Phys. Rev. D}\ }\textbf {\bibinfo {volume}
  {102}},\ \bibinfo {pages} {072004} (\bibinfo {year} {2020})},\ \Eprint
  {http://arxiv.org/abs/2006.09721} {arXiv:2006.09721 [hep-ex]} \BibitemShut
  {NoStop}%
\bibitem [{\citenamefont {Graham}\ \emph {et~al.}(2012)\citenamefont {Graham},
  \citenamefont {Kaplan}, \citenamefont {Rajendran},\ and\ \citenamefont
  {Walters}}]{Graham:2012su}%
  \BibitemOpen
  \bibfield  {author} {\bibinfo {author} {\bibfnamefont {Peter~W.}\
  \bibnamefont {Graham}}, \bibinfo {author} {\bibfnamefont {David~E.}\
  \bibnamefont {Kaplan}}, \bibinfo {author} {\bibfnamefont {Surjeet}\
  \bibnamefont {Rajendran}}, \ and\ \bibinfo {author} {\bibfnamefont
  {Matthew~T.}\ \bibnamefont {Walters}},\ }\bibfield  {title} {\enquote
  {\bibinfo {title} {{Semiconductor Probes of Light Dark Matter}},}\ }\href
  {\doibase 10.1016/j.dark.2012.09.001} {\bibfield  {journal} {\bibinfo
  {journal} {Phys. Dark Univ.}\ }\textbf {\bibinfo {volume} {1}},\ \bibinfo
  {pages} {32--49} (\bibinfo {year} {2012})},\ \Eprint
  {http://arxiv.org/abs/1203.2531} {arXiv:1203.2531 [hep-ph]} \BibitemShut
  {NoStop}%
\bibitem [{\citenamefont {Essig}\ \emph {et~al.}(2016)\citenamefont {Essig},
  \citenamefont {Fernandez-Serra}, \citenamefont {Mardon}, \citenamefont
  {Soto}, \citenamefont {Volansky},\ and\ \citenamefont {Yu}}]{Essig:2015cda}%
  \BibitemOpen
  \bibfield  {author} {\bibinfo {author} {\bibfnamefont {Rouven}\ \bibnamefont
  {Essig}}, \bibinfo {author} {\bibfnamefont {Marivi}\ \bibnamefont
  {Fernandez-Serra}}, \bibinfo {author} {\bibfnamefont {Jeremy}\ \bibnamefont
  {Mardon}}, \bibinfo {author} {\bibfnamefont {Adrian}\ \bibnamefont {Soto}},
  \bibinfo {author} {\bibfnamefont {Tomer}\ \bibnamefont {Volansky}}, \ and\
  \bibinfo {author} {\bibfnamefont {Tien-Tien}\ \bibnamefont {Yu}},\ }\bibfield
   {title} {\enquote {\bibinfo {title} {{Direct Detection of sub-GeV Dark
  Matter with Semiconductor Targets}},}\ }\href {\doibase
  10.1007/JHEP05(2016)046} {\bibfield  {journal} {\bibinfo  {journal} {JHEP}\
  }\textbf {\bibinfo {volume} {05}},\ \bibinfo {pages} {046} (\bibinfo {year}
  {2016})},\ \Eprint {http://arxiv.org/abs/1509.01598} {arXiv:1509.01598
  [hep-ph]} \BibitemShut {NoStop}%
\bibitem [{\citenamefont {Derenzo}\ \emph {et~al.}(2017)\citenamefont
  {Derenzo}, \citenamefont {Essig}, \citenamefont {Massari}, \citenamefont
  {Soto},\ and\ \citenamefont {Yu}}]{Derenzo:2016fse}%
  \BibitemOpen
  \bibfield  {author} {\bibinfo {author} {\bibfnamefont {Stephen}\ \bibnamefont
  {Derenzo}}, \bibinfo {author} {\bibfnamefont {Rouven}\ \bibnamefont {Essig}},
  \bibinfo {author} {\bibfnamefont {Andrea}\ \bibnamefont {Massari}}, \bibinfo
  {author} {\bibfnamefont {AdrÃ­an}\ \bibnamefont {Soto}}, \ and\ \bibinfo
  {author} {\bibfnamefont {Tien-Tien}\ \bibnamefont {Yu}},\ }\bibfield  {title}
  {\enquote {\bibinfo {title} {{Direct Detection of sub-GeV Dark Matter with
  Scintillating Targets}},}\ }\href {\doibase 10.1103/PhysRevD.96.016026}
  {\bibfield  {journal} {\bibinfo  {journal} {Phys. Rev.}\ }\textbf {\bibinfo
  {volume} {D96}},\ \bibinfo {pages} {016026} (\bibinfo {year} {2017})},\
  \Eprint {http://arxiv.org/abs/1607.01009} {arXiv:1607.01009 [hep-ph]}
  \BibitemShut {NoStop}%
\bibitem [{\citenamefont {Agnese}\ \emph {et~al.}(2018)\citenamefont {Agnese}
  \emph {et~al.}}]{Agnese:2018col}%
  \BibitemOpen
  \bibfield  {author} {\bibinfo {author} {\bibfnamefont {R.}~\bibnamefont
  {Agnese}} \emph {et~al.} (\bibinfo {collaboration} {SuperCDMS}),\ }\bibfield
  {title} {\enquote {\bibinfo {title} {{First Dark Matter Constraints from a
  SuperCDMS Single-Charge Sensitive Detector}},}\ }\href {\doibase
  10.1103/PhysRevLett.122.069901, 10.1103/PhysRevLett.121.051301} {\bibfield
  {journal} {\bibinfo  {journal} {Phys. Rev. Lett.}\ }\textbf {\bibinfo
  {volume} {121}},\ \bibinfo {pages} {051301} (\bibinfo {year} {2018})},\
  \bibinfo {note} {[erratum: Phys. Rev. Lett.122,no.6,069901(2019)]},\ \Eprint
  {http://arxiv.org/abs/1804.10697} {arXiv:1804.10697 [hep-ex]} \BibitemShut
  {NoStop}%
\bibitem [{\citenamefont {Kurinsky}\ \emph {et~al.}(2019)\citenamefont
  {Kurinsky}, \citenamefont {Yu}, \citenamefont {Hochberg},\ and\ \citenamefont
  {Cabrera}}]{Kurinsky:2019pgb}%
  \BibitemOpen
  \bibfield  {author} {\bibinfo {author} {\bibfnamefont {Noah~Alexander}\
  \bibnamefont {Kurinsky}}, \bibinfo {author} {\bibfnamefont {To~Chin}\
  \bibnamefont {Yu}}, \bibinfo {author} {\bibfnamefont {Yonit}\ \bibnamefont
  {Hochberg}}, \ and\ \bibinfo {author} {\bibfnamefont {Blas}\ \bibnamefont
  {Cabrera}},\ }\bibfield  {title} {\enquote {\bibinfo {title} {{Diamond
  Detectors for Direct Detection of Sub-GeV Dark Matter}},}\ }\href {\doibase
  10.1103/PhysRevD.99.123005} {\bibfield  {journal} {\bibinfo  {journal} {Phys.
  Rev. D}\ }\textbf {\bibinfo {volume} {99}},\ \bibinfo {pages} {123005}
  (\bibinfo {year} {2019})},\ \Eprint {http://arxiv.org/abs/1901.07569}
  {arXiv:1901.07569 [hep-ex]} \BibitemShut {NoStop}%
\bibitem [{\citenamefont {Aguilar-Arevalo}\ \emph {et~al.}(2019)\citenamefont
  {Aguilar-Arevalo} \emph {et~al.}}]{DAMIC:2019dcn}%
  \BibitemOpen
  \bibfield  {author} {\bibinfo {author} {\bibfnamefont {A.}~\bibnamefont
  {Aguilar-Arevalo}} \emph {et~al.} (\bibinfo {collaboration} {DAMIC}),\
  }\bibfield  {title} {\enquote {\bibinfo {title} {{Constraints on Light Dark
  Matter Particles Interacting with Electrons from DAMIC at SNOLAB}},}\ }\href
  {\doibase 10.1103/PhysRevLett.123.181802} {\bibfield  {journal} {\bibinfo
  {journal} {Phys. Rev. Lett.}\ }\textbf {\bibinfo {volume} {123}},\ \bibinfo
  {pages} {181802} (\bibinfo {year} {2019})},\ \Eprint
  {http://arxiv.org/abs/1907.12628} {arXiv:1907.12628 [astro-ph.CO]}
  \BibitemShut {NoStop}%
\bibitem [{\citenamefont {Arnaud}\ \emph {et~al.}(2020)\citenamefont {Arnaud}
  \emph {et~al.}}]{EDELWEISS:2020fxc}%
  \BibitemOpen
  \bibfield  {author} {\bibinfo {author} {\bibfnamefont {Q.}~\bibnamefont
  {Arnaud}} \emph {et~al.} (\bibinfo {collaboration} {EDELWEISS}),\ }\bibfield
  {title} {\enquote {\bibinfo {title} {{First germanium-based constraints on
  sub-MeV Dark Matter with the EDELWEISS experiment}},}\ }\href {\doibase
  10.1103/PhysRevLett.125.141301} {\bibfield  {journal} {\bibinfo  {journal}
  {Phys. Rev. Lett.}\ }\textbf {\bibinfo {volume} {125}},\ \bibinfo {pages}
  {141301} (\bibinfo {year} {2020})},\ \Eprint
  {http://arxiv.org/abs/2003.01046} {arXiv:2003.01046 [astro-ph.GA]}
  \BibitemShut {NoStop}%
\bibitem [{\citenamefont {Barak}\ \emph {et~al.}(2020)\citenamefont {Barak}
  \emph {et~al.}}]{SENSEI:2020dpa}%
  \BibitemOpen
  \bibfield  {author} {\bibinfo {author} {\bibfnamefont {Liron}\ \bibnamefont
  {Barak}} \emph {et~al.} (\bibinfo {collaboration} {SENSEI}),\ }\bibfield
  {title} {\enquote {\bibinfo {title} {{SENSEI: Direct-Detection Results on
  sub-GeV Dark Matter from a New Skipper-CCD}},}\ }\href {\doibase
  10.1103/PhysRevLett.125.171802} {\bibfield  {journal} {\bibinfo  {journal}
  {Phys. Rev. Lett.}\ }\textbf {\bibinfo {volume} {125}},\ \bibinfo {pages}
  {171802} (\bibinfo {year} {2020})},\ \Eprint
  {http://arxiv.org/abs/2004.11378} {arXiv:2004.11378 [astro-ph.CO]}
  \BibitemShut {NoStop}%
\bibitem [{\citenamefont {Griffin}\ \emph
  {et~al.}(2021{\natexlab{a}})\citenamefont {Griffin}, \citenamefont
  {Hochberg}, \citenamefont {Inzani}, \citenamefont {Kurinsky}, \citenamefont
  {Lin},\ and\ \citenamefont {Chin}}]{Griffin:2020lgd}%
  \BibitemOpen
  \bibfield  {author} {\bibinfo {author} {\bibfnamefont {Sin\'ead~M.}\
  \bibnamefont {Griffin}}, \bibinfo {author} {\bibfnamefont {Yonit}\
  \bibnamefont {Hochberg}}, \bibinfo {author} {\bibfnamefont {Katherine}\
  \bibnamefont {Inzani}}, \bibinfo {author} {\bibfnamefont {Noah}\ \bibnamefont
  {Kurinsky}}, \bibinfo {author} {\bibfnamefont {Tongyan}\ \bibnamefont {Lin}},
  \ and\ \bibinfo {author} {\bibfnamefont {To}~\bibnamefont {Chin}},\
  }\bibfield  {title} {\enquote {\bibinfo {title} {{Silicon carbide detectors
  for sub-GeV dark matter}},}\ }\href {\doibase 10.1103/PhysRevD.103.075002}
  {\bibfield  {journal} {\bibinfo  {journal} {Phys. Rev. D}\ }\textbf {\bibinfo
  {volume} {103}},\ \bibinfo {pages} {075002} (\bibinfo {year}
  {2021}{\natexlab{a}})},\ \Eprint {http://arxiv.org/abs/2008.08560}
  {arXiv:2008.08560 [hep-ph]} \BibitemShut {NoStop}%
\bibitem [{\citenamefont {Catena}\ \emph {et~al.}(2021)\citenamefont {Catena},
  \citenamefont {Emken}, \citenamefont {Matas}, \citenamefont {Spaldin},\ and\
  \citenamefont {Urdshals}}]{Catena:2021qsr}%
  \BibitemOpen
  \bibfield  {author} {\bibinfo {author} {\bibfnamefont {Riccardo}\
  \bibnamefont {Catena}}, \bibinfo {author} {\bibfnamefont {Timon}\
  \bibnamefont {Emken}}, \bibinfo {author} {\bibfnamefont {Marek}\ \bibnamefont
  {Matas}}, \bibinfo {author} {\bibfnamefont {Nicola~A.}\ \bibnamefont
  {Spaldin}}, \ and\ \bibinfo {author} {\bibfnamefont {Einar}\ \bibnamefont
  {Urdshals}},\ }\bibfield  {title} {\enquote {\bibinfo {title} {{Crystal
  responses to general dark matter-electron interactions}},}\ }\href {\doibase
  10.1103/PhysRevResearch.3.033149} {\bibfield  {journal} {\bibinfo  {journal}
  {Phys. Rev. Res.}\ }\textbf {\bibinfo {volume} {3}},\ \bibinfo {pages}
  {033149} (\bibinfo {year} {2021})},\ \Eprint
  {http://arxiv.org/abs/2105.02233} {arXiv:2105.02233 [hep-ph]} \BibitemShut
  {NoStop}%
\bibitem [{\citenamefont {Griffin}\ \emph
  {et~al.}(2021{\natexlab{b}})\citenamefont {Griffin}, \citenamefont {Inzani},
  \citenamefont {Trickle}, \citenamefont {Zhang},\ and\ \citenamefont
  {Zurek}}]{Griffin:2021znd}%
  \BibitemOpen
  \bibfield  {author} {\bibinfo {author} {\bibfnamefont {Sin\'ead~M.}\
  \bibnamefont {Griffin}}, \bibinfo {author} {\bibfnamefont {Katherine}\
  \bibnamefont {Inzani}}, \bibinfo {author} {\bibfnamefont {Tanner}\
  \bibnamefont {Trickle}}, \bibinfo {author} {\bibfnamefont {Zhengkang}\
  \bibnamefont {Zhang}}, \ and\ \bibinfo {author} {\bibfnamefont {Kathryn~M.}\
  \bibnamefont {Zurek}},\ }\bibfield  {title} {\enquote {\bibinfo {title}
  {{Extended Calculation of Dark Matter-Electron Scattering in Crystal
  Targets}},}\ }\href@noop {} {\  (\bibinfo {year} {2021}{\natexlab{b}})},\
  \Eprint {http://arxiv.org/abs/2105.05253} {arXiv:2105.05253 [hep-ph]}
  \BibitemShut {NoStop}%
\bibitem [{\citenamefont {Knapen}\ \emph {et~al.}(2021)\citenamefont {Knapen},
  \citenamefont {Kozaczuk},\ and\ \citenamefont {Lin}}]{Knapen:2021run}%
  \BibitemOpen
  \bibfield  {author} {\bibinfo {author} {\bibfnamefont {Simon}\ \bibnamefont
  {Knapen}}, \bibinfo {author} {\bibfnamefont {Jonathan}\ \bibnamefont
  {Kozaczuk}}, \ and\ \bibinfo {author} {\bibfnamefont {Tongyan}\ \bibnamefont
  {Lin}},\ }\bibfield  {title} {\enquote {\bibinfo {title} {{Dark
  matter-electron scattering in dielectrics}},}\ }\href {\doibase
  10.1103/PhysRevD.104.015031} {\bibfield  {journal} {\bibinfo  {journal}
  {Phys. Rev. D}\ }\textbf {\bibinfo {volume} {104}},\ \bibinfo {pages}
  {015031} (\bibinfo {year} {2021})},\ \Eprint
  {http://arxiv.org/abs/2101.08275} {arXiv:2101.08275 [hep-ph]} \BibitemShut
  {NoStop}%
\bibitem [{\citenamefont {Hochberg}\ \emph
  {et~al.}(2021{\natexlab{a}})\citenamefont {Hochberg}, \citenamefont {Kahn},
  \citenamefont {Kurinsky}, \citenamefont {Lehmann}, \citenamefont {Yu},\ and\
  \citenamefont {Berggren}}]{Hochberg:2021pkt}%
  \BibitemOpen
  \bibfield  {author} {\bibinfo {author} {\bibfnamefont {Yonit}\ \bibnamefont
  {Hochberg}}, \bibinfo {author} {\bibfnamefont {Yonatan}\ \bibnamefont
  {Kahn}}, \bibinfo {author} {\bibfnamefont {Noah}\ \bibnamefont {Kurinsky}},
  \bibinfo {author} {\bibfnamefont {Benjamin~V.}\ \bibnamefont {Lehmann}},
  \bibinfo {author} {\bibfnamefont {To~Chin}\ \bibnamefont {Yu}}, \ and\
  \bibinfo {author} {\bibfnamefont {Karl~K.}\ \bibnamefont {Berggren}},\
  }\bibfield  {title} {\enquote {\bibinfo {title} {{Determining
  Dark-Matter\textendash{}Electron Scattering Rates from the Dielectric
  Function}},}\ }\href {\doibase 10.1103/PhysRevLett.127.151802} {\bibfield
  {journal} {\bibinfo  {journal} {Phys. Rev. Lett.}\ }\textbf {\bibinfo
  {volume} {127}},\ \bibinfo {pages} {151802} (\bibinfo {year}
  {2021}{\natexlab{a}})},\ \Eprint {http://arxiv.org/abs/2101.08263}
  {arXiv:2101.08263 [hep-ph]} \BibitemShut {NoStop}%
\bibitem [{\citenamefont {Lasenby}\ and\ \citenamefont
  {Prabhu}(2022)}]{Lasenby:2021wsc}%
  \BibitemOpen
  \bibfield  {author} {\bibinfo {author} {\bibfnamefont {Robert}\ \bibnamefont
  {Lasenby}}\ and\ \bibinfo {author} {\bibfnamefont {Anirudh}\ \bibnamefont
  {Prabhu}},\ }\bibfield  {title} {\enquote {\bibinfo {title} {{Dark
  matter\textendash{}electron scattering in materials: Sum rules and
  heterostructures}},}\ }\href {\doibase 10.1103/PhysRevD.105.095009}
  {\bibfield  {journal} {\bibinfo  {journal} {Phys. Rev. D}\ }\textbf {\bibinfo
  {volume} {105}},\ \bibinfo {pages} {095009} (\bibinfo {year} {2022})},\
  \Eprint {http://arxiv.org/abs/2110.01587} {arXiv:2110.01587 [hep-ph]}
  \BibitemShut {NoStop}%
\bibitem [{\citenamefont {Chen}\ \emph {et~al.}(2022)\citenamefont {Chen},
  \citenamefont {Mitridate}, \citenamefont {Trickle}, \citenamefont {Zhang},
  \citenamefont {Bernardi},\ and\ \citenamefont {Zurek}}]{Chen:2022pyd}%
  \BibitemOpen
  \bibfield  {author} {\bibinfo {author} {\bibfnamefont {Hsiao-Yi}\
  \bibnamefont {Chen}}, \bibinfo {author} {\bibfnamefont {Andrea}\ \bibnamefont
  {Mitridate}}, \bibinfo {author} {\bibfnamefont {Tanner}\ \bibnamefont
  {Trickle}}, \bibinfo {author} {\bibfnamefont {Zhengkang}\ \bibnamefont
  {Zhang}}, \bibinfo {author} {\bibfnamefont {Marco}\ \bibnamefont {Bernardi}},
  \ and\ \bibinfo {author} {\bibfnamefont {Kathryn~M.}\ \bibnamefont {Zurek}},\
  }\bibfield  {title} {\enquote {\bibinfo {title} {{Dark matter direct
  detection in materials with spin-orbit coupling}},}\ }\href {\doibase
  10.1103/PhysRevD.106.015024} {\bibfield  {journal} {\bibinfo  {journal}
  {Phys. Rev. D}\ }\textbf {\bibinfo {volume} {106}},\ \bibinfo {pages}
  {015024} (\bibinfo {year} {2022})},\ \Eprint
  {http://arxiv.org/abs/2202.11716} {arXiv:2202.11716 [hep-ph]} \BibitemShut
  {NoStop}%
\bibitem [{\citenamefont {Hochberg}\ \emph {et~al.}(2016)\citenamefont
  {Hochberg}, \citenamefont {Zhao},\ and\ \citenamefont
  {Zurek}}]{Hochberg:2015pha}%
  \BibitemOpen
  \bibfield  {author} {\bibinfo {author} {\bibfnamefont {Yonit}\ \bibnamefont
  {Hochberg}}, \bibinfo {author} {\bibfnamefont {Yue}\ \bibnamefont {Zhao}}, \
  and\ \bibinfo {author} {\bibfnamefont {Kathryn~M.}\ \bibnamefont {Zurek}},\
  }\bibfield  {title} {\enquote {\bibinfo {title} {{Superconducting Detectors
  for Superlight Dark Matter}},}\ }\href {\doibase
  10.1103/PhysRevLett.116.011301} {\bibfield  {journal} {\bibinfo  {journal}
  {Phys. Rev. Lett.}\ }\textbf {\bibinfo {volume} {116}},\ \bibinfo {pages}
  {011301} (\bibinfo {year} {2016})},\ \Eprint
  {http://arxiv.org/abs/1504.07237} {arXiv:1504.07237 [hep-ph]} \BibitemShut
  {NoStop}%
\bibitem [{\citenamefont {Hochberg}\ \emph
  {et~al.}(2021{\natexlab{b}})\citenamefont {Hochberg}, \citenamefont
  {Lehmann}, \citenamefont {Charaev}, \citenamefont {Chiles}, \citenamefont
  {Colangelo}, \citenamefont {Nam},\ and\ \citenamefont
  {Berggren}}]{Hochberg:2021yud}%
  \BibitemOpen
  \bibfield  {author} {\bibinfo {author} {\bibfnamefont {Yonit}\ \bibnamefont
  {Hochberg}}, \bibinfo {author} {\bibfnamefont {Benjamin~V.}\ \bibnamefont
  {Lehmann}}, \bibinfo {author} {\bibfnamefont {Ilya}\ \bibnamefont {Charaev}},
  \bibinfo {author} {\bibfnamefont {Jeff}\ \bibnamefont {Chiles}}, \bibinfo
  {author} {\bibfnamefont {Marco}\ \bibnamefont {Colangelo}}, \bibinfo {author}
  {\bibfnamefont {Sae~Woo}\ \bibnamefont {Nam}}, \ and\ \bibinfo {author}
  {\bibfnamefont {Karl~K.}\ \bibnamefont {Berggren}},\ }\bibfield  {title}
  {\enquote {\bibinfo {title} {{New Constraints on Dark Matter from
  Superconducting Nanowires}},}\ }\href@noop {} {\  (\bibinfo {year}
  {2021}{\natexlab{b}})},\ \Eprint {http://arxiv.org/abs/2110.01586}
  {arXiv:2110.01586 [hep-ph]} \BibitemShut {NoStop}%
\bibitem [{\citenamefont {Hochberg}\ \emph {et~al.}(2018)\citenamefont
  {Hochberg}, \citenamefont {Kahn}, \citenamefont {Lisanti}, \citenamefont
  {Zurek}, \citenamefont {Grushin}, \citenamefont {Ilan}, \citenamefont
  {Griffin}, \citenamefont {Liu}, \citenamefont {Weber},\ and\ \citenamefont
  {Neaton}}]{Hochberg:2017wce}%
  \BibitemOpen
  \bibfield  {author} {\bibinfo {author} {\bibfnamefont {Yonit}\ \bibnamefont
  {Hochberg}}, \bibinfo {author} {\bibfnamefont {Yonatan}\ \bibnamefont
  {Kahn}}, \bibinfo {author} {\bibfnamefont {Mariangela}\ \bibnamefont
  {Lisanti}}, \bibinfo {author} {\bibfnamefont {Kathryn~M.}\ \bibnamefont
  {Zurek}}, \bibinfo {author} {\bibfnamefont {Adolfo~G.}\ \bibnamefont
  {Grushin}}, \bibinfo {author} {\bibfnamefont {Roni}\ \bibnamefont {Ilan}},
  \bibinfo {author} {\bibfnamefont {SinÃ©ad~M.}\ \bibnamefont {Griffin}},
  \bibinfo {author} {\bibfnamefont {Zhen-Fei}\ \bibnamefont {Liu}}, \bibinfo
  {author} {\bibfnamefont {Sophie~F.}\ \bibnamefont {Weber}}, \ and\ \bibinfo
  {author} {\bibfnamefont {Jeffrey~B.}\ \bibnamefont {Neaton}},\ }\bibfield
  {title} {\enquote {\bibinfo {title} {{Detection of sub-MeV Dark Matter with
  Three-Dimensional Dirac Materials}},}\ }\href {\doibase
  10.1103/PhysRevD.97.015004} {\bibfield  {journal} {\bibinfo  {journal} {Phys.
  Rev.}\ }\textbf {\bibinfo {volume} {D97}},\ \bibinfo {pages} {015004}
  (\bibinfo {year} {2018})},\ \Eprint {http://arxiv.org/abs/1708.08929}
  {arXiv:1708.08929 [hep-ph]} \BibitemShut {NoStop}%
\bibitem [{\citenamefont {Geilhufe}\ \emph {et~al.}(2019)\citenamefont
  {Geilhufe}, \citenamefont {Kahlhoefer},\ and\ \citenamefont
  {Winkler}}]{Geilhufe:2019ndy}%
  \BibitemOpen
  \bibfield  {author} {\bibinfo {author} {\bibfnamefont {R.~Matthias}\
  \bibnamefont {Geilhufe}}, \bibinfo {author} {\bibfnamefont {Felix}\
  \bibnamefont {Kahlhoefer}}, \ and\ \bibinfo {author} {\bibfnamefont
  {Martin~Wolfgang}\ \bibnamefont {Winkler}},\ }\bibfield  {title} {\enquote
  {\bibinfo {title} {{Dirac Materials for Sub-MeV Dark Matter Detection: New
  Targets and Improved Formalism}},}\ }\href@noop {} {\  (\bibinfo {year}
  {2019})},\ \Eprint {http://arxiv.org/abs/1910.02091} {arXiv:1910.02091
  [hep-ph]} \BibitemShut {NoStop}%
\bibitem [{\citenamefont {Coskuner}\ \emph {et~al.}(2021)\citenamefont
  {Coskuner}, \citenamefont {Mitridate}, \citenamefont {Olivares},\ and\
  \citenamefont {Zurek}}]{Coskuner2021Jan}%
  \BibitemOpen
  \bibfield  {author} {\bibinfo {author} {\bibfnamefont {Ahmet}\ \bibnamefont
  {Coskuner}}, \bibinfo {author} {\bibfnamefont {Andrea}\ \bibnamefont
  {Mitridate}}, \bibinfo {author} {\bibfnamefont {Andres}\ \bibnamefont
  {Olivares}}, \ and\ \bibinfo {author} {\bibfnamefont {Kathryn~M.}\
  \bibnamefont {Zurek}},\ }\bibfield  {title} {\enquote {\bibinfo {title}
  {{Directional dark matter detection in anisotropic Dirac materials}},}\
  }\href {\doibase 10.1103/PhysRevD.103.016006} {\bibfield  {journal} {\bibinfo
   {journal} {Phys. Rev. D}\ }\textbf {\bibinfo {volume} {103}},\ \bibinfo
  {pages} {016006} (\bibinfo {year} {2021})}\BibitemShut {NoStop}%
\bibitem [{\citenamefont {Hochberg}\ \emph {et~al.}(2017)\citenamefont
  {Hochberg}, \citenamefont {Kahn}, \citenamefont {Lisanti}, \citenamefont
  {Tully},\ and\ \citenamefont {Zurek}}]{Hochberg:2016ntt}%
  \BibitemOpen
  \bibfield  {author} {\bibinfo {author} {\bibfnamefont {Yonit}\ \bibnamefont
  {Hochberg}}, \bibinfo {author} {\bibfnamefont {Yonatan}\ \bibnamefont
  {Kahn}}, \bibinfo {author} {\bibfnamefont {Mariangela}\ \bibnamefont
  {Lisanti}}, \bibinfo {author} {\bibfnamefont {Christopher~G.}\ \bibnamefont
  {Tully}}, \ and\ \bibinfo {author} {\bibfnamefont {Kathryn~M.}\ \bibnamefont
  {Zurek}},\ }\bibfield  {title} {\enquote {\bibinfo {title} {{Directional
  detection of dark matter with two-dimensional targets}},}\ }\href {\doibase
  10.1016/j.physletb.2017.06.051} {\bibfield  {journal} {\bibinfo  {journal}
  {Phys. Lett.}\ }\textbf {\bibinfo {volume} {B772}},\ \bibinfo {pages}
  {239--246} (\bibinfo {year} {2017})},\ \Eprint
  {http://arxiv.org/abs/1606.08849} {arXiv:1606.08849 [hep-ph]} \BibitemShut
  {NoStop}%
\bibitem [{\citenamefont {Catena}\ \emph
  {et~al.}(2023{\natexlab{a}})\citenamefont {Catena}, \citenamefont {Emken},
  \citenamefont {Matas}, \citenamefont {Spaldin},\ and\ \citenamefont
  {Urdshals}}]{Catena:2023qkj}%
  \BibitemOpen
  \bibfield  {author} {\bibinfo {author} {\bibfnamefont {Riccardo}\
  \bibnamefont {Catena}}, \bibinfo {author} {\bibfnamefont {Timon}\
  \bibnamefont {Emken}}, \bibinfo {author} {\bibfnamefont {Marek}\ \bibnamefont
  {Matas}}, \bibinfo {author} {\bibfnamefont {Nicola~A.}\ \bibnamefont
  {Spaldin}}, \ and\ \bibinfo {author} {\bibfnamefont {Einar}\ \bibnamefont
  {Urdshals}},\ }\bibfield  {title} {\enquote {\bibinfo {title} {{Direct
  searches for general dark matter-electron interactions with graphene
  detectors: Part I. Electronic structure calculations}},}\ }\href {\doibase
  10.1103/PhysRevResearch.5.043257} {\bibfield  {journal} {\bibinfo  {journal}
  {Phys. Rev. Res.}\ }\textbf {\bibinfo {volume} {5}},\ \bibinfo {pages}
  {043257} (\bibinfo {year} {2023}{\natexlab{a}})},\ \Eprint
  {http://arxiv.org/abs/2303.15497} {arXiv:2303.15497 [hep-ph]} \BibitemShut
  {NoStop}%
\bibitem [{\citenamefont {Cavoto}\ \emph {et~al.}(2020)\citenamefont {Cavoto},
  \citenamefont {Betti}, \citenamefont {Mariani}, \citenamefont {Pandolfi},
  \citenamefont {Polosa}, \citenamefont {Rago},\ and\ \citenamefont
  {Ruocco}}]{Cavoto:2019flp}%
  \BibitemOpen
  \bibfield  {author} {\bibinfo {author} {\bibfnamefont {G.}~\bibnamefont
  {Cavoto}}, \bibinfo {author} {\bibfnamefont {M.~G.}\ \bibnamefont {Betti}},
  \bibinfo {author} {\bibfnamefont {C.}~\bibnamefont {Mariani}}, \bibinfo
  {author} {\bibfnamefont {F.}~\bibnamefont {Pandolfi}}, \bibinfo {author}
  {\bibfnamefont {A.~D.}\ \bibnamefont {Polosa}}, \bibinfo {author}
  {\bibfnamefont {I.}~\bibnamefont {Rago}}, \ and\ \bibinfo {author}
  {\bibfnamefont {A.}~\bibnamefont {Ruocco}},\ }\bibfield  {title} {\enquote
  {\bibinfo {title} {{Carbon nanotubes as anisotropic target for dark
  matter}},}\ }\href {\doibase 10.1088/1742-6596/1468/1/012232} {\bibfield
  {journal} {\bibinfo  {journal} {J. Phys. Conf. Ser.}\ }\textbf {\bibinfo
  {volume} {1468}},\ \bibinfo {pages} {012232} (\bibinfo {year} {2020})},\
  \Eprint {http://arxiv.org/abs/1911.01122} {arXiv:1911.01122
  [physics.ins-det]} \BibitemShut {NoStop}%
\bibitem [{\citenamefont {Catena}\ \emph
  {et~al.}(2023{\natexlab{b}})\citenamefont {Catena}, \citenamefont {Emken},
  \citenamefont {Matas}, \citenamefont {Spaldin},\ and\ \citenamefont
  {Urdshals}}]{Catena:2023awl}%
  \BibitemOpen
  \bibfield  {author} {\bibinfo {author} {\bibfnamefont {Riccardo}\
  \bibnamefont {Catena}}, \bibinfo {author} {\bibfnamefont {Timon}\
  \bibnamefont {Emken}}, \bibinfo {author} {\bibfnamefont {Marek}\ \bibnamefont
  {Matas}}, \bibinfo {author} {\bibfnamefont {Nicola~A.}\ \bibnamefont
  {Spaldin}}, \ and\ \bibinfo {author} {\bibfnamefont {Einar}\ \bibnamefont
  {Urdshals}},\ }\bibfield  {title} {\enquote {\bibinfo {title} {{Direct
  searches for general dark matter-electron interactions with graphene
  detectors: Part II. Sensitivity studies}},}\ }\href {\doibase
  10.1103/PhysRevResearch.5.043258} {\bibfield  {journal} {\bibinfo  {journal}
  {Phys. Rev. Res.}\ }\textbf {\bibinfo {volume} {5}},\ \bibinfo {pages}
  {043258} (\bibinfo {year} {2023}{\natexlab{b}})},\ \Eprint
  {http://arxiv.org/abs/2303.15509} {arXiv:2303.15509 [hep-ph]} \BibitemShut
  {NoStop}%
\bibitem [{\citenamefont {Knapen}\ \emph {et~al.}(2018)\citenamefont {Knapen},
  \citenamefont {Lin}, \citenamefont {Pyle},\ and\ \citenamefont
  {Zurek}}]{Knapen:2017ekk}%
  \BibitemOpen
  \bibfield  {author} {\bibinfo {author} {\bibfnamefont {Simon}\ \bibnamefont
  {Knapen}}, \bibinfo {author} {\bibfnamefont {Tongyan}\ \bibnamefont {Lin}},
  \bibinfo {author} {\bibfnamefont {Matt}\ \bibnamefont {Pyle}}, \ and\
  \bibinfo {author} {\bibfnamefont {Kathryn~M.}\ \bibnamefont {Zurek}},\
  }\bibfield  {title} {\enquote {\bibinfo {title} {{Detection of Light Dark
  Matter With Optical Phonons in Polar Materials}},}\ }\href {\doibase
  10.1016/j.physletb.2018.08.064} {\bibfield  {journal} {\bibinfo  {journal}
  {Phys. Lett.}\ }\textbf {\bibinfo {volume} {B785}},\ \bibinfo {pages}
  {386--390} (\bibinfo {year} {2018})},\ \Eprint
  {http://arxiv.org/abs/1712.06598} {arXiv:1712.06598 [hep-ph]} \BibitemShut
  {NoStop}%
\bibitem [{\citenamefont {Trickle}\ \emph
  {et~al.}(2020{\natexlab{a}})\citenamefont {Trickle}, \citenamefont {Zhang},
  \citenamefont {Zurek}, \citenamefont {Inzani},\ and\ \citenamefont
  {Griffin}}]{Trickle:2019nya}%
  \BibitemOpen
  \bibfield  {author} {\bibinfo {author} {\bibfnamefont {Tanner}\ \bibnamefont
  {Trickle}}, \bibinfo {author} {\bibfnamefont {Zhengkang}\ \bibnamefont
  {Zhang}}, \bibinfo {author} {\bibfnamefont {Kathryn~M.}\ \bibnamefont
  {Zurek}}, \bibinfo {author} {\bibfnamefont {Katherine}\ \bibnamefont
  {Inzani}}, \ and\ \bibinfo {author} {\bibfnamefont {Sin\'ead~M.}\
  \bibnamefont {Griffin}},\ }\bibfield  {title} {\enquote {\bibinfo {title}
  {{Multi-Channel Direct Detection of Light Dark Matter: Theoretical
  Framework}},}\ }\href {\doibase 10.1007/JHEP03(2020)036} {\bibfield
  {journal} {\bibinfo  {journal} {JHEP}\ }\textbf {\bibinfo {volume} {03}},\
  \bibinfo {pages} {036} (\bibinfo {year} {2020}{\natexlab{a}})},\ \Eprint
  {http://arxiv.org/abs/1910.08092} {arXiv:1910.08092 [hep-ph]} \BibitemShut
  {NoStop}%
\bibitem [{\citenamefont {Trickle}\ \emph
  {et~al.}(2020{\natexlab{b}})\citenamefont {Trickle}, \citenamefont {Zhang},\
  and\ \citenamefont {Zurek}}]{Trickle:2019ovy}%
  \BibitemOpen
  \bibfield  {author} {\bibinfo {author} {\bibfnamefont {Tanner}\ \bibnamefont
  {Trickle}}, \bibinfo {author} {\bibfnamefont {Zhengkang}\ \bibnamefont
  {Zhang}}, \ and\ \bibinfo {author} {\bibfnamefont {Kathryn~M.}\ \bibnamefont
  {Zurek}},\ }\bibfield  {title} {\enquote {\bibinfo {title} {{Detecting Light
  Dark Matter with Magnons}},}\ }\href {\doibase
  10.1103/PhysRevLett.124.201801} {\bibfield  {journal} {\bibinfo  {journal}
  {Phys. Rev. Lett.}\ }\textbf {\bibinfo {volume} {124}},\ \bibinfo {pages}
  {201801} (\bibinfo {year} {2020}{\natexlab{b}})},\ \Eprint
  {http://arxiv.org/abs/1905.13744} {arXiv:1905.13744 [hep-ph]} \BibitemShut
  {NoStop}%
\bibitem [{\citenamefont {Kahn}\ and\ \citenamefont
  {Lin}(2022)}]{Kahn:2021ttr}%
  \BibitemOpen
  \bibfield  {author} {\bibinfo {author} {\bibfnamefont {Yonatan}\ \bibnamefont
  {Kahn}}\ and\ \bibinfo {author} {\bibfnamefont {Tongyan}\ \bibnamefont
  {Lin}},\ }\bibfield  {title} {\enquote {\bibinfo {title} {{Searches for light
  dark matter using condensed matter systems}},}\ }\href {\doibase
  10.1088/1361-6633/ac5f63} {\bibfield  {journal} {\bibinfo  {journal} {Rept.
  Prog. Phys.}\ }\textbf {\bibinfo {volume} {85}},\ \bibinfo {pages} {066901}
  (\bibinfo {year} {2022})},\ \Eprint {http://arxiv.org/abs/2108.03239}
  {arXiv:2108.03239 [hep-ph]} \BibitemShut {NoStop}%
\bibitem [{\citenamefont {Fayet}(1980)}]{Fayet:1980ad}%
  \BibitemOpen
  \bibfield  {author} {\bibinfo {author} {\bibfnamefont {Pierre}\ \bibnamefont
  {Fayet}},\ }\bibfield  {title} {\enquote {\bibinfo {title} {{Effects of the
  Spin 1 Partner of the Goldstino (Gravitino) on Neutral Current
  Phenomenology}},}\ }\href {\doibase 10.1016/0370-2693(80)90488-8} {\bibfield
  {journal} {\bibinfo  {journal} {Phys. Lett. B}\ }\textbf {\bibinfo {volume}
  {95}},\ \bibinfo {pages} {285--289} (\bibinfo {year} {1980})}\BibitemShut
  {NoStop}%
\bibitem [{\citenamefont {Holdom}(1986)}]{Holdom:1985ag}%
  \BibitemOpen
  \bibfield  {author} {\bibinfo {author} {\bibfnamefont {Bob}\ \bibnamefont
  {Holdom}},\ }\bibfield  {title} {\enquote {\bibinfo {title} {{Two U(1)'s and
  Epsilon Charge Shifts}},}\ }\href {\doibase 10.1016/0370-2693(86)91377-8}
  {\bibfield  {journal} {\bibinfo  {journal} {Phys. Lett.}\ }\textbf {\bibinfo
  {volume} {166B}},\ \bibinfo {pages} {196--198} (\bibinfo {year}
  {1986})}\BibitemShut {NoStop}%
\bibitem [{\citenamefont {Boehm}\ and\ \citenamefont
  {Fayet}(2004)}]{Boehm:2003hm}%
  \BibitemOpen
  \bibfield  {author} {\bibinfo {author} {\bibfnamefont {C.}~\bibnamefont
  {Boehm}}\ and\ \bibinfo {author} {\bibfnamefont {Pierre}\ \bibnamefont
  {Fayet}},\ }\bibfield  {title} {\enquote {\bibinfo {title} {{Scalar dark
  matter candidates}},}\ }\href {\doibase 10.1016/j.nuclphysb.2004.01.015}
  {\bibfield  {journal} {\bibinfo  {journal} {Nucl. Phys. B}\ }\textbf
  {\bibinfo {volume} {683}},\ \bibinfo {pages} {219--263} (\bibinfo {year}
  {2004})},\ \Eprint {http://arxiv.org/abs/hep-ph/0305261}
  {arXiv:hep-ph/0305261} \BibitemShut {NoStop}%
\bibitem [{\citenamefont {Fayet}(2004)}]{Fayet:2004bw}%
  \BibitemOpen
  \bibfield  {author} {\bibinfo {author} {\bibfnamefont {Pierre}\ \bibnamefont
  {Fayet}},\ }\bibfield  {title} {\enquote {\bibinfo {title} {{Light spin 1/2
  or spin 0 dark matter particles}},}\ }\href {\doibase
  10.1103/PhysRevD.70.023514} {\bibfield  {journal} {\bibinfo  {journal} {Phys.
  Rev. D}\ }\textbf {\bibinfo {volume} {70}},\ \bibinfo {pages} {023514}
  (\bibinfo {year} {2004})},\ \Eprint {http://arxiv.org/abs/hep-ph/0403226}
  {arXiv:hep-ph/0403226} \BibitemShut {NoStop}%
\bibitem [{\citenamefont {Battaglieri}\ \emph {et~al.}(2017)\citenamefont
  {Battaglieri} \emph {et~al.}}]{Battaglieri:2017aum}%
  \BibitemOpen
  \bibfield  {author} {\bibinfo {author} {\bibfnamefont {Marco}\ \bibnamefont
  {Battaglieri}} \emph {et~al.},\ }\bibfield  {title} {\enquote {\bibinfo
  {title} {{US Cosmic Visions: New Ideas in Dark Matter 2017; Community
  Report}},}\ }\href@noop {} {\bibfield  {journal} {\bibinfo  {journal}
  {FERMILAB-CONF-17-282-AE-PPD-T}\ } (\bibinfo {year} {2017})},\ \Eprint
  {http://arxiv.org/abs/1707.04591} {arXiv:1707.04591 [hep-ph]} \BibitemShut
  {NoStop}%
\bibitem [{\citenamefont {Kopp}\ \emph {et~al.}(2009)\citenamefont {Kopp},
  \citenamefont {Niro}, \citenamefont {Schwetz},\ and\ \citenamefont
  {Zupan}}]{Kopp:2009et}%
  \BibitemOpen
  \bibfield  {author} {\bibinfo {author} {\bibfnamefont {Joachim}\ \bibnamefont
  {Kopp}}, \bibinfo {author} {\bibfnamefont {Viviana}\ \bibnamefont {Niro}},
  \bibinfo {author} {\bibfnamefont {Thomas}\ \bibnamefont {Schwetz}}, \ and\
  \bibinfo {author} {\bibfnamefont {Jure}\ \bibnamefont {Zupan}},\ }\bibfield
  {title} {\enquote {\bibinfo {title} {{DAMA/LIBRA and leptonically interacting
  Dark Matter}},}\ }\href {\doibase 10.1103/PhysRevD.80.083502} {\bibfield
  {journal} {\bibinfo  {journal} {Phys. Rev.}\ }\textbf {\bibinfo {volume}
  {D80}},\ \bibinfo {pages} {083502} (\bibinfo {year} {2009})},\ \Eprint
  {http://arxiv.org/abs/0907.3159} {arXiv:0907.3159 [hep-ph]} \BibitemShut
  {NoStop}%
\bibitem [{\citenamefont {Pandey}\ \emph {et~al.}(2018)\citenamefont {Pandey},
  \citenamefont {Singh}, \citenamefont {Wu}, \citenamefont {Chen},
  \citenamefont {Chi}, \citenamefont {Hsieh}, \citenamefont {Liu},\ and\
  \citenamefont {Wong}}]{Pandey:2018esq}%
  \BibitemOpen
  \bibfield  {author} {\bibinfo {author} {\bibfnamefont {Mukesh~K.}\
  \bibnamefont {Pandey}}, \bibinfo {author} {\bibfnamefont {Lakhwinder}\
  \bibnamefont {Singh}}, \bibinfo {author} {\bibfnamefont {Chih-Pan}\
  \bibnamefont {Wu}}, \bibinfo {author} {\bibfnamefont {Jiunn-Wei}\
  \bibnamefont {Chen}}, \bibinfo {author} {\bibfnamefont {Hsin-Chang}\
  \bibnamefont {Chi}}, \bibinfo {author} {\bibfnamefont {Chung-Chun}\
  \bibnamefont {Hsieh}}, \bibinfo {author} {\bibfnamefont {C.~P}\ \bibnamefont
  {Liu}}, \ and\ \bibinfo {author} {\bibfnamefont {Henry~T.}\ \bibnamefont
  {Wong}},\ }\bibfield  {title} {\enquote {\bibinfo {title} {{Constraints on
  spin-independent dark matter scattering off electrons with germanium and
  xenon detectors}},}\ }\href@noop {} {\  (\bibinfo {year} {2018})},\ \Eprint
  {http://arxiv.org/abs/1812.11759} {arXiv:1812.11759 [hep-ph]} \BibitemShut
  {NoStop}%
\bibitem [{\citenamefont {Roberts}\ \emph {et~al.}(2016)\citenamefont
  {Roberts}, \citenamefont {Dzuba}, \citenamefont {Flambaum}, \citenamefont
  {Pospelov},\ and\ \citenamefont {Stadnik}}]{Roberts:2016xfw}%
  \BibitemOpen
  \bibfield  {author} {\bibinfo {author} {\bibfnamefont {B.~M.}\ \bibnamefont
  {Roberts}}, \bibinfo {author} {\bibfnamefont {V.~A.}\ \bibnamefont {Dzuba}},
  \bibinfo {author} {\bibfnamefont {V.~V.}\ \bibnamefont {Flambaum}}, \bibinfo
  {author} {\bibfnamefont {M.}~\bibnamefont {Pospelov}}, \ and\ \bibinfo
  {author} {\bibfnamefont {Y.~V.}\ \bibnamefont {Stadnik}},\ }\bibfield
  {title} {\enquote {\bibinfo {title} {{Dark matter scattering on electrons:
  Accurate calculations of atomic excitations and implications for the DAMA
  signal}},}\ }\href {\doibase 10.1103/PhysRevD.93.115037} {\bibfield
  {journal} {\bibinfo  {journal} {Phys. Rev.}\ }\textbf {\bibinfo {volume}
  {D93}},\ \bibinfo {pages} {115037} (\bibinfo {year} {2016})},\ \Eprint
  {http://arxiv.org/abs/1604.04559} {arXiv:1604.04559 [hep-ph]} \BibitemShut
  {NoStop}%
\bibitem [{\citenamefont {Dreyer}\ \emph {et~al.}(2023)\citenamefont {Dreyer},
  \citenamefont {Essig}, \citenamefont {Fernandez-Serra}, \citenamefont
  {Singal},\ and\ \citenamefont {Zhen}}]{Dreyer:2023ovn}%
  \BibitemOpen
  \bibfield  {author} {\bibinfo {author} {\bibfnamefont {Cyrus~E.}\
  \bibnamefont {Dreyer}}, \bibinfo {author} {\bibfnamefont {Rouven}\
  \bibnamefont {Essig}}, \bibinfo {author} {\bibfnamefont {Marivi}\
  \bibnamefont {Fernandez-Serra}}, \bibinfo {author} {\bibfnamefont {Aman}\
  \bibnamefont {Singal}}, \ and\ \bibinfo {author} {\bibfnamefont {Cheng}\
  \bibnamefont {Zhen}},\ }\bibfield  {title} {\enquote {\bibinfo {title}
  {{Fully ab-initio all-electron calculation of dark matter--electron
  scattering in crystals with evaluation of systematic uncertainties}},}\
  }\href@noop {} {\  (\bibinfo {year} {2023})},\ \Eprint
  {http://arxiv.org/abs/2306.14944} {arXiv:2306.14944 [hep-ph]} \BibitemShut
  {NoStop}%
\bibitem [{\citenamefont {Catena}\ \emph
  {et~al.}(2023{\natexlab{c}})\citenamefont {Catena}, \citenamefont {Cole},
  \citenamefont {Emken}, \citenamefont {Matas}, \citenamefont {Spaldin},
  \citenamefont {Tarantino},\ and\ \citenamefont {Urdshals}}]{Catena:2022fnk}%
  \BibitemOpen
  \bibfield  {author} {\bibinfo {author} {\bibfnamefont {Riccardo}\
  \bibnamefont {Catena}}, \bibinfo {author} {\bibfnamefont {Daniel}\
  \bibnamefont {Cole}}, \bibinfo {author} {\bibfnamefont {Timon}\ \bibnamefont
  {Emken}}, \bibinfo {author} {\bibfnamefont {Marek}\ \bibnamefont {Matas}},
  \bibinfo {author} {\bibfnamefont {Nicola}\ \bibnamefont {Spaldin}}, \bibinfo
  {author} {\bibfnamefont {Walter}\ \bibnamefont {Tarantino}}, \ and\ \bibinfo
  {author} {\bibfnamefont {Einar}\ \bibnamefont {Urdshals}},\ }\bibfield
  {title} {\enquote {\bibinfo {title} {{Dark matter-electron interactions in
  materials beyond the dark photon model}},}\ }\href {\doibase
  10.1088/1475-7516/2023/03/052} {\bibfield  {journal} {\bibinfo  {journal}
  {JCAP}\ }\textbf {\bibinfo {volume} {03}},\ \bibinfo {pages} {052} (\bibinfo
  {year} {2023}{\natexlab{c}})},\ \Eprint {http://arxiv.org/abs/2210.07305}
  {arXiv:2210.07305 [hep-ph]} \BibitemShut {NoStop}%
\bibitem [{\citenamefont {Fan}\ \emph {et~al.}(2010)\citenamefont {Fan},
  \citenamefont {Reece},\ and\ \citenamefont {Wang}}]{Fan:2010gt}%
  \BibitemOpen
  \bibfield  {author} {\bibinfo {author} {\bibfnamefont {JiJi}\ \bibnamefont
  {Fan}}, \bibinfo {author} {\bibfnamefont {Matthew}\ \bibnamefont {Reece}}, \
  and\ \bibinfo {author} {\bibfnamefont {Lian-Tao}\ \bibnamefont {Wang}},\
  }\bibfield  {title} {\enquote {\bibinfo {title} {{Non-relativistic effective
  theory of dark matter direct detection}},}\ }\href {\doibase
  10.1088/1475-7516/2010/11/042} {\bibfield  {journal} {\bibinfo  {journal}
  {JCAP}\ }\textbf {\bibinfo {volume} {1011}},\ \bibinfo {pages} {042}
  (\bibinfo {year} {2010})},\ \Eprint {http://arxiv.org/abs/1008.1591}
  {arXiv:1008.1591 [hep-ph]} \BibitemShut {NoStop}%
\bibitem [{\citenamefont {Anand}\ \emph {et~al.}(2014)\citenamefont {Anand},
  \citenamefont {Fitzpatrick},\ and\ \citenamefont {Haxton}}]{Anand:2013yka}%
  \BibitemOpen
  \bibfield  {author} {\bibinfo {author} {\bibfnamefont {Nikhil}\ \bibnamefont
  {Anand}}, \bibinfo {author} {\bibfnamefont {A.~Liam}\ \bibnamefont
  {Fitzpatrick}}, \ and\ \bibinfo {author} {\bibfnamefont {W.~C.}\ \bibnamefont
  {Haxton}},\ }\bibfield  {title} {\enquote {\bibinfo {title} {{Weakly
  interacting massive particle-nucleus elastic scattering response}},}\ }\href
  {\doibase 10.1103/PhysRevC.89.065501} {\bibfield  {journal} {\bibinfo
  {journal} {Phys. Rev.}\ }\textbf {\bibinfo {volume} {C89}},\ \bibinfo {pages}
  {065501} (\bibinfo {year} {2014})},\ \Eprint {http://arxiv.org/abs/1308.6288}
  {arXiv:1308.6288 [hep-ph]} \BibitemShut {NoStop}%
\bibitem [{\citenamefont
  {S{\ifmmode\acute{o}\else\'{o}\fi}lyom}(2007)}]{Solyom2007}%
  \BibitemOpen
  \bibfield  {author} {\bibinfo {author} {\bibfnamefont
  {Jen{\ifmmode\mbox{\H{o}}\else\H{o}\fi}}\ \bibnamefont
  {S{\ifmmode\acute{o}\else\'{o}\fi}lyom}},\ }\href
  {https://link.springer.com/book/10.1007/978-3-540-72600-5} {\emph {\bibinfo
  {title} {{Fundamentals of the Physics of Solids}}}}\ (\bibinfo  {publisher}
  {Springer},\ \bibinfo {address} {Berlin, Germany},\ \bibinfo {year}
  {2007})\BibitemShut {NoStop}%
\bibitem [{\citenamefont {Weissker}\ \emph {et~al.}(2010)\citenamefont
  {Weissker}, \citenamefont {Serrano}, \citenamefont {Huotari}, \citenamefont
  {Luppi}, \citenamefont {Cazzaniga}, \citenamefont {Bruneval}, \citenamefont
  {Sottile}, \citenamefont {Monaco}, \citenamefont {Olevano},\ and\
  \citenamefont {Reining}}]{Weissker2010Feb}%
  \BibitemOpen
  \bibfield  {author} {\bibinfo {author} {\bibfnamefont {Hans-Christian}\
  \bibnamefont {Weissker}}, \bibinfo {author} {\bibfnamefont {Jorge}\
  \bibnamefont {Serrano}}, \bibinfo {author} {\bibfnamefont {Simo}\
  \bibnamefont {Huotari}}, \bibinfo {author} {\bibfnamefont {Eleonora}\
  \bibnamefont {Luppi}}, \bibinfo {author} {\bibfnamefont {Marco}\ \bibnamefont
  {Cazzaniga}}, \bibinfo {author} {\bibfnamefont {Fabien}\ \bibnamefont
  {Bruneval}}, \bibinfo {author} {\bibfnamefont {Francesco}\ \bibnamefont
  {Sottile}}, \bibinfo {author} {\bibfnamefont {Giulio}\ \bibnamefont
  {Monaco}}, \bibinfo {author} {\bibfnamefont {Valerio}\ \bibnamefont
  {Olevano}}, \ and\ \bibinfo {author} {\bibfnamefont {Lucia}\ \bibnamefont
  {Reining}},\ }\bibfield  {title} {\enquote {\bibinfo {title} {{Dynamic
  structure factor and dielectric function of silicon for finite momentum
  transfer: Inelastic x-ray scattering experiments and ab initio
  calculations}},}\ }\href {\doibase 10.1103/PhysRevB.81.085104} {\bibfield
  {journal} {\bibinfo  {journal} {Phys. Rev. B}\ }\textbf {\bibinfo {volume}
  {81}},\ \bibinfo {pages} {085104} (\bibinfo {year} {2010})}\BibitemShut
  {NoStop}%
\bibitem [{\citenamefont {Baxter}\ \emph {et~al.}(2021)\citenamefont {Baxter}
  \emph {et~al.}}]{Baxter:2021pqo}%
  \BibitemOpen
  \bibfield  {author} {\bibinfo {author} {\bibfnamefont {D.}~\bibnamefont
  {Baxter}} \emph {et~al.},\ }\bibfield  {title} {\enquote {\bibinfo {title}
  {{Recommended conventions for reporting results from direct dark matter
  searches}},}\ }\href {\doibase 10.1140/epjc/s10052-021-09655-y} {\bibfield
  {journal} {\bibinfo  {journal} {Eur. Phys. J. C}\ }\textbf {\bibinfo {volume}
  {81}},\ \bibinfo {pages} {907} (\bibinfo {year} {2021})},\ \Eprint
  {http://arxiv.org/abs/2105.00599} {arXiv:2105.00599 [hep-ex]} \BibitemShut
  {NoStop}%
\bibitem [{\citenamefont {Catena}\ and\ \citenamefont
  {Ullio}(2010)}]{Catena:2009mf}%
  \BibitemOpen
  \bibfield  {author} {\bibinfo {author} {\bibfnamefont {Riccardo}\
  \bibnamefont {Catena}}\ and\ \bibinfo {author} {\bibfnamefont {Piero}\
  \bibnamefont {Ullio}},\ }\bibfield  {title} {\enquote {\bibinfo {title} {{A
  novel determination of the local dark matter density}},}\ }\href {\doibase
  10.1088/1475-7516/2010/08/004} {\bibfield  {journal} {\bibinfo  {journal}
  {JCAP}\ }\textbf {\bibinfo {volume} {1008}},\ \bibinfo {pages} {004}
  (\bibinfo {year} {2010})},\ \Eprint {http://arxiv.org/abs/0907.0018}
  {arXiv:0907.0018 [astro-ph.CO]} \BibitemShut {NoStop}%
\bibitem [{\citenamefont {Knapen}\ \emph {et~al.}(2022)\citenamefont {Knapen},
  \citenamefont {Kozaczuk},\ and\ \citenamefont {Lin}}]{Knapen:2021bwg}%
  \BibitemOpen
  \bibfield  {author} {\bibinfo {author} {\bibfnamefont {Simon}\ \bibnamefont
  {Knapen}}, \bibinfo {author} {\bibfnamefont {Jonathan}\ \bibnamefont
  {Kozaczuk}}, \ and\ \bibinfo {author} {\bibfnamefont {Tongyan}\ \bibnamefont
  {Lin}},\ }\bibfield  {title} {\enquote {\bibinfo {title} {{python package for
  dark matter scattering in dielectric targets}},}\ }\href {\doibase
  10.1103/PhysRevD.105.015014} {\bibfield  {journal} {\bibinfo  {journal}
  {Phys. Rev. D}\ }\textbf {\bibinfo {volume} {105}},\ \bibinfo {pages}
  {015014} (\bibinfo {year} {2022})},\ \Eprint
  {http://arxiv.org/abs/2104.12786} {arXiv:2104.12786 [hep-ph]} \BibitemShut
  {NoStop}%
\bibitem [{\citenamefont {Boyd}\ \emph {et~al.}(2023)\citenamefont {Boyd},
  \citenamefont {Hochberg}, \citenamefont {Kahn}, \citenamefont {Kramer},
  \citenamefont {Kurinsky}, \citenamefont {Lehmann},\ and\ \citenamefont
  {Yu}}]{Boyd:2022tcn}%
  \BibitemOpen
  \bibfield  {author} {\bibinfo {author} {\bibfnamefont {Christian}\
  \bibnamefont {Boyd}}, \bibinfo {author} {\bibfnamefont {Yonit}\ \bibnamefont
  {Hochberg}}, \bibinfo {author} {\bibfnamefont {Yonatan}\ \bibnamefont
  {Kahn}}, \bibinfo {author} {\bibfnamefont {Eric~David}\ \bibnamefont
  {Kramer}}, \bibinfo {author} {\bibfnamefont {Noah}\ \bibnamefont {Kurinsky}},
  \bibinfo {author} {\bibfnamefont {Benjamin~V.}\ \bibnamefont {Lehmann}}, \
  and\ \bibinfo {author} {\bibfnamefont {To~Chin}\ \bibnamefont {Yu}},\
  }\bibfield  {title} {\enquote {\bibinfo {title} {{Directional detection of
  dark matter with anisotropic response functions}},}\ }\href {\doibase
  10.1103/PhysRevD.108.015015} {\bibfield  {journal} {\bibinfo  {journal}
  {Phys. Rev. D}\ }\textbf {\bibinfo {volume} {108}},\ \bibinfo {pages}
  {015015} (\bibinfo {year} {2023})},\ \Eprint
  {http://arxiv.org/abs/2212.04505} {arXiv:2212.04505 [hep-ph]} \BibitemShut
  {NoStop}%
\bibitem [{\citenamefont {Hubbard}(1958)}]{Hubbard1958Jan}%
  \BibitemOpen
  \bibfield  {author} {\bibinfo {author} {\bibfnamefont {J.}~\bibnamefont
  {Hubbard}},\ }\bibfield  {title} {\enquote {\bibinfo {title} {{The
  description of collective motions in terms of many-body perturbation theory.
  II. The correlation energy of a free-electron gas}},}\ }\href {\doibase
  10.1098/rspa.1958.0003} {\bibfield  {journal} {\bibinfo  {journal} {Proc. R.
  Soc. London A - Math. Phys. Sci.}\ }\textbf {\bibinfo {volume} {243}},\
  \bibinfo {pages} {336--352} (\bibinfo {year} {1958})}\BibitemShut {NoStop}%
\bibitem [{\citenamefont
  {S{\ifmmode\acute{o}\else\'{o}\fi}lyom}(2010{\natexlab{a}})}]{Solyom2010Dec}%
  \BibitemOpen
  \bibfield  {author} {\bibinfo {author} {\bibfnamefont
  {Jen{\ifmmode\ddot{o}\else\"{o}\fi}}\ \bibnamefont
  {S{\ifmmode\acute{o}\else\'{o}\fi}lyom}},\ }\href
  {https://books.google.se/books/about/Fundamentals_of_the_Physics_of_Solids.html?id=rL5eGGiY1WUC&redir_esc=y}
  {\emph {\bibinfo {title} {{Fundamentals of the Physics of Solids: Volume 3 -
  Normal, Broken-Symmetry, and Correlated Systems}}}}\ (\bibinfo  {publisher}
  {Springer},\ \bibinfo {address} {Berlin, Germany},\ \bibinfo {year}
  {2010})\BibitemShut {NoStop}%
\bibitem [{\citenamefont
  {S{\ifmmode\acute{o}\else\'{o}\fi}lyom}(2010{\natexlab{b}})}]{Solyom2010}%
  \BibitemOpen
  \bibfield  {author} {\bibinfo {author} {\bibfnamefont
  {Jen{\ifmmode\mbox{\H{o}}\else\H{o}\fi}}\ \bibnamefont
  {S{\ifmmode\acute{o}\else\'{o}\fi}lyom}},\ }\href
  {https://link.springer.com/book/10.1007/978-3-642-04518-9} {\emph {\bibinfo
  {title} {{Fundamentals of the Physics of Solids, Volume 3}}}}\ (\bibinfo
  {publisher} {Springer},\ \bibinfo {address} {Berlin, Germany},\ \bibinfo
  {year} {2010})\BibitemShut {NoStop}%
\bibitem [{\citenamefont {Urdshals}\ and\ \citenamefont
  {Matas}(2021)}]{Urdshals2021May}%
  \BibitemOpen
  \bibfield  {author} {\bibinfo {author} {\bibfnamefont {Einar}\ \bibnamefont
  {Urdshals}}\ and\ \bibinfo {author} {\bibfnamefont {Marek}\ \bibnamefont
  {Matas}},\ }\bibfield  {title} {\enquote {\bibinfo {title} {{QEdark-EFT}},}\
  }\href {\doibase 10.5281/zenodo.4739187} {\bibfield  {journal} {\bibinfo
  {journal} {Zenodo}\ } (\bibinfo {year} {2021}),\
  10.5281/zenodo.4739187}\BibitemShut {NoStop}%
\bibitem [{\citenamefont {Giannozzi}\ \emph {et~al.}(2009)\citenamefont
  {Giannozzi}, \citenamefont {Baroni}, \citenamefont {Bonini}, \citenamefont
  {Calandra}, \citenamefont {Car}, \citenamefont {Cavazzoni}, \citenamefont
  {Ceresoli}, \citenamefont {Chiarotti}, \citenamefont {Cococcioni},
  \citenamefont {Dabo}, \citenamefont {Dal~Corso}, \citenamefont
  {de~Gironcoli}, \citenamefont {Fabris}, \citenamefont {Fratesi},
  \citenamefont {Gebauer}, \citenamefont {Gerstmann}, \citenamefont
  {Gougoussis}, \citenamefont {Kokalj}, \citenamefont {Lazzeri}, \citenamefont
  {Martin-Samos}, \citenamefont {Marzari}, \citenamefont {Mauri}, \citenamefont
  {Mazzarello}, \citenamefont {Paolini}, \citenamefont {Pasquarello},
  \citenamefont {Paulatto}, \citenamefont {Sbraccia}, \citenamefont {Scandolo},
  \citenamefont {Sclauzero}, \citenamefont {Seitsonen}, \citenamefont
  {Smogunov}, \citenamefont {Umari},\ and\ \citenamefont
  {Wentzcovitch}}]{Giannozzi2009Sep}%
  \BibitemOpen
  \bibfield  {author} {\bibinfo {author} {\bibfnamefont {Paolo}\ \bibnamefont
  {Giannozzi}}, \bibinfo {author} {\bibfnamefont {Stefano}\ \bibnamefont
  {Baroni}}, \bibinfo {author} {\bibfnamefont {Nicola}\ \bibnamefont {Bonini}},
  \bibinfo {author} {\bibfnamefont {Matteo}\ \bibnamefont {Calandra}}, \bibinfo
  {author} {\bibfnamefont {Roberto}\ \bibnamefont {Car}}, \bibinfo {author}
  {\bibfnamefont {Carlo}\ \bibnamefont {Cavazzoni}}, \bibinfo {author}
  {\bibfnamefont {Davide}\ \bibnamefont {Ceresoli}}, \bibinfo {author}
  {\bibfnamefont {Guido~L.}\ \bibnamefont {Chiarotti}}, \bibinfo {author}
  {\bibfnamefont {Matteo}\ \bibnamefont {Cococcioni}}, \bibinfo {author}
  {\bibfnamefont {Ismaila}\ \bibnamefont {Dabo}}, \bibinfo {author}
  {\bibfnamefont {Andrea}\ \bibnamefont {Dal~Corso}}, \bibinfo {author}
  {\bibfnamefont {Stefano}\ \bibnamefont {de~Gironcoli}}, \bibinfo {author}
  {\bibfnamefont {Stefano}\ \bibnamefont {Fabris}}, \bibinfo {author}
  {\bibfnamefont {Guido}\ \bibnamefont {Fratesi}}, \bibinfo {author}
  {\bibfnamefont {Ralph}\ \bibnamefont {Gebauer}}, \bibinfo {author}
  {\bibfnamefont {Uwe}\ \bibnamefont {Gerstmann}}, \bibinfo {author}
  {\bibfnamefont {Christos}\ \bibnamefont {Gougoussis}}, \bibinfo {author}
  {\bibfnamefont {Anton}\ \bibnamefont {Kokalj}}, \bibinfo {author}
  {\bibfnamefont {Michele}\ \bibnamefont {Lazzeri}}, \bibinfo {author}
  {\bibfnamefont {Layla}\ \bibnamefont {Martin-Samos}}, \bibinfo {author}
  {\bibfnamefont {Nicola}\ \bibnamefont {Marzari}}, \bibinfo {author}
  {\bibfnamefont {Francesco}\ \bibnamefont {Mauri}}, \bibinfo {author}
  {\bibfnamefont {Riccardo}\ \bibnamefont {Mazzarello}}, \bibinfo {author}
  {\bibfnamefont {Stefano}\ \bibnamefont {Paolini}}, \bibinfo {author}
  {\bibfnamefont {Alfredo}\ \bibnamefont {Pasquarello}}, \bibinfo {author}
  {\bibfnamefont {Lorenzo}\ \bibnamefont {Paulatto}}, \bibinfo {author}
  {\bibfnamefont {Carlo}\ \bibnamefont {Sbraccia}}, \bibinfo {author}
  {\bibfnamefont {Sandro}\ \bibnamefont {Scandolo}}, \bibinfo {author}
  {\bibfnamefont {Gabriele}\ \bibnamefont {Sclauzero}}, \bibinfo {author}
  {\bibfnamefont {Ari~P.}\ \bibnamefont {Seitsonen}}, \bibinfo {author}
  {\bibfnamefont {Alexander}\ \bibnamefont {Smogunov}}, \bibinfo {author}
  {\bibfnamefont {Paolo}\ \bibnamefont {Umari}}, \ and\ \bibinfo {author}
  {\bibfnamefont {Renata~M.}\ \bibnamefont {Wentzcovitch}},\ }\bibfield
  {title} {\enquote {\bibinfo {title} {{QUANTUM ESPRESSO: a modular and
  open-source software project for quantum}},}\ }\href {\doibase
  10.1088/0953-8984/21/39/395502} {\bibfield  {journal} {\bibinfo  {journal}
  {J. Phys.: Condens. Matter}\ }\textbf {\bibinfo {volume} {21}},\ \bibinfo
  {pages} {395502} (\bibinfo {year} {2009})}\BibitemShut {NoStop}%
\bibitem [{\citenamefont {Mortensen}\ \emph {et~al.}(2005)\citenamefont
  {Mortensen}, \citenamefont {Hansen},\ and\ \citenamefont
  {Jacobsen}}]{Mortensen2005Jan}%
  \BibitemOpen
  \bibfield  {author} {\bibinfo {author} {\bibfnamefont {J.~J.}\ \bibnamefont
  {Mortensen}}, \bibinfo {author} {\bibfnamefont {L.~B.}\ \bibnamefont
  {Hansen}}, \ and\ \bibinfo {author} {\bibfnamefont {K.~W.}\ \bibnamefont
  {Jacobsen}},\ }\bibfield  {title} {\enquote {\bibinfo {title} {{Real-space
  grid implementation of the projector augmented wave method}},}\ }\href
  {\doibase 10.1103/PhysRevB.71.035109} {\bibfield  {journal} {\bibinfo
  {journal} {Phys. Rev. B}\ }\textbf {\bibinfo {volume} {71}},\ \bibinfo
  {pages} {035109} (\bibinfo {year} {2005})}\BibitemShut {NoStop}%
\bibitem [{\citenamefont {Del~Nobile}\ \emph {et~al.}(2013)\citenamefont
  {Del~Nobile}, \citenamefont {Gelmini}, \citenamefont {Gondolo},\ and\
  \citenamefont {Huh}}]{DelNobile:2013cva}%
  \BibitemOpen
  \bibfield  {author} {\bibinfo {author} {\bibfnamefont {Eugenio}\ \bibnamefont
  {Del~Nobile}}, \bibinfo {author} {\bibfnamefont {Graciela}\ \bibnamefont
  {Gelmini}}, \bibinfo {author} {\bibfnamefont {Paolo}\ \bibnamefont
  {Gondolo}}, \ and\ \bibinfo {author} {\bibfnamefont {Ji-Haeng}\ \bibnamefont
  {Huh}},\ }\bibfield  {title} {\enquote {\bibinfo {title} {{Generalized Halo
  Independent Comparison of Direct Dark Matter Detection Data}},}\ }\href
  {\doibase 10.1088/1475-7516/2013/10/048} {\bibfield  {journal} {\bibinfo
  {journal} {JCAP}\ }\textbf {\bibinfo {volume} {1310}},\ \bibinfo {pages}
  {048} (\bibinfo {year} {2013})},\ \Eprint {http://arxiv.org/abs/1306.5273}
  {arXiv:1306.5273 [hep-ph]} \BibitemShut {NoStop}%
\bibitem [{\citenamefont {Del~Nobile}\ \emph {et~al.}(2014)\citenamefont
  {Del~Nobile}, \citenamefont {Gelmini}, \citenamefont {Gondolo},\ and\
  \citenamefont {Huh}}]{DelNobile:2014eta}%
  \BibitemOpen
  \bibfield  {author} {\bibinfo {author} {\bibfnamefont {Eugenio}\ \bibnamefont
  {Del~Nobile}}, \bibinfo {author} {\bibfnamefont {Graciela~B.}\ \bibnamefont
  {Gelmini}}, \bibinfo {author} {\bibfnamefont {Paolo}\ \bibnamefont
  {Gondolo}}, \ and\ \bibinfo {author} {\bibfnamefont {Ji-Haeng}\ \bibnamefont
  {Huh}},\ }\bibfield  {title} {\enquote {\bibinfo {title} {{Direct detection
  of Light Anapole and Magnetic Dipole DM}},}\ }\href {\doibase
  10.1088/1475-7516/2014/06/002} {\bibfield  {journal} {\bibinfo  {journal}
  {JCAP}\ }\textbf {\bibinfo {volume} {06}},\ \bibinfo {pages} {002} (\bibinfo
  {year} {2014})},\ \Eprint {http://arxiv.org/abs/1401.4508} {arXiv:1401.4508
  [hep-ph]} \BibitemShut {NoStop}%
\bibitem [{\citenamefont {Inzani}\ \emph {et~al.}(2021)\citenamefont {Inzani},
  \citenamefont {Faghaninia},\ and\ \citenamefont {Griffin}}]{Inzani:2020szg}%
  \BibitemOpen
  \bibfield  {author} {\bibinfo {author} {\bibfnamefont {Katherine}\
  \bibnamefont {Inzani}}, \bibinfo {author} {\bibfnamefont {Alireza}\
  \bibnamefont {Faghaninia}}, \ and\ \bibinfo {author} {\bibfnamefont
  {Sin\'ead~M.}\ \bibnamefont {Griffin}},\ }\bibfield  {title} {\enquote
  {\bibinfo {title} {{Prediction of Tunable Spin-Orbit Gapped Materials for
  Dark Matter Detection}},}\ }\href {\doibase 10.1103/PhysRevResearch.3.013069}
  {\bibfield  {journal} {\bibinfo  {journal} {Phys. Rev. Res.}\ }\textbf
  {\bibinfo {volume} {3}},\ \bibinfo {pages} {013069} (\bibinfo {year}
  {2021})},\ \Eprint {http://arxiv.org/abs/2008.05062} {arXiv:2008.05062
  [cond-mat.mtrl-sci]} \BibitemShut {NoStop}%
\bibitem [{\citenamefont {Tran}\ and\ \citenamefont
  {Blaha}(2009)}]{PhysRevLett.102.226401}%
  \BibitemOpen
  \bibfield  {author} {\bibinfo {author} {\bibfnamefont {Fabien}\ \bibnamefont
  {Tran}}\ and\ \bibinfo {author} {\bibfnamefont {Peter}\ \bibnamefont
  {Blaha}},\ }\bibfield  {title} {\enquote {\bibinfo {title} {Accurate band
  gaps of semiconductors and insulators with a semilocal exchange-correlation
  potential},}\ }\href {\doibase 10.1103/PhysRevLett.102.226401} {\bibfield
  {journal} {\bibinfo  {journal} {Phys. Rev. Lett.}\ }\textbf {\bibinfo
  {volume} {102}},\ \bibinfo {pages} {226401} (\bibinfo {year}
  {2009})}\BibitemShut {NoStop}%
\bibitem [{\citenamefont {Trickle}\ \emph {et~al.}(2022)\citenamefont
  {Trickle}, \citenamefont {Zhang},\ and\ \citenamefont
  {Zurek}}]{Trickle:2020oki}%
  \BibitemOpen
  \bibfield  {author} {\bibinfo {author} {\bibfnamefont {Tanner}\ \bibnamefont
  {Trickle}}, \bibinfo {author} {\bibfnamefont {Zhengkang}\ \bibnamefont
  {Zhang}}, \ and\ \bibinfo {author} {\bibfnamefont {Kathryn~M.}\ \bibnamefont
  {Zurek}},\ }\bibfield  {title} {\enquote {\bibinfo {title} {{Effective field
  theory of dark matter direct detection with collective excitations}},}\
  }\href {\doibase 10.1103/PhysRevD.105.015001} {\bibfield  {journal} {\bibinfo
   {journal} {Phys. Rev. D}\ }\textbf {\bibinfo {volume} {105}},\ \bibinfo
  {pages} {015001} (\bibinfo {year} {2022})},\ \Eprint
  {http://arxiv.org/abs/2009.13534} {arXiv:2009.13534 [hep-ph]} \BibitemShut
  {NoStop}%
\end{thebibliography}
%

\end{document}